\documentclass[11pt,oneside,openany]{book}

\usepackage{pifont}
\usepackage{color}

\makeatletter
\def\secretfootnote{\gdef\@thefnmark{}\@footnotetext}
\makeatother

\usepackage[top=3cm,bottom=3cm,left=3cm,right=3cm,headsep=10pt,a4paper]{geometry} 

\usepackage{graphicx} 
\graphicspath{{./}} 

\usepackage{lipsum} 

\usepackage[english]{babel} 

\usepackage{enumitem} 
\setlist{nolistsep} 

\usepackage{booktabs} 

\usepackage[svgnames]{xcolor} 
\definecolor{ocre}{RGB}{179,120,211} 

\usepackage{tikz} 

\usepackage{nicefrac} 

\usepackage{pdfpages} 

\usepackage{color}
\usepackage{pifont}

\usepackage{avant} 
\usepackage{mathptmx} 

\usepackage{microtype} 
\usepackage[utf8]{inputenc} 
\usepackage[T1]{fontenc} 

\usepackage[compress]{natbib}
\setcitestyle{square,numbers,citesep={,}}

\usepackage{titletoc} 

\contentsmargin{0cm} 

\titlecontents{part}[0cm]
{\addvspace{20pt}\centering\large\bfseries}
{}
{}
{}

\titlecontents{chapter}[1.25cm] 
{\addvspace{12pt}\large\sffamily\bfseries} 
{\color{ocre!60}\contentslabel[\Large\thecontentslabel]{1.25cm}\color{ocre}} 
{\color{ocre}}  
{\color{ocre!60}\normalsize\;\titlerule*[.5pc]{.}\;\thecontentspage} 

\titlecontents{section}[1.25cm] 
{\addvspace{3pt}\sffamily\bfseries} 
{\contentslabel[\thecontentslabel]{1.25cm}} 
{}
{\hfill\color{black}\thecontentspage} 
[]

\titlecontents{subsection}[1.25cm] 
{\addvspace{1pt}\sffamily\small} 
{\contentslabel[\thecontentslabel]{1.25cm}} 
{}
{\ \titlerule*[.5pc]{.}\;\thecontentspage} 
[]

\titlecontents{figure}[0em]
{\addvspace{-5pt}\sffamily}
{\thecontentslabel\hspace*{1em}}
{}
{\ \titlerule*[.5pc]{.}\;\thecontentspage}
[]

\titlecontents{table}[0em]
{\addvspace{-5pt}\sffamily}
{\thecontentslabel\hspace*{1em}}
{}
{\ \titlerule*[.5pc]{.}\;\thecontentspage}
[]

\titlecontents{lchapter}[0em] 
{\addvspace{15pt}\large\sffamily\bfseries} 
{\color{ocre}\contentslabel[\Large\thecontentslabel]{1.25cm}\color{ocre}} 
{}  
{\color{ocre}\normalsize\sffamily\bfseries\;\titlerule*[.5pc]{.}\;\thecontentspage} 

\titlecontents{lsection}[0em] 
{\sffamily\small} 
{\contentslabel[\thecontentslabel]{1.25cm}} 
{}
{}

\titlecontents{lsubsection}[.5em] 
{\normalfont\footnotesize\sffamily} 
{}
{}
{}

\usepackage{fancyhdr} 

\renewcommand{\chaptermark}[1]{\markboth{\sffamily\normalsize\bfseries #1}{}}
\fancypagestyle{main}
{
\fancyhf{} \fancyhead[LE,RO]{\sffamily\normalsize\thepage} 
\fancyhead[LO]{\rightmark} 
\fancyhead[RE]{\leftmark} 
\addtolength{\headheight}{2.5pt} 
}
\fancypagestyle{plain}{\fancyhead{}} 

\fancypagestyle{simple}
{
\fancyhf{} \fancyhead[LE,RO]{\sffamily\normalsize\thepage} 
\addtolength{\headheight}{2.5pt} 
}

\makeatletter
\renewcommand{\cleardoublepage}{
\clearpage\ifodd\c@page\else
\hbox{}
\vspace*{\fill}
\thispagestyle{empty}
\newpage
\fi}

\usepackage{amsmath,amsfonts,amssymb,amsthm} 

\newtheoremstyle{ocrenumbox}
{0pt}
{0pt}
{\normalfont}
{}
{\small\bf\sffamily\color{ocre}}
{\;}
{0.25em}
{\small\sffamily\color{ocre}\thmname{#1}\nobreakspace\thmnumber{\@ifnotempty{#1}{}\@upn{#2}}
\thmnote{\nobreakspace\the\thm@notefont\sffamily\bfseries\color{black}---\nobreakspace#3.}} 

\newtheoremstyle{ocrebox}
{0pt}
{0pt}
{\itshape}
{0pt}
{\small\bf\sffamily\color{ocre}}
{}
{0.0em}
{\small\sffamily\color{ocre}
} 

\newtheoremstyle{blacknumex}
{5pt}
{5pt}
{\normalfont}
{} 
{\small\bf\sffamily}
{\;}
{0.25em}
{\small\sffamily{\tiny\ensuremath{\blacksquare}}\nobreakspace\thmname{#1}\nobreakspace\thmnumber{\@ifnotempty{#1}{}\@upn{#2}}
\thmnote{\nobreakspace\the\thm@notefont\sffamily\bfseries---\nobreakspace#3.}}

\newtheoremstyle{blacknumbox} 
{0pt}
{0pt}
{\normalfont}
{}
{\small\bf\sffamily}
{\;}
{0.25em}
{\small\sffamily\thmname{#1}\nobreakspace\thmnumber{\@ifnotempty{#1}{}\@upn{#2}}
\thmnote{\nobreakspace\the\thm@notefont\sffamily\bfseries---\nobreakspace#3.}}

\newtheoremstyle{blackbox} 
{0pt}
{0pt}
{\normalfont}
{0pt}
{\small\bf\sffamily}
{}
{0.0em}

\newtheoremstyle{ocrenum}
{5pt}
{5pt}
{\normalfont}
{}
{\small\bf\sffamily\color{ocre}}
{\;}
{0.25em}
{\small\sffamily\color{ocre}\thmname{#1}\nobreakspace\thmnumber{\@ifnotempty{#1}{}\@upn{#2}}
\thmnote{\nobreakspace\the\thm@notefont\sffamily\bfseries\color{black}---\nobreakspace#3.}} 
\makeatother

\newcounter{dummy} 
\numberwithin{dummy}{section}
\theoremstyle{ocrenumbox}
\newtheorem{theoremeT}[dummy]{Theorem}

\newtheorem{exerciseT}{Exercise}[chapter]
\theoremstyle{ocrebox}
\newtheorem{quoteT}{}[]
\theoremstyle{blacknumex}
\newtheorem{exampleT}{Example}[chapter]
\theoremstyle{blacknumbox}

\newtheorem{definitionT}{Definition}[section]
\newtheorem{corollaryT}[dummy]{Corollary}
\theoremstyle{ocrenum}

\theoremstyle{blackbox}
\newtheorem{newexampleT}[]{}

\RequirePackage[framemethod=default]{mdframed} 

\newmdenv[skipabove=7pt,
skipbelow=7pt,
backgroundcolor=black!5,
linecolor=ocre,
linewidth=2pt,
innerleftmargin=5pt,
innerrightmargin=5pt,
innertopmargin=5pt,
leftmargin=0cm,
rightmargin=0cm,
innerbottommargin=5pt]{tBox}

\newmdenv[skipabove=7pt,
skipbelow=7pt,
rightline=false,
leftline=true,
topline=false,
bottomline=false,
backgroundcolor=ocre!10,
linecolor=ocre,
innerleftmargin=5pt,
innerrightmargin=5pt,
innertopmargin=5pt,
innerbottommargin=5pt,
leftmargin=0cm,
rightmargin=0cm,
linewidth=4pt]{eBox}	

\newmdenv[skipabove=7pt,
skipbelow=7pt,
rightline=false,
leftline=true,
topline=false,
bottomline=false,
linecolor=ocre,
innerleftmargin=5pt,
innerrightmargin=5pt,
innertopmargin=0pt,
leftmargin=0cm,
rightmargin=0cm,
linewidth=4pt,
innerbottommargin=0pt]{dBox}	

\newmdenv[skipabove=7pt,
skipbelow=7pt,
rightline=false,
leftline=true,
topline=false,
bottomline=false,
linecolor=gray,
backgroundcolor=black!5,
innerleftmargin=5pt,
innerrightmargin=5pt,
innertopmargin=5pt,
leftmargin=0cm,
rightmargin=0cm,
linewidth=4pt,
innerbottommargin=5pt]{cBox}

\newenvironment{newquote}[2]
{\def\reference{#1}
 \def\reflink{#2}
 \begin{tBox}\begin{quoteT}}
{\hfill{\sffamily\normalfont\color{black}---\nobreakspace
{\if\relax\detokenize{\reflink}\relax
  \reference
\else
  \href{\reflink}{\reference}
\fi}
\color{ocre} \nobreakspace\tiny\ensuremath { \blacksquare}}
\end{quoteT}\end{tBox}}

\newenvironment{newexample}{\begin{cBox}\begin{newexampleT}}{\end{newexampleT}\end{cBox}}

\makeatletter
\renewcommand{\@seccntformat}[1]{\llap{\textcolor{ocre}{\csname the#1\endcsname}\hspace{1em}}}                    
\renewcommand{\section}{\@startsection{section}{1}{\z@}
{-4ex \@plus -1ex \@minus -.4ex}
{1ex \@plus.2ex }
{\normalfont\Large\scshape\sffamily\bfseries}}
\renewcommand{\subsection}{\@startsection {subsection}{2}{\z@}
{-3ex \@plus -0.1ex \@minus -.4ex}
{0.5ex \@plus.2ex }
{\normalfont\scshape\sffamily\bfseries}}
\renewcommand{\subsubsection}{\@startsection {subsubsection}{3}{\z@}
{-2ex \@plus -0.1ex \@minus -.2ex}
{.2ex \@plus.2ex }
{\normalfont\scshape\sffamily}}                        
\renewcommand{\paragraph}{\@startsection{paragraph}{4}{\z@}
{-2ex \@plus-.2ex \@minus .2ex}
{.1ex}
{\normalfont\small\sffamily\bfseries}}

\newcommand{\@mypartnumtocformat}[2]{%
\setlength\fboxsep{0pt}%
\noindent\colorbox{ocre!20}{\strut\parbox[c][.7cm]{\ecart}{\color{ocre!70}\Large\sffamily\bfseries\centering#1}}\hskip\esp\colorbox{ocre!40}{\strut\parbox[c][.7cm]{\linewidth-\ecart-\esp}{\Large\sffamily\centering#2}}}%
\newcommand{\@myparttocformat}[1]{%
\setlength\fboxsep{0pt}%
\noindent\colorbox{ocre!40}{\strut\parbox[c][.7cm]{\linewidth}{\Large\sffamily\centering#1}}}%
\newlength\esp
\newlength\ecart
\def\@part[#1]#2{%
\ifnum \c@secnumdepth >-2\relax%
\refstepcounter{part}%
\addcontentsline{toc}{part}{\texorpdfstring{\protect\@mypartnumtocformat{\thepart}{#1}}{\partname~\thepart\ ---\ #1}}
\else%
\addcontentsline{toc}{part}{\texorpdfstring{\protect\@myparttocformat{#1}}{#1}}%
\fi%
\startcontents%
\markboth{}{}%
{\thispagestyle{empty}%
\begin{tikzpicture}[remember picture,overlay]%
\node at (current page.north west){\begin{tikzpicture}[remember picture,overlay]%
\fill[ocre!20](0cm,0cm) rectangle (\paperwidth,-\paperheight);
\node[anchor=north] at (4cm,-3.25cm){\color{ocre!40}\fontsize{220}{100}\sffamily\bfseries\thepart}; 
\node[anchor=south east] at (\paperwidth-1cm,-\paperheight+1cm){\parbox[t][][t]{8.5cm}{
\printcontents{l}{0}{\setcounter{tocdepth}{1}}%
}};
\node[anchor=north east] at (\paperwidth-1.5cm,-3.25cm){\parbox[t][][t]{15cm}{\strut\raggedleft\color{white}\fontsize{30}{30}\sffamily\bfseries#2}};
\end{tikzpicture}};
\end{tikzpicture}}%
\@endpart}
\def\@spart#1{%
\startcontents%
\phantomsection
{\thispagestyle{empty}%
\begin{tikzpicture}[remember picture,overlay]%
\node at (current page.north west){\begin{tikzpicture}[remember picture,overlay]%
\fill[ocre!20](0cm,0cm) rectangle (\paperwidth,-\paperheight);
\node[anchor=north east] at (\paperwidth-1.5cm,-3.25cm){\parbox[t][][t]{15cm}{\strut\raggedleft\color{white}\fontsize{30}{30}\sffamily\bfseries#1}};
\end{tikzpicture}};
\end{tikzpicture}}
\addcontentsline{toc}{part}{\texorpdfstring{%
\setlength\fboxsep{0pt}%
\noindent\protect\colorbox{ocre!40}{\strut\protect\parbox[c][.7cm]{\linewidth}{\Large\sffamily\protect\centering #1\quad\mbox{}}}}{#1}}%
\@endpart}
\def\@endpart{\vfil\newpage
\if@twoside
\if@openright
\null
\thispagestyle{empty}%
\newpage
\fi
\fi
\if@tempswa
\twocolumn
\fi}

\newif\ifusechapterimage
\usechapterimagetrue
\newcommand{\thechapterimage}{}%
\newcommand{\chapterimage}[1]{\ifusechapterimage\renewcommand{\thechapterimage}{#1}\fi}%
\newcommand{\autodot}{.}
\def\@makechapterhead#1{%
{\parindent \z@ \raggedright \normalfont
\ifnum \c@secnumdepth >\m@ne
\if@mainmatter
\begin{tikzpicture}[remember picture,overlay]
\node at (current page.north west)
{\begin{tikzpicture}[remember picture,overlay]
\node[anchor=north west,inner sep=0pt] at (0,0) {\ifusechapterimage\includegraphics[width=\paperwidth]{\thechapterimage}\fi};
\draw[anchor=west] (\Gm@lmargin,-9cm) node [line width=2pt,rounded corners=15pt,draw=ocre,fill=white,fill opacity=0.5,inner sep=15pt]{\strut\makebox[22cm]{}};
\draw[anchor=west] (\Gm@lmargin+.3cm,-9cm) node {\huge\sffamily\bfseries\color{black}\thechapter\autodot~#1\strut};
\end{tikzpicture}};
\end{tikzpicture}
\else
\begin{tikzpicture}[remember picture,overlay]
\node at (current page.north west)
{\begin{tikzpicture}[remember picture,overlay]
\node[anchor=north west,inner sep=0pt] at (0,0) {\ifusechapterimage\includegraphics[width=\paperwidth]{\thechapterimage}\fi};
\draw[anchor=west] (\Gm@lmargin,-9cm) node [line width=2pt,rounded corners=15pt,draw=ocre,fill=white,fill opacity=0.5,inner sep=15pt]{\strut\makebox[22cm]{}};
\draw[anchor=west] (\Gm@lmargin+.3cm,-9cm) node {\huge\sffamily\bfseries\color{black}#1\strut};
\end{tikzpicture}};
\end{tikzpicture}
\fi\fi\par\vspace*{270\p@}}}

\def\@makeschapterhead#1{%
\begin{tikzpicture}[remember picture,overlay]
\node at (current page.north west)
{\begin{tikzpicture}[remember picture,overlay]
\node[anchor=north west,inner sep=0pt] at (0,0) {\ifusechapterimage\includegraphics[width=\paperwidth]{\thechapterimage}\fi};
\draw[anchor=west] (\Gm@lmargin,-9cm) node [line width=2pt,rounded corners=15pt,draw=ocre,fill=white,fill opacity=0.5,inner sep=15pt]{\strut\makebox[22cm]{}};
\draw[anchor=west] (\Gm@lmargin+.3cm,-9cm) node {\huge\sffamily\bfseries\color{black}#1\strut};
\end{tikzpicture}};
\end{tikzpicture}
\par\vspace*{270\p@}}
\makeatother

\usepackage[obeyspaces]{url}
\PassOptionsToPackage{obeyspaces}{url}
\usepackage{hyperref}
\hypersetup{hidelinks,backref=true,pagebackref=true,hyperindex=true,colorlinks=false,
breaklinks=true,urlcolor=ocre,bookmarks=true,bookmarksopen=false,pdftitle={Title},pdfauthor={Author}
}
\usepackage{bookmark}
\bookmarksetup{
open,
numbered,
addtohook={%
\ifnum\bookmarkget{level}=0 
\bookmarksetup{bold}%
\fi
\ifnum\bookmarkget{level}=-1 
\bookmarksetup{color=ocre,bold}%
\fi
}
}

\usepackage{lineno}

\input{tags.sty}

\begin{document}
\includepdf{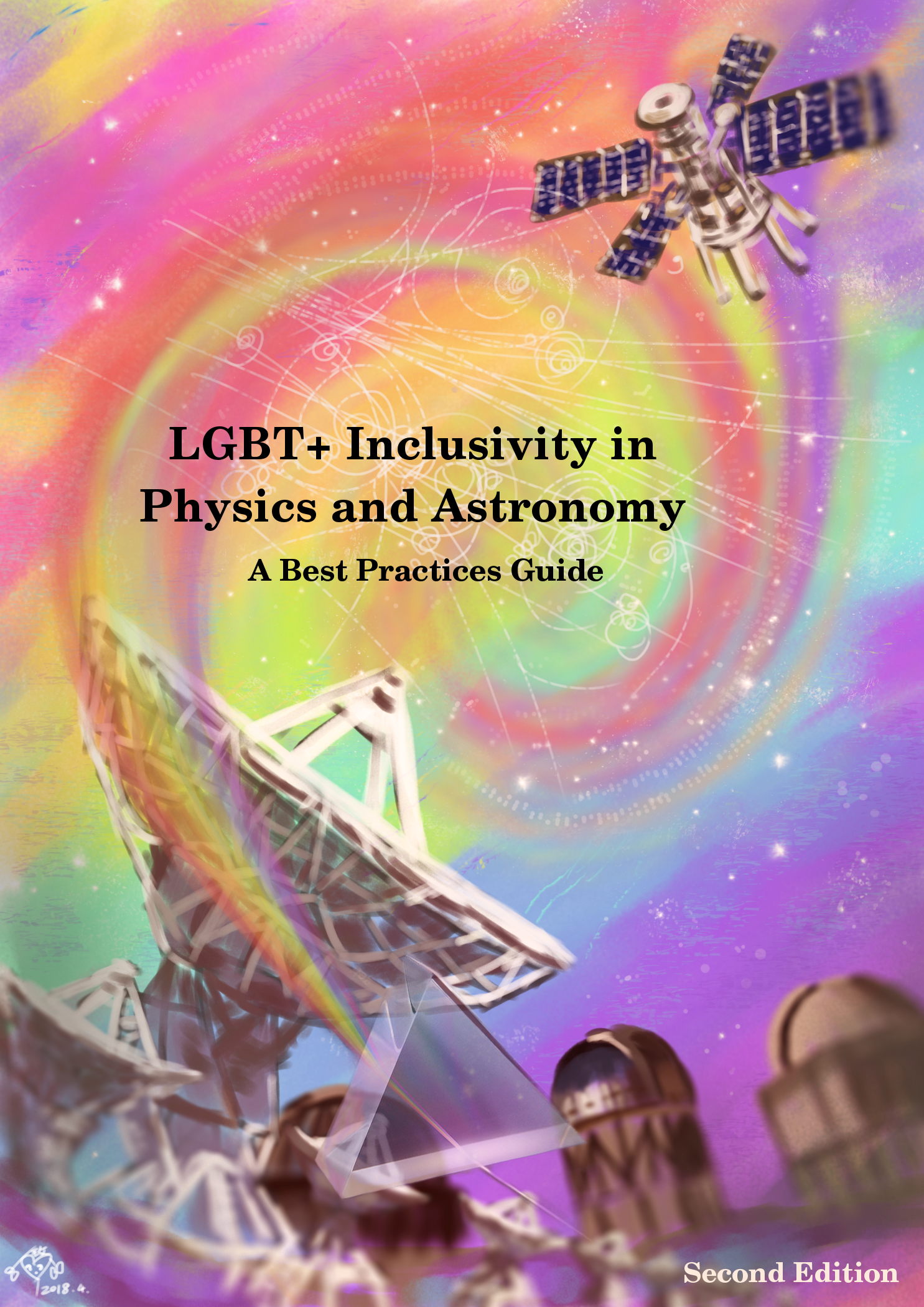}\clearpage 

\usechapterimagefalse
 \pagestyle{empty} 
 \clearpage
\vspace*{\fill}
\begin{center}
\vspace{-4cm}
\large\textit{A Joint Publication of} \\
\bigskip
LGBT+ Physicists \\ \medskip and \\ \medskip
The AAS Committee for Sexual and Gender Minorities in Astronomy

\end{center}
\vfill
\clearpage

 \tableofcontents 
 \pagestyle{main} 

\frontmatter
\usechapterimagetrue
\chapterimage{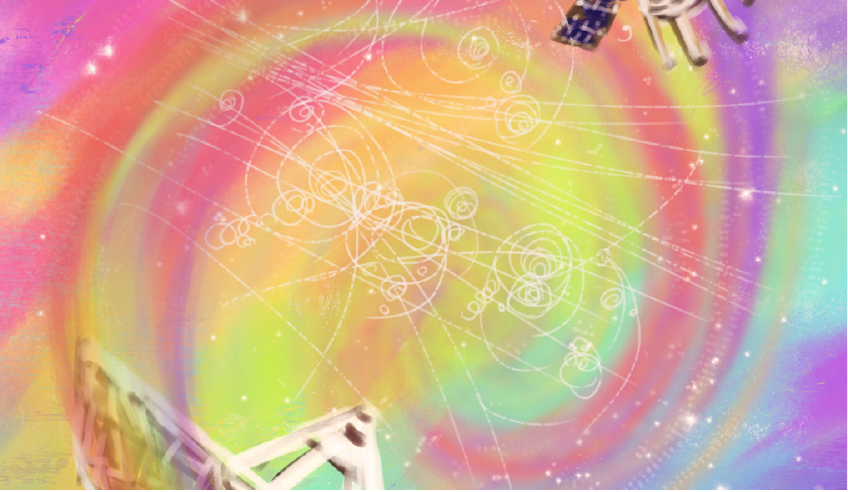}
\chapter{Introduction}
\label{ch:introduction}
\usechapterimagefalse

\section{Introduction to the Second Edition}
As much as we might like to think that physics and astronomy, and science in general, are a meritocracy neutral to a person’s individual identity, institutional policies and anecdotal evidence show a contrary truth. LGBT+ physicists\footnote{In this document, we use the phrase ``LGBT+'' to represent the entire community of lesbian, gay, bisexual, transgender, queer, questioning, intersex, asexual, and other persons with marginalized gender identities or sexual orientations.} face a range of concerns from isolation to exclusion and structural barriers. Some of these challenges and burdens are institutional; some are unintended, and others originate with explicit hostility towards LGBT+ people. Given the variety of structural and interpersonal factors that affect LGBT+ scientists, including institutional policies, healthcare and insurance, implicit bias, exclusionary behavior, and even physical safety concerns, a strong support network can be critical to thriving in academia. 

Some LGBT+ people have the option of selectively being ``out''; e.g. an LGBT+ person may try to keep their personal life separate from their professional life so as to shield themselves from any discrimination at work. As a result, isolation can be a common experience among LGBT+ scientists, who may have a hard time connecting with mentors and peers who share similar experiences, or even connecting with those who identify as allies. Hiding one's true identity can take a severe toll in the form of extra emotional effort that saps energy and impacts productivity. The consequent invisibility can be alienating, amplifying isolation and feelings of lack of support and leaving people without the resources they need to succeed. Among LGBT+ students, those who are newly arrived at the institution or recently self-identified as LGBT+ may have a harder time feeling welcome and finding mentors to help them. Without clear community support, there is a further strong incentive to hide or downplay one's LGBT+ identity at work and in the classroom, making it even harder to raise concerns or find help navigating barriers. 

Examples of positive change range from institutional efforts, such as the creation of an LGBT+ non-discrimination policy, to establishing social norms to encourage a diverse and inclusive workplace. However, these initiatives can be complicated by the intersection of a person's LGBT+ identity with their race, class, gender, disability, national origin, and other identities. We encourage our readers to learn and educate themselves on how these identities intersect, and how to support all members of our professional community. \textbf{This guide is written for those who want to become aware of the concerns and extra burdens facing members of the LGBT+ community in their professional lives, and to provide a set of best practices to mitigate them, creating a safe and welcoming work environment.}

In the four years since the First Edition of this Best Practices Guide was released in 2014, the conversation around LGBT+ inclusion has changed dramatically on the national, local, and institutional levels. To take one particularly high-profile example, the First Edition grappled with a confusing and often-changing landscape of state-level laws concerning same-gender marriage, civil unions, and domestic partnerships -- and one year later, on June 26, 2015, the United States Supreme Court established same-gender marriage throughout the country in \textit{Obergefell v. Hodges}. However, such equality has not been consistent across all aspects of law. For instance, between 2010 and 2016 the US Department of Education issued guidance indicating that transgender students are protected from discrimination under Title IX -- guidance which was rescinded in 2017. At the regional level, members of the LGBT+ community still face a patchwork of local and state laws on matters ranging from protection against employment discrimination to the simple ability to use the bathroom in peace. This patchwork necessarily affects the decisions that LGBT+ scientists -- and non-LGBT+ scientists with LGBT+ people in their households -- make about where to live, work, travel and collaborate.

Since 2014, new data have emerged about the experiences of LGBT+ people in STEM fields, and also more specifically in physics and astronomy. At about the same time the First Edition was published, Patridge, Barthelemy and Rankin published the first quantitative study of the experiences of lesbian, gay, and bisexual STEM faculty members~\cite{patridge:2014}. Results from the \textit{Queer in STEM} project, the first broad national survey of LGBT+ graduate students, researchers, and industry professionals in STEM fields, were published in 2016~\cite{queerinstem:2016}. 
 
In 2014, the American Physical Society formed an Ad-Hoc Committee on LGBT Issues (C-LGBT), charged with investigating LGBT+ representation in physics, assessing the climate experienced by LGBT+ physics students and physicists, and making recommendations for improving LGBT+ inclusion in the field. The 2016 report issued by the C-LGBT~\cite{LGBTClimateInPhysics:2016} included the first published survey results specifically on the experiences of LGBT+ people in physics; many anonymous responses to this survey are reproduced in this Guide. LGBT+ roundtable sessions have become a popular event at the APS spring meetings. The Working Group on LGBTIQ Equality (WGLE), established by the American Astronomical Society (AAS) in January 2012, was chartered by the AAS Council into the Committee for Sexual-Orientation and Gender Minorities in Astronomy (SGMA) in August 2015. The committee is charged with improving the professional climate for LGBT+ astronomers by promoting equitable hiring and compensation practices, creating networking and mentoring opportunities, and serving as a platform to give voice to the LGBT+ community in astronomy. SGMA has organized informational booths and networking dinners at the summer and winter AAS meetings and Facebook groups to connect the community, as well as interviews with LGBT+ astronomers to increase LGBT+ visibility within and outside astronomy.

We have prepared this Second Edition in order to address new research, the changing political landscape, and evolving terminology and best practices from the LGBT+ community. The new Second Edition, prepared by physicists and astronomers ranging from undergraduate students to tenured faculty, has been restructured to highlight particular contexts in which educational and scientific work is performed and facilitated: the department (Chapter~\ref{ch:department}), the classroom (Chapter~\ref{ch:classroom}), the mentoring and advising relationship (Chapter~\ref{ch:mentor-advise}), hiring and promotions (Chapter~\ref{ch:hire-promote}), travel and hosting (Chapter~\ref{ch:travel-host}), and the academic institution as a whole (Chapter~\ref{ch:university}). Of course, many issues apply in multiple contexts, and so each section and recommendation is tagged with a letter for relevant roles and contexts that one might not expect based on its chapter placement. A brief explanation of the tag system may be found in Section~\ref{expl:tags}.

We have chosen not to use the phrase ``under-represented'' to describe LGBT+ people in physics, for two important reasons. First, there is a dearth of research on LGBT+ people in physics, and what studies do exist seldom make direct comparisons between LGBT+ representation in physics against that in the general population. Second, the phrase ``under-represented'' has a weighty history in the context of racial and ethnic minorities' struggle for equitable access to education in the United States. We do not wish to co-opt or confuse the common (and legal) use of this phrase, regardless of whether it is technically accurate for LGBT+ persons in physics. This is especially important given the majority-white character of the US LGBT+ physicist community.

In seeking alternative terminology, the authors considered several phrases which emphasize unmet needs, including ``under-supported'' and ``under-served.'' These descriptions are accurate, and evoke a way forward through increased interpersonal and institutional support. However, in the end, our consensus was to use the broader term ``marginalized.'' When we say that LGBT+ people are marginalized in physics, we speak of a broad pattern of social forces pushing LGBT+ physicists ``to the margins.'' These include social exclusion, structural barriers, and other factors which are not strictly due to a lack of resources. LGBT+ physicists who are marginalized in multiple ways are even more at risk in unsupportive environments. For instance, LGBT+ scientists who are also members of a racial minority may feel even more pressure to hide their LGBT+ status, so as to avoid compounding the exclusion they already face. It is therefore important to think broadly about LGBT+ concerns in physics, and to keep in mind that each individual will face a different set of circumstances.

As with the First Edition, this document primarily addresses US institutions and research done in the US; the authors are members of the American Physical Society and the American Astronomical Society, and thus this is where our expertise lies. We acknowledge and celebrate the international character of physics (and, indeed, science in general) as well as of LGBT+ scientists, and it is our hope that science institutions in other countries will find aspects of this Guide helpful. At the same time, US-based physicists hail from and do their work all over the world, and we strive to be conscious of this fact in our recommendations. Similarly, these recommendations are very much oriented toward the physics and astronomy communities (which we often abbreviate ``physics'' for brevity), but most are applicable to other fields in general, and especially to STEM fields. We have drawn on research relating not only to the physics community, but also to engineering, chemistry, biology, mathematics, and other STEM disciplines.

It is our intention to make this Guide an accessible reference for researchers, faculty and staff in a range of administrative, technical, teaching and research roles -- whether or not the reader identifies as a member of the LGBT+ community. We include a glossary of terms that may be unfamiliar (Sec.~\ref{glossary}), as well as a partial list of resources that may be helpful to department members (Sec.~\ref{resources}). Of course, this Guide cannot explain or treat LGBT+ issues comprehensively, but it does offer a starting point to those wishing to make their departments, and institutions of higher learning in general, more welcoming places. We hope that you will find it useful.
\clearpage
\pagestyle{simple}
\section{Preface to the First Edition (2014)}
When physicists and astronomers discuss issues related to diversity or broadening participation in the field, the focus is typically on creating support mechanisms for women or people of color.  However, scientists who identify as lesbian, gay, bisexual, or transgender (LGBT) are also a minority within the physics and astronomy communities and can find themselves marginalized in a variety of ways.  This document aims to highlight opportunities for making academic departments more welcoming for LGBT+ students, staff, and faculty.  (In this document, we add a plus symbol to remind ourselves that not everyone fits neatly into the LGBT constructs, and some may identify differently). 

Even if you consider your department to be a safe and welcoming space, it is important to note that the campus environment is often perceived differently by different groups: A 2003 study of campus climate~\cite{rankin:2003} found that, while 90\% of heterosexual students classified their campuses as friendly, 74\% of LGBT students rated the campus climate as homophobic. LGBT staff and faculty routinely report that their campuses are more homophobic than reported by students. A 2010 study~\cite{rankin:2010} found that 23\% of LGB respondents had been harassed within the past year, with even higher rates (31--39\%) for transgender individuals. Around half of all LGB students and two-thirds of transgender students had avoided disclosing their identity to avoid harassment. 

The argument for inclusion is simple:  Science advances fastest when scientists are free to apply their intelligence and imagination to the exploration of the universe without limits and without fear.  Sometimes, those scientists are LGBT+.  Institutions that are viewed as unfriendly to LGBT+ people quickly find themselves at a competitive disadvantage.  When LGBT+ scientists leave our departments to work at other institutions, our students, our scholarly communities, and our own research suffer.  Furthermore, a more inclusive workplace has advantages for all of us: greater flexibility to perform our work, greater support for work/life issues, and greater freedom to be ourselves. 

Recent studies suggest that, over the next decade, the U.S.\ will need a million more college graduates in the science, technology, engineering, and mathematics (STEM) fields than previously expected.  This gap could largely be filled simply by increasing the retention rate of undergraduates majoring in STEM fields from 40 to 50\%~\cite{pcast-engagetoexcel}.  Past initiatives to increase the participation of underrepresented groups in STEM fields have ignored LGBT+ people.  Given our need to remain competitive in a global economy, it seems prudent to increase the recruitment and retention of STEM talent from all demographics~\cite{patridge:2014}. 

Research has shown that the presence of ``difference'' on campus is important to the intellectual and social development of the majority~\cite{reason:2010,smith:2010}. Specifically, students who interact with people different from themselves show more developed critical and creative thinking, improved cross-cultural relationships, and increased volunteerism and civil activism. Students and staff who encounter contradictions in their own thinking around LGBT+ issues can develop greater levels of cognitive sophistication~\cite{love:1997}. The advantages of engaging with ``difference'' appear to continue beyond college and into the workforce~\cite{smith:2010}. A diverse department can thus improve your students' performance, their ability to function in a multicultural workforce, and the reputation of your department.

Best practices for the inclusion of LGBT+ people on campus have been proposed by several authors~\cite{rankin:2010,massachusetts-cgly:1993,campusprideindex}. In this document, we limit ourselves to recommendations that are particularly relevant to faculty and department chairs (as opposed to university administrators). After a brief glossary of terms, we offer both short-term and long-term department-level suggestions, then address recruitment and personnel issues.  We conclude with recommendations for university-level policies that may guide conversations with institutional administrators. Lists of useful external resources and the authors of this document are provided as appendices.
 \clearpage
\section{General considerations for employing this Guide}

\paragraph{Keep the scope of all diversity, equity, and inclusion efforts broad.} The issues addressed in this Guide are only one part of a larger conversation on diversity, equity, and inclusion in STEM. Many members of the LGBT+ community are marginalized 
in other ways as well, 
including (but not limited to) 
their race, gender, class, dis/ability status, national origin, 
and citizenship. 
These attributes, and their social impact, cannot be separated from an individual's experience as an LGBT+ person.

While we strive to highlight the intersections of our recommendations with factors beyond LGBT+ identity, we also ask that the reader approach this Guide as a starting point, and not an ending point. It is by no means true that solely adhering to the recommendations herein will create a welcoming environment for all members of the LGBT+ community. In fact, if these guidelines are followed naively (without any consideration of race, gender, class, dis/ability, etc), their benefit will naturally accrue primarily to those LGBT+ people who \textit{already} face the fewest professional and personal barriers to success. 

\paragraph{Consider the differences between LGB+ and T+ experiences and needs.} Among LGBT+ people, members of the transgender community (broadly defined) face particularly significant challenges. It is important to remember that different measures are needed to support different populations within the community. For example, rules around birth certificates and legal identification can be a source of serious trouble for \textit{T} (trans) people, but not for cisgender \textit{LGB} people. A welcoming atmosphere for same-gender partnerships is critically important for partnered \textit{LG} (lesbian and gay) people and for many partnered \textit{T} and \textit{B} (bisexual) people, but can be less of an issue for heterosexual \textit{T} persons, or for \textit{B} people with partners of a different gender. The LGBT+ community is a diverse one, with varied needs.

\paragraph{Adjust to changes in language and terminology.}  Well-intentioned people, including members of the LGBT+ community, may find that the terminology they employ has been replaced with words that are considered more inclusive.  In this Guide we have endeavored to use preferred terminology, but some words and phrases may be soon out of date or considered offensive within certain communities.  There are many cases where there is not a unanimous opinion on the best terminology, for instance, many people dislike the phrase "preferred pronouns" as it implies a sense of voluntary choice, rather than intrinsic identity.  As an advocate or ally, you will need to be open and responsive to individuals providing feedback on the language you use and you should be proactive in staying up-to-date on the language used within and about the LGBT+ community.  

\paragraph{Seek education, partnerships, and ideas outside of this Guide.} This Guide is not intended primarily as an educational resource on the experiences, vocabularies, and cultures of various LGBT+ communities, nor is it meant to supplant the role of active partnership and consultation with actual LGBT+ people and organizations. We encourage all readers of this Guide to read and listen widely on related subjects, especially to the lived experiences of LGBT+ people, and most especially to those who are marginalized and/or minoritized in multiple ways.

\paragraph{Implement changes proactively, not reactively.} Many LGBT+ people keep their identities purposefully ``invisible" in professional settings, to avoid scrutiny and potential negative effects on their careers. Even those who are openly LGBT+ may not report any problems they encounter, for similar reasons. Furthermore, many of the recommendations in this Guide require thought and planning, and cannot be implemented easily on the fly. It is therefore critical to consider and make changes \textit{before} any specific person requests them.
 \clearpage
\section{How to use this Guide}



\subsection{Chapters}
The primary organizational principle of this Guide is in the form of chapters, spanning suggestions at the level of individual academic departments in Chapter \ref{ch:department}, to those at the university or institutional level in Chapter \ref{ch:university}. Along the way, we provide suggestions for improving the climate in classrooms (Chapter \ref{ch:classroom}), for mentors and advisors (Chapter \ref{ch:mentor-advise}), hiring (Chapter \ref{ch:hire-promote}) and travel (Chapter \ref{ch:travel-host}).
If you are seeking information on a particular topic or issue, use the main table of contents to find it. The topic-based organization creates a logical flow of the Guide and allows multiple related recommendations to be grouped under the same motivational and framing sections.

\subsection{Tags}
\label{expl:tags}
We recognize that certain recommendations may apply across chapters and scenarios. We also expect the Guide to be read by persons in specific roles within their institutional context. We therefore provide an additional means of navigating this Guide through the use of context-specific \textit{tags}, represented by a single letter in a color-coded square in the margins. Not all sections have associated tags printed, and certain sections have multiple tags. The primary tag for a section is suppressed within its chapter -- so, for instance, there are no ``[H]iring'' tags in the chapter dedicated to hiring and promotions (Chapter \ref{ch:hire-promote}). However, sections elsewhere in the Guide related to hiring LGBT+ persons (such as Section \ref{positions-of-power}) have a corresponding [H] tag.

In addition to the context-specific tags, we provide two additional role-based tags, for \textit{Staff} and \textit{Research collaborations}, respectively. Whereas there are no single chapters dedicated to LGBT+ specific issues in these contexts, nevertheless persons who are members of one of these groups may find additional information tagged specifically for them by searching the [S] and [R] tags. Similar role-based tags exist for [A]dvisors and mentors outside of Chapter \ref{ch:mentor-advise}, and for educators in [C]lassrooms outside of chapter \ref{ch:classroom}. 
These tags provide a separate and complementary organization system, and are intended to let readers consolidate a list of advice most relevant to their position within STEM academia. 

\begin{itemize}
  \item $[D]$epartment level: Advice for fostering a welcoming and supportive department including 
guidelines for department chairs, administrators, individual members, and staff. \BPGtag{D}
  \item $[C]$lassroom level: Advice for improving the climate in classrooms and other sites of instruction. \BPGtag{C}
  \item $[A]$dvisors and Mentors: Advising and mentoring relationships between faculty, 
postdoctoral research associates, graduate and undergraduate students, and staff. \BPGtag{A}
  \item $[H]$iring and promotions. \BPGtag{H}
  \item $[T]$ravel: Advice regarding travel, including to regions with different laws and customs. \BPGtag{T}
  \item $[I]$nstitution level: Advice for promoting healthy institutional practices, including 
guidelines for administrators and deans. \BPGtag{I}
  \item $[S]$taff: Advice for staff for establishing a welcoming environment within their institution. \BPGtag{S}
  \item $[R]$esearch collaborations: Advice for sustaining healthy research groups and 
collaborations, including guidelines for group leaders and principal investigators. \BPGtag{R}
\end{itemize}

%
%
 \clearpage
\section{Executive summary}

Science advances fastest when scientists are free to apply their intelligence and imagination to the exploration of the universe without limits and without fear. This Guide is written for anyone who wants to become aware of the concerns, extra burdens, and impediments faced by LGBT+ physicists and astronomers in realizing this ideal. It aims to provide a set of best practices to address these issues and for creating a safe, welcoming, and supportive scientific work environment.
 
Following the inaugural Best Practices Guide in 2014, this second edition seeks to incorporate the results of new research, broader societal developments, and on-going work in the area of climate and diversity. The authors seek to provide members of physics and astronomy departments in academic and laboratory settings, along with staff and administrators in their host institutions, with concrete, practical steps toward a more inclusive and supportive environment. These steps are intended to address the experiences of invisibility and isolation that are often endemic to LGBT+ physicists and astronomers, as well as the experiences of implicit and overt bias and discrimination that affect members of marginalized groups more broadly.
 
While we encourage you to read the entire document, we provide here a quick-start guide: a summary of key themes for action and representative examples, along with pointers to the more detailed discussion.
 
\noindent{\normalfont\small\sffamily\bfseries Assess and address:} Seek to determine the climate for LGBT+ members of your department through participating in or conducting a climate survey (\ref{demographics}, \ref{univ-surveys}), collecting demographic information (\ref{univ-identify}), carrying out classroom climate assessments (\ref{class:monitorclimate}). Respond by establishing a departmental climate committee and/or liaison (\ref{liaison}) and explicit LGBT+ supportive policies.

\smallskip 

\noindent{\normalfont\small\sffamily\bfseries Break the silence and invisibility:} Initiate department-wide discussions of LGBT+ concerns (\ref{discuss_climate_advisees}), highlight the scientific contributions of LGBT+ department members at all levels (\ref{recognize-achievements}), include welcoming language and non-discrimination policies in your syllabus (\ref{course-expectations}) and on your departmental website (\ref{gender-language}), join an ``Out List'' as an ally or LGBT+ scientist (\ref{outlist}), identify LGBT+ supportive mentors (\ref{sec:general_mentoring}), invite LGBT+ speakers to campus (\ref{lgbt-speakers}).

\smallskip

\noindent{\normalfont\small\sffamily\bfseries Educate and advocate:} Participate in LGBT+ friendly climate and anti-bias training (\ref{diversity-training}, \ref{sec:biastraining}), identify LGBT+ support services on campus and in the broader scientific community (\ref{find-resources}), work for campus-wide LGBT+ supportive practices such as supportive first responders (\ref{firstresponders}) and gender-inclusive restrooms and accompanying signage (\ref{restroom_access_dept}, \ref{hosting-conferences}).
 
\smallskip 
 
\noindent{\normalfont\small\sffamily\bfseries Set the example and expectations:} include preferred pronouns in your email signature, invite students and/or meeting participants to share their preferred pronouns (\ref{sec:pronouns}, \ref{gender-language}, \ref{subsec:precoursecurvey}), articulate classroom environment expectations on the first day of class (\ref{class:firstday}), speak up in response to discriminatory behavior and report where appropriate (\ref{sec:serious_situations}, \ref{offensive-language}, \ref{class:intervene}, \ref{univ-titleix}).
 
\smallskip 
 
\noindent{\normalfont\small\sffamily\bfseries Support and include:} Plan gender-neutral and inclusive social events (\ref{social-events}), create LGBT+ safe spaces in your department (\ref{safe-spaces}), provide equal restroom access (\ref{restroom_access_dept}), include LGBT+ faculty in positions of authority, e.g.~as committee chairs (\ref{positions-of-power}), provide support for participation in LGBT+ networking events (\ref{networking}, \ref{lgbt-conferences}), actively recruit LGBT+ students in undergraduate and graduate admissions (\ref{recruit-students}), ensure LGBT+ needs are considered in dual-career hires (\ref{dual-career}), family-friendly policies (\ref{families}), and benefits (\ref{health-insurance}, \ref{partner_benefits}).\\

The foregoing highlights are by no means exhaustive, and the remainder of this Guide provides additional steps along with background material and detailed discussion. We hope that you will find it both helpful and practical; that you will share it with your colleagues; and that it will catalyze further thought and action about climate, diversity, and inclusion in your scientific workplace.
 \clearpage
\section{About the Authors}

\textit{Authors are listed in alphabetical order by surname. The text was written collaboratively with all authors involved in discussion, writing and revision.}

\subsection{Authors, Second Edition}

\begin{table*}[htp]
\begin{center}
\begin{tabular}{ll}
\textbf{Nicole Ackerman} & Assistant Professor of Physics \\ 
& Agnes Scott College \\
&  Decatur, GA, USA \\[0.2cm]
\textbf{Timothy Atherton} & Associate Professor of Physics \\
& Tufts University \\
& Boston, MA, USA \\[0.2cm]
\textbf{Adrian Ray Avalani} & Undergraduate student \\
& California Institute of Technology \\
& Pasadena, CA, USA \\[0.2cm]
\textbf{Christine A. Berven} & Associate Professor of Physics \\
& University of Idaho \\
& Moscow, ID, USA \\[0.2cm]
\textbf{Tanmoy Laskar$^*$} & Jansky Postdoctoral Fellow \\
& National Radio Astronomy Observatory \\
& Charlottesville, VA, USA \\[0.2cm]
\textbf{Ansel Neunzert$^*$} & PhD student in Physics \\
& University of Michigan \\ 
& Ann Arbor, MI, USA \\[0.2cm]
\textbf{Diana S. Parno$^{\dag}$} & Assistant Research Professor of Physics \\
& Carnegie Mellon University \\
& Pittsburgh, PA, USA \\[0.2cm]
\textbf{Michael Ramsey-Musolf} & Professor of Physics \\
& University of Massachusetts \\
& Amherst, MA, USA
\end{tabular}

\end{center}
\end{table*}

\noindent$^{\dag}$Chair\\
$^*$Editor

\clearpage

\subsection{Authors, First Edition (2014)}

\begin{table*}[htp]
\begin{center}
\begin{tabular}{ll}
\textbf{Timothy Atherton} & Assistant Professor of Physics \\
& Tufts University \\
& Boston, MA, USA \\[0.2cm]
\textbf{Ram\'{o}n Barthelemy} & PhD student in Science Education \\
& Western Michigan University \\
& Kalamazoo, MI, USA \\[0.2cm]
\textbf{Carolyn Brinkworth} & Associate Staff Scientist \\
& California Institute of Technology \\
& Pasadena, CA, USA \\[0.2cm]
\textbf{Wouter Deconinck} & Assistant Professor of Physics \\
& College of William and Mary \\
& Williamsburg, VA, USA \\[0.2cm]
\textbf{Van Dixon} & Scientist \\
& Space Telescope Science Institute \\
& Baltimore, MD, USA \\[0.2cm]
\textbf{Elena Long} & Postdoctoral Researcher in Physics \\
& University of New Hampshire \\
& Durham, NH, USA \\[0.2cm]
\textbf{Merav Opher} & Associate Professor \\
& Boston University \\
& Boston, MA, USA \\[0.2cm]
\textbf{Diana Parno} & Acting Assistant Professor of Physics \\
& University of Washington \\
& Seattle, WA, USA \\[0.2cm]
\textbf{Michael Ramsey-Musolf} & Professor of Physics \\
& University of Massachusetts \\
& Amherst, MA, USA \\[0.2cm]
\textbf{Jane Rigby} & Astrophysicist \\
& NASA Goddard Space Flight Center \\ 
& Greenbelt, MD, USA \\[0.2cm]
\textbf{Elizabeth H. Simmons} & Professor of Physics \\
& Michigan State University \\
& East Lansing, MI, USA \\[0.2cm]
\end{tabular}
\end{center}
\end{table*}%
\clearpage
\section{Acknowledgments and Credits}
\label{acknowledgments}

This document represents a significant amount of work invested over several years, and the authors gratefully acknowledge the support they have received during this effort. The organizations LGBT+ Physicists and SGMA (Sexual and Gender Minorities in Astronomy) have supported the Best Practices Guide from its earliest inception. The American Physical Society and the American Astronomical Society publicized the First Edition in 2014, bringing this important discussion to a much wider portion of the scientific community. Cover artwork for the second edition was produced by physicist and artist, Kitty Yeung\secretfootnote{See more of Kitty Yeung's work at www.artbyphysicistkittyyeung.com}. Students and scientists from different backgrounds and different fields have shared their stories, which have guided our recommendations and which often appear in pull quotes to illustrate the problems we hope to address. Finally, we are tremendously grateful for the input of our beta readers -- including Veronica Berglyd Olsen, Sarah Bergman, Alex Brown, Van Dixon, Adam Iaizzi, Hannah Klion, Kristen Larson, Danielle Leonard, Giampiero Mancinelli, Laura McCullough, Alexander Nguyen, Cristina Olds, Quinton Singer, Beck E. Strauss, and Diane Turnshek -- who have helped immeasurably in the final editing and polishing of the document.

\mainmatter
\pagestyle{main}

\usechapterimagetrue
\chapterimage{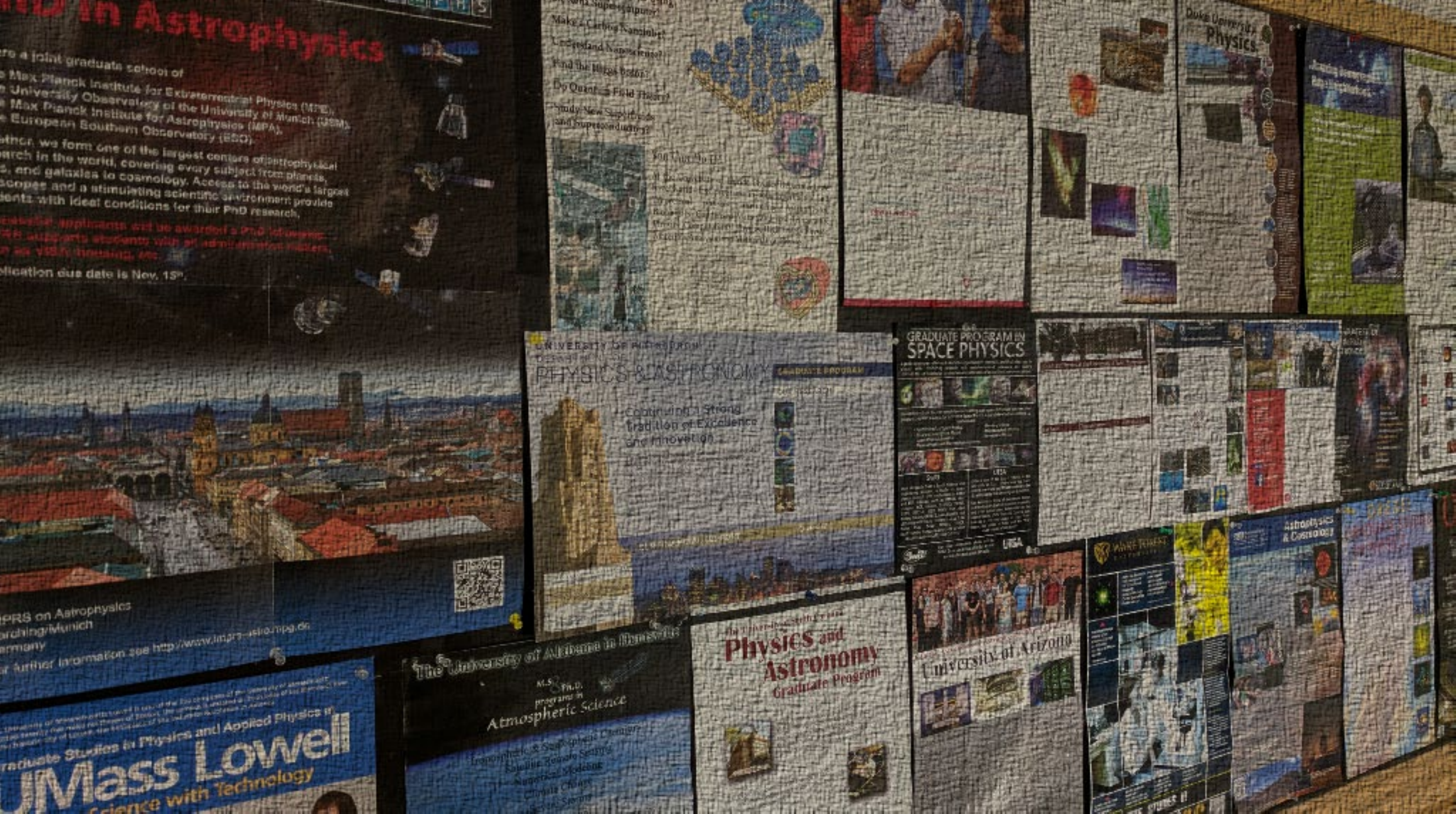}
\chapter{Toward a welcoming department}
\label{ch:department}
\usechapterimagefalse
\begin{newquote}{Yoder and Mattheis, J. Homosexuality \textbf{63} 1 (2016)~\cite{queerinstem:2016}}{}
Individual identity factors are often considered inconsequential or irrelevant to STEM professional achievement, but research suggests that being part of a marginalized or minoritized group can hamper job satisfaction, career success, and workplace productivity.
\end{newquote}

\medskip\noindent
The department is the academic home of researchers, educators, and students, and is thus the most fertile ground for grass-roots efforts toward improving equity and inclusion. An inclusive and equitable department is trained to recognize and mitigate systematic bias; respects the privacy of its marginalized individuals; welcomes members of all marginalized groups; provides, recognizes, and rewards mentorship and advocacy work; allocates resources toward improving diversity and inclusion; and advocates for visibility and diversity in the field within the institution and in a global context. In this section, we provide suggestions by which departments may achieve leadership in their role of providing a dynamic and inclusive workplace for their LGBT+ members. A well-equipped department will be Aware, Welcoming, Advocating, and REsourceful, or AWARE. Here, we will refer to the ``academic home'' where scholars spend the greatest amount of their time as the ``department'', regardless of how this maps directly to the structure at a given institution; the guidelines given in this section apply even in those instances whose members may not identify directly with a traditional department structure.

As with any other marginalized group, the presence of openly LGBT+ department members does not mean that the department is fully inclusive. True inclusivity involves a continuous, iterative process of evaluation, assessment, and improvement. 
    \section{Aware}
      \subsection {Understand departmental demographics}
\label{demographics}
\BPGtag{A} \BPGtag{H} \BPGtag{R} \BPGtag{S} \BPGtag{P} 
Consider how internal demographic information and/or demographic information from job applicants and 
prospective students may be collected in an inclusive way. Does the department's demographic form 
include a question about sexual orientation and a question about gender identity? Can respondents 
list a domestic partnership as a marital status? Are respondents limited to a binary 
identification, or can they write in how they self-identify? Sample survey questions are 
presented at the end of this Guide in Appendix~\ref{survey}. You may also wish to include 
definitions of terms with the survey; these may be particularly valuable to international respondents.

As with other diversity questions, responses to such queries provide valuable statistical 
information, but can also pose risks for the respondent. Survey responses should be anonymized; 
forms with these questions should clearly indicate what will happen to the data so that respondents 
can feel confident about how their answers will be used. If it is possible for respondents to modify or 
withdraw their submissions without compromising anonymity, be sure to communicate how they may do so. 
All such data should be separated from any 
decision-making related to hiring, awards, or promotions. Demographic data should be passed to 
Human Resources or to a designated collator and kept separately from other materials. Do not collect 
this information if you cannot both prevent its misuse and assure respondents of their privacy.

      \subsection{Encourage accurate pronoun use}
\label{sec:pronouns}
\BPGtag{A} \BPGtag{C} \BPGtag{R} \BPGtag{S} \BPGtag{P} 

\subsubsection{Motivation}
Not everyone goes by the pronouns which their colleagues typically assume. Some people are 
frequently \emph{misgendered} (called by the incorrect pronouns). This is often a deeply negative 
experience, and difficult to handle in a professional context, as it requires either correcting a 
colleague or allowing the error to persist indefinitely. It can also lead to awkwardness for the 
person who made the mistake.

Many instances of misgendering could be avoided by simply making one's pronouns known up front. But 
this, too, is more difficult than it appears. Unlike names, there are few existing 
socially-accepted opportunities for a person to specify their pronouns to colleagues. These 
opportunities need to be created and \emph{normalized} (used by many people, so that it is not 
considered startling or unusual) in order to be accessible to all members of the community.

\begin{newquote}{Alex Hanna, sociologist, in ``Being Transgender on the Job Market''. 
Inside Higher Ed., July 15, 2016~\cite{hanna:2016}}{
https://www.insidehighered.com/advice/2016/07/15/challenge-being-transgender-academic-job-market-ess
ay}
 I recently had a nightmare interview in a sociology department at a college in a major metropolitan 
city in the Northeast. At the job talk, the ... search chair ... introduced me with the wrong 
pronouns, which meant the rest of the audience took his lead. He knew my correct pronouns, but he 
didn't seem to be able to get them right at all ... His misgendering was constant, and the other 
faculty members were repeating it. It annoyed me less that he was consistent in doing it but that 
the other faculty members were acquiescent in it. I remember at one point, he said something to 
which he wanted to engender a big response or laugh but managed to misgender me in the statement. 
While everyone else laughed, I just looked down and felt humiliated.
\end{newquote}

\subsubsection{Ways to normalize pronoun specification}

\paragraph{Name tags for events and conferences}
Name tags are used for many events, and may specify information besides names (such as home 
institution). Another field can easily be added for pronouns. Alternatively, pronoun stickers, buttons or ribbons, appropriately sized for the event badges, may be provided at the registration desk. Materials should be provided for attendees to write their pronouns onto these add-ons; recall that ``he/his'' and ``she/her'' are not a sufficiently broad range of options.

\paragraph{Email signatures}
Email signatures often contain information about the sender which may be of use to the recipient. This 
includes name, title/position, institution, contact information, etc. Pronouns are a natural fit 
for an email signature. Specifying your own pronouns in your email signature models the behavior 
for others.

\paragraph{Course surveys}
Many instructors distribute surveys at the beginning of the term, as a way to get to know their 
students. For instance, these surveys might ask about a student's name, major, academic 
preparation, reason for taking the course, concerns or questions, etc. Pronouns are a natural fit 
for a course survey question.

\paragraph{Class rosters}
Some schools allow students to specify their pronouns for class rosters, along with their names. If 
your school has this option, take advantage of the data by reading over it in advance, and 
encourage students to use the feature.

Absent institutional infrastructure, an instructor can still include in their syllabus a note that 
they will accept requests regarding names and pronouns not listed on the official class roster, for 
use on assignments and in class. This can significantly lower the barrier for students reaching out 
to communicate their preferences directly.

\paragraph{Self introductions}
Simply specifying your own pronouns when you introduce yourself signals that you are familiar with 
the concept and open for others to do the same. For example, when teaching a course, you could 
introduce yourself on the first day with your pronouns, in addition to your name and other general 
information.

\subsubsection{Principles and caveats}

\paragraph{Encourage, don't require}
It is a good idea to encourage all members of the scientific community to state their own pronouns, 
for the purpose of normalizing the practice. However, it is not generally a good idea to require 
people to state their pronouns. Some people are simply not comfortable specifying their pronouns. 
This includes some people who are transgender but don't feel safe coming out, and some people who 
are questioning.

A common piece of advice to faculty allies is to start (small) courses by asking students to 
introduce themselves using both name and pronouns. \emph{This advice should be treated with 
caution}. It does help to normalize the practice, but it may put actual transgender and 
gender-nonconforming students in a difficult position, as the attitudes of classmates are likely to 
be unknown.
Alternatives include (a) stating explicitly that it is optional to specify your pronouns, (b) 
modeling the idea by specifying your own pronouns, and (c) requesting pronouns in a less public 
way, such as via survey.

\paragraph{Be prepared to educate, and to set professional standards}
When you introduce a new mode of pronoun specification for widespread use in some group (e.g. 
pronouns on rosters, classrooms, conference badges, or surveys) be sure that you are prepared to handle 
confusion or backlash from the members of that group. This preparation does not have to be 
extensive. You may choose to provide specific educational materials (e.g. \cite{pronoun_spec}), or direct people to some of the resources listed in Appendix \ref{resources}. 
Regardless of your approach to education, it is critical to set guidelines for professional behavior around 
pronouns, lest the trans and gender-nonconforming members of your group end up bearing the full 
burden of education and explanation.

In particular, everyone should be aware that:
\begin{itemize}
\item A person's pronouns are not up for debate.
\item A person's correct pronouns should be used consistently, whether or not that person is present.
\item Making a mistake can be dealt with by a simple apology and correct use in all future 
interactions.
\item A person's pronouns are never an invitation to ask invasive questions about their gender 
identity or life experience, nor to quiz them about transgender or gender issues without invitation.
\end{itemize}

\paragraph{Be cautious with defaults}
It can be difficult to remember pronouns, just as it is often difficult to remember names, especially in 
the context of large groups. In this situation, it can be tempting to default to calling everyone "they".
However, if someone has clearly specified a pronoun other than "they," it is not polite to persist in
using the default. (One reason: unfortunately, some people use "they" as a convenient way to avoid calling 
trans women "she" and trans men "he," when the core issue is not forgetfulness but an objection to 
acknowledging that trans women are women and trans men are men. This may not be your intent in using 
"they", but it is the reality of many trans people's experience). If you forget someone's pronoun in the 
moment, it may be best to simply avoid pronouns (e.g. using their name as a substitute) until you can 
remind yourself, which you should do as soon as possible.

\begin{newquote}{Anonymous professor of physics, as told to the authors}{}
After my transition (male to female) I experienced discriminatory behavior from two of my colleagues. After many years of trying to get my colleague to stop misgendering me, I finally filed a formal complaint. The worst part was later when I found myself on trial during his appeal of the finding that he had discriminated against me. On top of that I had another one of my colleagues take his side. The emotional cost was very extreme due to the fear that he might win his appeal and I'd have to work alongside him for the rest of my career if he won his appeal.
\end{newquote}

      \subsection {Participate in surveys exploring LGBT+ experiences}
\label{surveys}
\BPGtag{A} \BPGtag{R}  \BPGtag{S} \BPGtag{P} 

Data collection is a vital component of diversity efforts. For any individual department or 
organization, it is necessary to evaluate the effects of existing policies and identify areas where 
improvement is required. For the larger academic community, an extensive, reliable data set allows 
constructive comparisons among departments and institutions, which may guide policy-making or even 
career decisions. The inclusion of LGBT+ demographic information and LGBT+ experiences in data 
collection is thus an essential element of formulating policies that are friendly to LGBT+ students, 
faculty and staff.

An important aspect of being a supportive chair is helping with the dissemination of research 
surveys. There is a dearth of data on the numbers of LGBT+ physicists and astronomers and their 
experiences within the academy. When you receive an email message asking you and your department 
members (students, faculty and/or staff) to participate in a survey, it is important that you 
distribute the email widely. This will help the greater community to collect the data necessary to 
understand what is actually happening and what issues need to be addressed.

\begin{newquote}{P.J. Carlino, comment on 2017 article \textit{Is Science Too Straight?}~\cite{moran:2017}}{}
As a gay STEM graduate I faced considerable discrimination in my first years of work, so much so that I quit my job as a chemist and went back to school. Having to re-start my career has had long-term and deep repercussions in my life. The study of the experience of LGBTQ professionals in STEM is helpful -- it might also be helpful for someone to survey LGBTQ professionals who received STEM degrees, but are no longer working in the field, to determine why they left.
\end{newquote}

      \subsection {Encourage faculty and staff to pursue diversity training}
\label{diversity-training}
\BPGtag{A} \BPGtag{R}  \BPGtag{S} \BPGtag{P} 
In seeking to develop an inclusive and supportive climate for LGBT+ members of one's department, it 
may be helpful to seek the assistance of a diversity training professional. This individual may 
provide specific sensitivity training for members of the department or offer other helpful 
resources. While department members may have the best intentions regarding inclusivity, some may not 
be fully aware of unconscious assumptions or biases that, when inadvertently expressed, can 
contribute to an adverse or exclusionary climate. A diversity training session or workshop can help 
alert department members to such potentially unconscious biases and signals, provide a forum for 
educating department members about best practices, and offer an opportunity for discussion regarding 
LGBT+ inclusivity.
Some universities offer professional diversity training on campus. Alternatively, a LGBT+ community 
center in the local community may provide contacts. These centers can be located through Center 
Link. Webinars are available through Campus Pride (see Appendix~\ref{resources}).

\begin{newquote}{Result from APS Report on LGBT Climate in Physics~\cite{LGBTClimateInPhysics:2016}}{}
Over one-third of LGBT survey respondents ``considered leaving their workplace or school in the past year'' after experiencing or observing harassment or discrimination.
\end{newquote}

      \subsection {Protect students' privacy}
\label{privacy}
\BPGtag{A} \BPGtag{R}  \BPGtag{S} \BPGtag{P} 
While departments are responsible for keeping all students’ records private, transgender students 
have additional privacy concerns that may not be obvious to well-meaning faculty members.
A student may be out in the college community, but not at home.  Disclosing a student’s transgender 
status to their family may risk their safety and/or financial support.  

Be especially mindful of 
any references to students that are visible to visiting family or that are published to electronic 
or print media.  
Verify that students are comfortable with the names and pronouns used in:
\begin{itemize}
 \item Departmental or college awards
 \item Listings of majors and minors (online and on bulletin boards)
 \item Photo captions in departmental or college newsletters
 \item Articles and photos on the college, departmental or research-group websites
\end{itemize}
You may need to intentionally work with the communications division of the college to ensure that 
names and pronouns are not modified to match outdated editorial policies.
\footnote{Most major style guides now affirm that people should be called by the name and pronouns 
of their choice (including, in many cases, explicit support for singular 'they' in the case of 
specific known persons who are neither male nor female). However, not all publications have updated 
their policies to match.} 

      \subsection{Support transitioning individuals}
\label{sec:support_transitioning}
\BPGtag{A} \BPGtag{R}  \BPGtag{S} \BPGtag{P} 

As an administrator, fellow faculty member, or fellow student there is much that one can do to 
support someone who is in the process of gender transition. The specifics of what a 
person can do to help will depend on what role they have in their department.
For everyone, the most fundamental advice is to always show respect for the person who is 
transitioning. The basics are by doing the following:
\begin{itemize}
 \item Use the name and pronouns that they prefer.
 \item If you mistakenly use the incorrect name or pronoun, quickly apologize and correct yourself, and move on.
 \item Do not make reference to their previous name or gender to anyone.
 \item Accept them as who they are. Do not judge them on their presentation, to their face or to others.
 \item Make it clear that you will support them even if they face discrimination from other members of the community (e.g. consider what students evaluations will look like)
 \item Do not out them without their permission, even to people who do not know them (i.e. do not mention their previous name or pronouns, or explicitly mention that they are trans).
 \item Correct others who are showing disrespect for the person who is transitioning and be a 
positive example of how to treat the person who is transitioning.
\end{itemize}
An internet search for ``how to respect a transgender person'' will provide many more 
examples of proper and respectful behavior. 
Take advantage of the LGBT+ support services on campus, supplemented by those listed in Appendix~\ref{resources} on page \pageref{resources}, as resources to learn more regarding how to best to support your colleague or 
student.
 
As a faculty member dealing with students, you can model these proper behaviors to your 
colleagues and your students. You may communicate privately with the person who is transitioning to 
inquire about anything you can do that would facilitate their transition. An example is to use a 
student's preferred name and pronouns in class even if their name has not yet been changed in 
University records.
 
As an administrator, you can help navigate challenges such as name changes on department records or 
department communications, and be an example to the rest of the institution on how to respect and
support the person who is transitioning. You might also find yourself 
contacted by those who are uncomfortable with one of your colleagues or students who is 
transitioning. For this, be prepared to model how to respect the person and show support 
for them. For example, students might protest taking a class from a professor who has transitioned. 
In such situations, you should communicate to the students that the status of the person who has 
transitioned or is transitioning has no bearing on their qualifications to teach or participate in 
department activities.

\begin{newquote}{Anonymous graduate student, as told to the authors}{}
When I came out for the first time, I only asked my colleagues and mentors for one thing: to use my name and pronouns correctly. I didn't want a lot of attention. Those who simply honored the request made my life a lot easier. They also earned my trust, making it easier to approach them for help when other issues came up.
\end{newquote} 

\begin{newquote}{Paige Flanagan, academic librarian, in ``Becoming a Woman.'' Chronicle of Higher 
Education, November 15, 2017~\cite{flanagan:2017}}{ 
http://www.chronicle.com/article/Becoming-a-Woman/241766?cid=wcontentgrid_hp_2}
My favorite moment: the faculty member who reacted to my transition almost as if I'd just informed 
her that I was getting married. She could hardly contain her excitement. That was an amazing feeling 
for me. If you are a cis-person reading this: How you react when someone initially tells you 
something this deeply personal can have a lasting impact. Go a little overboard, I promise you that 
we can use all the enthusiasm we can get during this process.
\end{newquote} 

\begin{newquote}{Anonymous cisgender professor, as told to the authors}{}
The evening after a colleague announced her transition via email, I went to a private place and practiced talking about her and her research using her new-to-me name and pronouns. This helped me avoid public mistakes that would have been awkward for me and painful for her.
\end{newquote} 

    \section{Welcoming}
      \subsection {Include everyone in social events}
\label{social-events}
\BPGtag{A} \BPGtag{R}  \BPGtag{S} \BPGtag{P} 
Department social events, whether on or off campus, are important opportunities for faculty, staff 
and students not only to network, but also to form a real community. Ensure that LGBT+ department 
members and their spouses, partners, and children are explicitly and implicitly invited to and 
welcome at these events, in the same way as their heterosexual and cisgender peers. For example, 
instead of inviting ``wives and husbands", invite ``spouses and partners" or ``significant others". This 
practice is especially important for new department members and for newly-out department members. Similarly, ensure that the same people are not always asked to host at recruitment dinners for prospective faculty or students; all department members of the appropriate level should have an opportunity to participate~\cite{bilimoria:2009}.
\begin{newquote}{Anonymous interviewee, APS report on LGBT Climate in Physics~\cite{LGBTClimateInPhysics:2016}}{}
It's ``don't ask, don't tell,'' [which leads to a] hard time networking because [my] mostly male colleagues [are] uncomfortable to invite [a] gay couple for outings etc. It's a subtle form of discrimination. Inability to network makes it difficult to join group grant proposals.
\end{newquote} 
When social events take place off-campus, ensure that the venues are welcoming of all department 
members. It is good practice to host social events at a variety of locations so as not to exclude 
department members based on the environment or activity. For example, bowling nights or drinks at a 
local bar should be supplemented by ice-cream outings or coffeehouse meetups. A diversity of events helps avoid a situation in which department members are inadvertently yet systematically excluded based on aspects of their identity. For example, evening events are difficult for department members with children to attend; meetups at bars can be unwelcoming to department members with religious, medical or personal reasons for not drinking alcohol; and many locations are not accessible for department members with mobility impairments.

\begin{newquote}{Sara, bisexual graduate student in engineering~\cite{cech:2011}}{}
When you're a minority, and especially when you are a discriminated-against minority, it would be great if I wouldn't have to worry if my professors were weirded out because I brought my girlfriend to social hour. If you're queer, just getting to that point ... is a lot of work ... It's just really hard to be isolated somewhere where people just don't understand what you had to do to get where you are in your life.
\end{newquote}


      \subsection {Use gender-neutral and inclusive language}
\label{gender-language}
\BPGtag{A} \BPGtag{R}  \BPGtag{S} \BPGtag{P} 
While it is true that most people in our society are heterosexual and cisgender, not everyone is. 
The heterosexual and cisgender norm is often unwittingly reinforced through our use of language. 
Just as the use of male pronouns as a default has a negative impact on women's motivation, and their
sense of belonging and identification with an organization~\cite{stout:2011}, assumptions in 
language can leave LGBT+ people feeling excluded. Here are some suggestions for gender-neutral and 
inclusive language:

\begin{itemize}
	\item {Use gender-neutral pronouns and phrasing} such as ``Bring your partner" instead of 
``Bring your wife", or ``Each student should bring their laptop" instead of ``Each student should 
bring his laptop".
	\item Avoid binary gender constructions such as ``he or she'' or ``ladies and gentlemen.'' Instead of being gender-neutral, these constructions exclude nonbinary persons. Instead, use ungendered default terms such as ``they,'' ``everyone,'' or ``students.''
	\item {Always use the name and pronoun of a person's choosing}. If you are unsure which 
pronoun a person uses, try to avoid using one until you can ask the person in private, ``How would 
you prefer to be addressed?" or ``What is your preferred gender pronoun?''  At the beginning of a 
semester, distribute a form to all students which asks for their name and pronouns along with any 
other information that an instructor might need (such as whether the student is on a sports team). 
Please see Section~\ref{sec:pronouns} for further advice on pronouns.
    \item Inform others if they do not use the correct pronouns for someone - many will have good intentions but simply not know or forget someone's pronouns.  Be careful about situations in which someone may not be ``out" to all people and always respect people's privacy.   
	\item {Avoid terms that sustain gender biases when describing titles or professions}. For 
example, use ``chair" instead of ``chairman", and ``custodian" instead of ``cleaning lady".
	\item Avoid using ``gay" or ``homosexual" as umbrella terms.  Use ``LGBT" or ``gender and 
sexual minorities'' to refer to a broad LGBT+ community.
	\item {Do not assume that all people are heterosexual and cisgender}, even in your own department.
	\item Remember that the term ``sexual orientation" is preferred over ``sexual preference."
The latter suggests a degree of voluntary choice.
	\item Do not split the class by gender, e.g., ``All the guys stand up.''
	\item Check with speakers and other visitors to find out what pronouns they prefer. It 
is especially important that this information goes to the person who introduces a speaker to the 
audience.
\end{itemize}
      
      \subsection {Allow name and gender changes on departmental records}
\label{name-changes}
\BPGtag{A} \BPGtag{H} \BPGtag{S} \BPGtag{P} 
Students, faculty members, and staff members sometimes change their names and/or genders from those 
originally given at enrollment or hiring, for reasons including gender transition and marriage. They 
may not pursue a legal change for various reasons, including (but not limited to) concerns about 
family disclosure. Ensuring that an up-to-date, preferred name\footnote{For examples of flexible 
policies on preferred names, see the University of Michigan \citep{namechanges:michigan}
and the University of Vermont \citep{namechanges:vermont}. Both institutions allow staff and 
students to specify a preferred name, for use across most university systems.}
and gender are used for departmental records -- including directories, awards, office nameplates, 
and letters of reference -- is an especially vital practical concern for transgender department 
members, who may face discrimination in applications for employment or for further education. 
Establish a simple way for individuals to change their names and genders in departmental files, and 
stress to faculty members that they should confidentially check with the student to determine what 
name and pronoun to use in reference letters. Always check with an individual before changing a name 
on any record, especially those that are publicly accessible. Be careful about access to, and storage of, 
historical records: an old name can reveal extremely sensitive information.

For more discussion of this issue from an institutional perspective, see Sec.~\ref{institutional-paper-trail} and~\ref{univ-username}.

      \subsection{Avoid gender assumptions for nominations and opportunities}
\label{gender_noms}
\BPGtag{A}  \BPGtag{R} \BPGtag{S} 

As part of both local and global efforts to mitigate the well-known gender disparity in physics and astrophysics, there is a wide range of opportunities especially earmarked for women, including prizes, conferences, and fellowships. If you are considering nominating a student or colleague for such an honor, please remember that their public, professional expression of gender may differ from their actual gender. A ``Women in Physics'' award could be received as an offensive and unwelcome surprise by a trans man (since it conflicts with his gender) or by a trans woman who is only ``out'' within her department. It is a good practice to check with a potential nominee \textit{before} nominating that person for an identity-based honor or opportunity. At the same time, try casting a wide net for women-in-physics conferences and networking opportunities in order to avoid excluding people who would benefit from them. Instead of advertising these opportunities only to a short list of people whom you believe to be women, send notices more generally to the whole department or to all students. This will also ensure no worthy and appropriate candidates are overlooked by accident.

\begin{newquote}{Anonymous transgender student, as told to the authors}{}
There was a group of us all hanging out during lunch. Someone walked by and asked each of us what we were interested in. ``Wow, so many girls in the sciences, that's weird,'' was their response. Not only was I hurt at the surprise in this person's voice, but I also felt like a fake. I'm a trans man. I'm not someone increasing the number of women in the sciences, I'm a man.
\end{newquote}

      \subsection{Set expectations for professional behavior}
\label{set-expectations}
\BPGtag{A} \BPGtag{R}  \BPGtag{S} \BPGtag{P} 
Departments, classrooms, and research groups are small communities, and every community has norms and expectations for behavior. For example, expectations regarding laboratory safety, plagiarism, group work, and exam-taking protocols are regularly communicated to students, due to the importance of these factors to the learning and research environment. Professional behavior is equally important. Slurs, harassment, and other exclusionary behaviors have dramatic and hostile effects not only on the targets, but also on observers. A recent analysis of survey data for STEM faculty identifying as gay, lesbian, bisexual or queer found that those who experienced exclusionary behavior were 4.9 times as likely to consider leaving the institution as those who did not, and those who merely observed exclusionary behavior were 3.4 times as likely to consider leaving~\cite{patridge:2014}. Both effects were statistically significant. A separate study of undergraduate students identifying as gay, lesbian, or bisexual found a statistically significant correlation between frequently hearing the derogatory (but undirected) phrase ``That's so gay'', and feeling left out on campus; suffering from more frequent headaches; and experiencing poor appetite~\cite{woodford:2012}.
In addition to the effects of such exclusionary behavior on the campus community, student interns or alumni who display unprofessional behavior in the workplace can damage the reputation of their home department.

Set department expectations for professional conduct, making clear that harassment based on race, 
ethnicity, sex, gender identity or expression, sexual orientation, age, national origin, disability, 
religion or class is unacceptable. Clarify that this includes all department spaces: classrooms, 
laboratories, study groups, and department-related chat or messaging channels. Communicate these 
department standards frequently -- for example, to incoming majors, in classroom syllabi, or at the 
beginning of each academic term. Enforce these standards by flagging inappropriate remarks or 
conduct and imposing proportionate consequences.

Many sections of this Guide contain specific advice for professional behavior toward the LGBT+ community, 
especially in situations perhaps unfamiliar to people outside the community. In particular, you may wish to consult:

\begin{itemize}
	\item Sec.~\ref{sec:pronouns} for pronoun use
	\item Sec.~\ref{sec:support_transitioning} for supporting individuals in transition
	\item Sec.~\ref{gender-language} for tips on inclusive language
	\item Sec.~\ref{offensive-language} for suggested classroom policies
\end{itemize}

\begin{newquote}{Savannah Garmon, as told to Physics 
Today~\cite{feder:2015}}{http://physicstoday.scitation.org/do/10.1063/PT.5.9034/full/}
 In the context of larger climate issues, it would be good if physics departments would think about 
the issues, and talk openly about them, so that people create an inclusive environment. We don't 
want an office environment that is completely sterile, but there is a difference between collegial 
banter, and things that border on someone being isolated. That message can be communicated in a lot 
of subtle ways, and the person communicating it may not even be aware that's the message they are 
sending out.
\end{newquote}

      \subsection{Reduce isolation in the department}
\BPGtag{R} \BPGtag{S} \BPGtag{P} 

Provide regular social and professional activities for students and postdocs to reduce isolation between research groups. Events such as poster sessions and cookie breaks provide junior researchers with valuable opportunities to network with department members outside their research groups.

For members of marginalized groups within the department, opportunities for collegial fellowship and self-expression are particularly important~\cite{marsh:2016}. Encourage local chapters of national organizations such as oSTEM \footnote{Out in STEM (oSTEM) is a national organization for LGBTQ+ people in science, technology, engineering, and mathematics. It has chapters at many different institutions, and hosts an annual national conference.} or Women in Physics. Consider subsidizing cookies or other refreshments for a departmental reading or discussion group on diversity and inclusion.

Make a practice of routinely encouraging discussion between visiting colloquium and seminar speakers, and students and postdocs. This might take the form of a student-centered question-and-answer session or a pizza lunch with junior department members. If students are asked to lead lunch outings with visitors or show them around campus, make sure that this honor is shared fairly among students, and that students who belong to marginalized groups are not overlooked. Likewise, make sure that all visitors have the chance to meet with students, not just visitors who happen to belong to marginalized groups themselves.

    \section{Advocating}
      \subsection {Join an Out List as an LGBT+ scientist or as an ally}
\label{outlist}
\BPGtag{A} \BPGtag{R}  \BPGtag{S} \BPGtag{P} 
Finding a mentor who is knowledgeable about and can address the concerns of an LGBT+ student can be 
a difficult process. When the LGBT+ Physicists group was created in 2010, one of the major concerns 
raised by attendees and members was networking and finding other LGBT+ people in the field. Before 
these conversations began, most LGBT+ physicists had not met another LGBT+ person in the field 
during their career.  Similarly, when LGBT+ astronomers began meeting informally at scientific 
conferences in 1992, their principal motivation was simply to develop a community.

Another way to raise visibility and provide targeted mentorship is to place one's name on a public 
Out List either as an LGBT+ scientist or as an ally. These lists allow students to find mentors and 
show leadership from allies. Some institutions already have such lists in place. There are also 
national lists, such as the LGBT+ Physicists Out List~\cite{lgbtphys_outlist}
and the 
Astronomy and Astrophysics Outlist~\cite{astro-outlist}. Signatories to these lists have stated a commitment to work 
against bias and discrimination in the fields of physics and astronomy.

As a potential mentor and ally, familiarize yourself with societies and organizations that work on 
behalf of LGBT+ people, both on your campus and beyond, so that you may recommend them to students, 
staff and faculty members who ask.  A list of these can be found in Appendix \ref{resources}.

      \subsection {Increase LGBT+ visibility within the department}
\label{visibility}
\BPGtag{A} \BPGtag{H} \BPGtag{R} \BPGtag{S} \BPGtag{P} 
Visibility and awareness are important aspects of promoting a positive departmental climate for 
LGBT+ people. Visibility and awareness of LGBT+ policies and even department members fosters an 
atmosphere of inclusion. This is particularly useful for students and faculty who worry about 
disclosure of their identities within the department. Visibility and awareness set the department's 
tone to be one of acceptance, encouragement, and focus on intellectual growth, regardless of 
identity. There are many easy ways to increase visibility and awareness within 
departments.

The most important step is distributing university anti-discrimination policies early, often, and 
widely. This can be done through postings in faculty lounges and in student common areas, as well as 
notes in graduate student and faculty offer letters and other correspondence. In their course 
syllabi, instructors can include both information on academic integrity and links to the 
non-discrimination policies of the university. A visiting weekend for graduate students is a great 
time to have a representative talk about diversity issues within the department and the larger 
community while also addressing the department's commitment to inclusion. Faculty candidate 
interviews are an appropriate time to include handouts about inclusionary policies such as health 
care policies, same-gender partner hiring, and all-gender restroom options.

If your university does not have official inclusionary policies, the department can draft its own 
statement to explain its stance and approach to diversity and inclusion. This will help to attract 
the best candidates by showing a strong and supportive community.

\begin{newquote}{Manil Suri in ``Why Is Science So Straight?'', op-ed in New York Times 4 September 2015~\cite{nyt:sciencesostraight}}{https://www.nytimes.com/2015/09/05/opinion/manil-suri-why-is-science-so-straight.html}
L.G.B.T. workers in STEM-related fields report significantly lower job satisfaction, both when compared to other STEM workers and to L.G.B.T. workers in other fields.
Such discontent -- and invisibility -- can contribute to a field's reputation for being unwelcoming. As a result, young lesbians and gays might sell themselves short, aiming for occupations with little use for their talents, but in which they see more people like themselves.\end{newquote}

      \subsection {Recognize and reward significant achievements}
\label{recognize-achievements}
\BPGtag{A} \BPGtag{R} \BPGtag{P} 
Recognizing significant achievements of LGBT+ department members communicates that their 
contributions are valued as highly as those of the rest of the department. Such recognition might 
include mention in a departmental newsletter or on a college or institutional website, nomination 
for a university or external prize, or an invitation to present a departmental colloquium.

The key point is that one needs to make sure LGBT+ department members are fairly considered for such 
recognitions, alongside all other department members. Any short list for an honor or recognition should be 
developed using as broad an applicant pool as possible; for a departmental award, a systematic review of all eligible candidates will ensure that no one is accidentally overlooked. For example, when identifying candidates for a graduate teaching award, the graduate chair might look over 
the teaching evaluations of all teaching assistants (TAs), or solicit suggestions from supervisors of all TAs above a certain experience level. 
This practice has been shown to be produce more generally inclusive shortlists than a simple brainstorming process, which often causes candidates from marginalized groups to be overlooked due to unconscious bias. The Mathematical Association of America's guidelines for award selections~\cite{maa-awards-bestpractices}, prepared in concert with the Association for Women in Science, has additional useful tips.

      \subsection {Evaluate service fairly and account for cultural taxation}
\label{cultural-taxation}
\BPGtag{H} \BPGtag{P} 
Academic members of marginalized groups often find themselves performing tasks above and 
beyond those expected of their colleagues owing to their marginalized status in the field. They may be 
asked to sit on committees to increase the diversity of perspective, be called upon (repeatedly) to 
educate individuals in the majority group about diversity, perform unofficial and/or extra mentoring 
of students and other colleagues, serve on affirmative action committees and task forces, act as 
liaisons between the administration and members of the marginalized community, give public lectures on 
diversity, work as troubleshooters or negotiators for disagreements due to sociocultural differences 
within the organization, or be asked to translate official documents or letters to clients or serve 
as interpreters \citep{pad94,bs09,jun15,ssfn17}. This additional sense of responsibility 
can be exacerbated by the strong desire such individuals often feel to give back to their 
communities, trapping them against the tide given the faster diversification of the student 
population under affirmative action, in contrast to the slower growth in available faculty advisors 
and mentors \citep{bla88,bol07,dallasnews:facultydiversification,tah+10}. 

\begin{newquote}{Audrey Williams June in The Chronicle of Higher Education \cite{jun15}}{}
The hands-on attention that many minority professors willingly provide is an unheralded linchpin 
in institutional efforts to create an inclusive learning environment and to keep students enrolled. 
That invisible labor reflects what has been described as cultural taxation: the pressure faculty 
members of color feel to serve as role models, mentors, even surrogate parents to minority 
students, and to meet every institutional need for ethnic representation.
\end{newquote}

Whereas each aspect of such service by itself is extremely important to improve the climate and 
diversity in academia (by ensuring fair hiring and evaluations of other faculty, for instance), it 
also presents an extra burden under which majority colleagues do not labor. \emph{Critically, such 
work and service often go unrecognized in formal assessments}, a phenomenon of 
essential, but ``invisible work'' \citep{ssfn17,grollman:invisiblelabor}. Even more insidiously, 
\emph{such service can actually harm the career of the faculty in question}, since service and 
mentoring has traditionally not been evaluated on the same footing as research and teaching at 
educational institutions \citep{bae99}. Collectively, this ``cultural taxation'' (also sometimes called a ``diversity tax'') leads to 
negative consequences for those doing such service, who then have less time overall to perform research. 
This harm is further compounded when members of marginalized groups such as women and faculty of color are held to higher 
standards than their white colleagues with regard to expectations for tenure and promotion \citep{pad94,ban00,blw04,mowa10,jh11,mat16}, 
regardless of allocations of effort in their position descriptions. 

We strongly emphasize the sentiment that  \emph{``service, especially that which furthers social 
justice, is important and meritorious, and should be rewarded''} \citep{bae99}, as it is important 
for the health of the the department, the field, and the institution. Specifically:

\begin{itemize}
\item Consider ways of rewarding extraordinary service. For example, your department might decide 
that documented, extraordinary contributions to advising and mentoring merit a reduction in 
teaching load or in other departmental service.
\item When considering a faculty member's service portfolio, work towards improving diversity or 
making the department climate more inclusive -- including for LGBT+ students, staff and faculty -- 
should be counted in the same way as any other professional service.
\item When evaluating the professional service of a professor or student, remember that those who 
are LGBT+, or who belong to other marginalized groups, will often do service focused on helping 
other community members in and out of academia.  These acts of service may be hard to quantify, 
and should be considered along with the candidate's portfolio in hiring and evaluations as a matter of explicitly stated policy~\citep{ymngh12}.
\end{itemize}

\begin{newquote}{Social Sciences Feminist Network Research Interest Group in the
Humboldt Journal of Social Relations \citep{ssfn17}}{}
Faculty of color, queer faculty, and faculty from working class backgrounds together [spend] a 
disproportionate amount of their time on the `invisible' work of academia, leaving them less time 
for the work that matters for tenure and promotion.
\end{newquote}


\begin{newquote}{Benjamin Baez in The NEA Higher Education Journal \cite{bae99}}{}
Service that seeks to improve the status of people of color in academia is essential. Without it, 
our faculties will remain predominantly white. The demand ... [should therefore be] not for 
deemphasizing service requirements but for reconceptualizing merit ...
Service, especially when it helps to eliminate barriers preventing some groups from fully participating in 
society, must be considered by institutions to be meritorious.
\end{newquote}

      \subsection {Include LGBT+ people in positions of power}
\label{positions-of-power}
\BPGtag{H} \BPGtag{P}  
As with other marginalized groups in physics and astronomy, LGBT+ scientists may encounter 
barriers to their academic or career advancement by virtue of exclusion from positions of power or 
opportunities for recognition - a phenomenon known as the ``lavender ceiling". While an atmosphere 
of ``tolerance" or friendliness may exist on an individual or interpersonal level, full inclusivity 
can only occur when LGBT+ persons have {equal representation in structures} that provide access to 
power, resources, and recognition. Indeed, it has been documented that the experience or observation 
of exclusionary behavior within a department is significantly correlated with an LGBT+ faculty 
member's likelihood to leave their institution for an appointment 
elsewhere~\cite{patridge:2014}.

Thus, an important ``best practice" closely related to visibility is the inclusion of openly LGBT+ 
members of a department in positions of authority and power. Such positions include department 
chair, assistant chair, or chairs of key committees that affect departmental governance (e.g., 
hiring, strategic planning, graduate admissions) as well as other ad hoc roles that could enable 
LGBT+ persons to have equal voices within the department.

    \section{Resourceful}
      \subsection {Help department members find resources}
\label{find-resources}
\BPGtag{A} \BPGtag{R}  \BPGtag{S} \BPGtag{P} 
Academic bureaucracy is notoriously complex, and it is often difficult for students and faculty to 
locate the information and resources they need to be effective -- especially when they belong to marginalized populations. 
Research shows that these scientists are less likely to be part of the 
informal information-sharing networks through which those in the majority gain much of their 
information about how to survive and thrive in the profession~\citep{sh08}.

Learn what resources are available on campus and then publicize 
them in a way that helps other faculty become part of the effort to be inclusive. As a starting 
point, consult the website of your local LGBT+ Resource Center (if one exists) or of the campus 
Diversity Office. Arrange a meeting with the director of the center or office to learn more about how 
your campus is working to support LGBT+ faculty and students and how your department can join these 
efforts. Engage the department leadership in distributing this information by inviting the director to make a 
brief presentation at a faculty meeting or meet with interested student groups such as oSTEM or Women in Science and Engineering 
(WISE). Ensure that campus resources that would be of use to students and faculty are publicized in a 
prominent section of the departmental website or graduate student handbook -- one visible 
to prospective as well as current department members.

      \subsection {Appoint a diversity liaison or committee}
\label{liaison}
\BPGtag{S} \BPGtag{P} 
It can be valuable for a department to appoint a faculty member, or a small committee of faculty 
members, as a climate/diversity liaison, to be a confidential advisor and listener for faculty and 
students who may be experiencing inclusion issues within the department. These liaisons could be listed 
alongside policy postings and in syllabi, and be introduced at student gatherings and welcome 
events. The liaison should receive training for the role, e.g., through the local LGBT+ Center or 
Women's Resource Center, to ensure that they know how to be effective, how to maintain 
confidentiality, and how to steer people to appropriate campus or community resources. 

Such a diversity liaison needs to be seen as available and approachable for department members. They 
should initially introduce themselves to faculty and students and let people know how to reach them, 
and also renew these conversations over time so people remain mindful of their role. They should 
also send out regular communications (e.g., via e-mail or department newsletter) that emphasize the 
department's ongoing commitment to inclusion and share useful campus resources. Moreover, they 
should be proactive in seeking input on diversity issues from the faculty and students, and in 
communicating general trends or concerns to the chair to ensure that these issues receive timely 
attention. The diversity liaison should keep conversations confidential, except where the law or 
university policy requires disclosure; when speaking with a student, staff member, or faculty member, 
the liaison should always make the limits of their confidentiality promise clear. Note that a diversity liaison 
must be able to bridge significant gaps of power and privilege, and not all faculty members will be able to fill the role well, 
even if they care deeply about diversity. The person or persons in this role must be carefully chosen.

      \subsection {Create safe spaces within the department}
\label{safe-spaces}
\BPGtag{A} \BPGtag{S} \BPGtag{P} 
Navigating campus life poses at least some difficulties for most students, but is even more 
challenging for members of sexual and gender minorities. Having a safe space 
for support and advice can make a big difference; just having such spaces available sends a powerful 
message of welcome and inclusion. Additionally, many well-meaning cisgender, heterosexual
faculty and staff members are interested in assisting LGBT+ students and colleagues, but worry about 
saying or doing the wrong thing.

To address these concerns, many colleges and universities operate {Safe Zone} programs. These vary 
from institution to institution, but participants typically receive diversity training, a briefing 
on university resources available for LGBT+ students, and a sticker with which to prominently mark 
their offices as safe areas for people wishing to discuss LGBT+ issues. If a Safe Zone program is 
available on your campus, encourage faculty and staff -- especially those with administrative 
responsibilities, such as the department chair, student liaisons, and staff -- to undergo training and work 
to make their offices safe spaces. This type of program has been shown to have a significant positive impact on climate
perception by LGBT+ students, faculty and staff, while increasing awareness and openness among cisgender, heterosexual allies~\cite{evans:2002}.

\begin{newquote}{Anonymous interviewee, APS report on LGBT Climate in Physics~\cite{LGBTClimateInPhysics:2016}}{}
I've identified two professors at [University] who are okay working with queer, LGBTQ people and one of them was actually my thesis advisor. And the reason I was able to identify him was because he had a little rainbow sticker on his window ... I was able to kind of receive whatever critique it was that he was giving me in terms of workstyle or homework sets whatever without having the stigma of being stereotyped for being queer or making him feeling uncomfortable because I might present something that may be queer or whatever. 
\end{newquote}


      \subsection {Provide equal access to restrooms}
\label{restroom_access_dept}
\BPGtag{S} \BPGtag{P} 
It is critical to provide non-gendered (a.k.a.\,gender-inclusive, gender-neutral, or all-gender) 
restrooms wherever physicists work and learn, and to have inclusive policies for access to 
gendered restrooms. Many transgender, intersex, and gender-nonconforming people are unsafe or 
uncomfortable in public restrooms. The US Transgender Survey~\cite{ustranssurvey:2016} revealed 
that in 2015, 26\% of its respondents had been denied access to a restroom, questioned or 
challenged about their presence, verbally harassed, or assaulted in a restroom. Many of these 
incidents occurred at work or school. The majority of respondents avoided public restrooms (69\%) 
and many limited their eating and drinking to do so (32\%). Cisgender people whose appearance does 
not match social expectations may also be harassed in gendered restrooms. In addition, a lack of 
non-gendered restrooms can pose challenges for families with young children, and/or individuals who 
need assistance in the restroom due to a disability.

All employees and students should have access to, and be aware of, a non-gendered restroom which is 
reasonably convenient to their office, classroom, or lab. If your department or workplace lacks 
non-gendered restrooms, contact a facilities administrator to request that one be converted. This 
may be a long-term advocacy project. Where possible, the simplest solution is to convert existing 
single-occupancy restrooms. Always keep in mind the need to preserve (or establish) equity in terms 
of restroom access for men and women. Avoid converting only one of a paired set of restrooms to 
non-gendered status unless there is a clear reason to do so.

\paragraph{Accessibility and amenities} 
Make sure that the available non-gendered restroom(s) include(s) an accessible stall, and is 
otherwise compliant with the Americans with Disabilities Act (ADA) (including having an accessible route for access). It is a good idea to 
provide changing tables and menstrual supplies in all bathrooms, not only those designated for 
women. Menstrual supplies are a basic need (on par with toilet paper) for most people who have a 
pre-menopausal uterus, including many trans men and non-binary persons.

\paragraph{Signage and maps} 
In order to be accessible, restrooms must be visible. Clearly post signs with directions to 
non-gendered restrooms in high-traffic areas. Signs may also be placed near gendered restrooms with 
directions to the nearest non-gendered alternative. All directional signs should include Braille 
lettering wherever possible. In the case of a conference or other event, distribute a map of the 
venue to attendees and be sure that non-gendered restrooms (temporary or permanent) are clearly 
marked.

\paragraph{Trans-inclusive policy for gendered restrooms}
While some individuals have a need or preference for non-gendered restrooms, this does not justify 
excluding trans people from available gendered restrooms. If a building or event has any gendered 
restrooms, there should be a clear (and clearly communicated) policy that trans individuals are 
free to use the restrooms which correspond to their gender identity. The campus human rights or 
diversity office can provide written policies to event organizers so that they will be able to 
address concerns regarding these issues with any of their attendees. If no explicit policy exists, 
event organizers can use the event as a prompt for their campus to create an inclusive policy on 
restroom use. 

\paragraph{Education}
Some individuals in your department or at your event may be confused by the need for a non-gendered 
restroom or an inclusive restroom access policy, or may even be upset by these things. Be prepared to 
explain the motivation, and know that you may need to handle objections. 

\paragraph{Legal concerns}
Be aware of state and local laws, as well as institutional policies, which affect restroom access. 
Depending on your goal (venue selection, restroom conversion, etc), different rules may affect the 
process. These rules include:
\begin{itemize}
	\item State and local ``bathroom bills''~\cite{tracker-bathroombill}
	\item State and local non-discrimination policies~\cite{tracker-employmentdiscrimination}
	\item Institutional non-discrimination policies
	\item Building codes
	\item Institutional facilities policies
\end{itemize}

If a discriminatory rule is in place which prevents you from securing appropriate facilities for 
your workplace or event, consider working to change it. This may be a long-term advocacy project. 
However, please note that visible political action around restroom access can also spark backlash 
against transgender individuals. For this reason, it is a good idea to consult and coordinate with 
campus or local LGBT+ organizations when pushing for a policy change.

\begin{newquote}{JC Salevan, as told to Physics Today~\cite{feder:2015}}
{http://physicstoday.scitation.org/do/10.1063/PT.5.9034/full/}
It bothers me that there are not gender-neutral facilities on the campus. I don't want to have to 
make awkward decisions about bodily functions when I could be thinking about the experimental 
problem I am trying to solve.
\end{newquote}

      \subsection {Increase networking opportunities}
\label{networking}
\BPGtag{A} \BPGtag{R} \BPGtag{P} 
The importance of forming effective networks for marginalized groups in physics and astronomy is 
well understood. Scientific networks provide access to mentoring, job opportunities, material and 
emotional support, potential collaborators, and recognition and dissemination of work. They can also 
be catalysts for instituting beneficial changes in policy. These same benefits hold true for LGBT+ 
people. While many institutions have valuable assets such as a Gay-Straight Alliance or an LGBT 
center, these groups are rarely able to support an individual simultaneously in their sexual and gender 
identity and also in their role as a scientist. Moreover, these organizations do not always cater to graduate 
students, postdocs and faculty. There's a need, therefore, for networks that explicitly address all 
aspects of identity at an appropriate level for a person's career stage.

Supportive department heads should reach out to the Gay-Straight Alliance or LGBT Center, if these 
exist at their institutions, to identify resources and networks that already exist, and should 
suggest to the leaders of those groups that it is necessary to make LGBT+ people feel welcome and 
supported explicitly in their scientific context. Make students and faculty aware of national 
networking organizations such as oSTEM and the National Organization of Gay and Lesbian Scientists 
and Technical Professionals (NOGLSTP). Provide travel support for LGBT+ students to attend relevant 
networking meetings such as OUT for Work, NOGLSTP's OUT to Innovate conference, and oSTEM's national 
conference, as is already best practice for networking conferences focused on scientists who are female, 
African-American, and/or Hispanic~\cite{rosa:2016}. Finally, if your school lacks an oSTEM chapter, encourage the 
formation of such a group.

Groups like oSTEM often have a hard time raising money, because they are not affiliated 
with the campus LGBT center and may have only one or two members from any particular department.  
Department chairs can help by donating money and encouraging the other STEM department chairs 
to do likewise. A commitment to ongoing support is even more valuable than a one-time donation.
      \subsection{Mitigate the negative effects of poor advisor-advisee relationships}
\label{sec:mitigate_hostileadvisor}
\BPGtag{A} \BPGtag{P}  
For a student or postdoc in a group that they perceive as unwelcoming, unsympathetic, or even hostile, the social pain is magnified by its professional consequences. An unsympathetic or hostile advisor may hinder degree progress, write a poor letter of recommendation, or fail to provide the normal mentorship that should smooth their advisee's path. A department leadership can mitigate this damaging isolation by taking steps to broaden all students' and postdocs' professional networks, providing additional potential letter-writers and mentors.

PhD students should be encouraged to form committees relatively early in their studies, and to meet periodically with committee members besides their advisors. Such a meeting requirement could be formalized in an annual review process for PhD candidates. Students should also be encouraged to form closer professional relationships with particular members of their committees by designating a formal Faculty Mentor distinct from the advisor. In many departments, some faculty administrators (such as a chair of graduate or undergraduate studies, or a diversity liaison) are natural points of contact for a student facing trouble in the group. Persons in such student-facing roles should be aware that diversity, inclusion, and dispute resolution are critical parts of their jobs. Such administrators should take diversity training (including SafeZone training, see Sec.~\ref{diversity-training}) and take steps to make themselves welcoming to persons from all demographics.

For postdoctoral researchers, there is even less formal structure for exposure to the larger department. Encourage postdocs to form connections outside their individual research groups wherever possible -- for example, by supporting a postdoc lunch group or by facilitating regular get-togethers between research groups in the same general area. Minor administrative responsibilities -- for example, managing a seminar series -- can help reduce isolation without having an undue impact on research.

\begin{newquote}{Anonymous interviewee in Bilimoria and Stewart, 2009~\cite{bilimoria:2009}}{}
{[}O{]}ne person's mentor had said of another person, ``I think she's a lesbian; I'd never trust her data.''
\end{newquote}

      \subsection{Take harassment, discrimination, and hostility seriously}
\label{sec:serious_situations}
\BPGtag{A} \BPGtag{R}  \BPGtag{S} \BPGtag{P} 
When resolving disputes, department administration should ensure that the resolution does not punish the victim, and that consequences do fall on harassers. For example, it is unfair to separate a harasser and their target by restricting the latter's access to the laboratory or to a useful course. If the victim is a student, be mindful of how the inherent power imbalance between the student and a professor puts the student at a disadvantage. A pattern of unjust resolutions tends to discourage victims from lodging complaints, and allows hostile and unsafe situations to fester over time. In addition to considering fairness issues when resolving a complaint, the department administration can proactively demonstrate their seriousness by following the institution's non-discrimination policy and by ensuring good connections with the institution's Title IX office (Sec.~\ref{univ-titleix}), including communicating the office's role to students and postdocs who may not be familiar with it.

It is sometimes tempting for advisors or administrators to allow a bad situation to slide, especially if the perpetrator is a faculty member, or a student close to graduation. Keep in mind that harassment and discrimination are highly unprofessional behavior, and that other members of the group and department are watching and learning from the actions (or inaction) of their leaders. Just as good work by alumni reflects well on the department, unprofessional acts by alumni will damage the department's reputation with their employers and colleagues, and may negatively affect the opportunities available to future graduates. Most importantly, tolerating unprofessional behavior creates a hostile environment that prevents students, staff and faculty from thriving and doing their best work.

\begin{newquote}{Anonymous interviewee, APS report on LGBT Climate in Physics~\cite{LGBTClimateInPhysics:2016}}{}
I was sexually harassed for multiple years within my physics department. Despite repeated attempts on my part to discuss the matter with other students, faculty, and the department head, I was consistently shut down, told that I was overreacting, or misinterpreting the other student who was consistently given the benefit of the doubt in contrast.
\end{newquote}

\begin{newquote}{Barthelemy et al. 2016~\cite{barthelemy:2016}, summarizing interviews of female grad students in physics and 
astronomy).}

The [lab] space ... was not one of just science. It was one where she had to clean, organize, and 
do tasks the men in her group were never assigned. For [other women] this was also true. When they 
attempted to conduct research at a telescope and national lab, respectively, male graduate students 
attempted to block their access to equipment and data. For [another female interviewee], her 
physical safety was in jeopardy whenever she entered the physics building ... The physical space 
became one of non-science-related work, confrontation, and bodily danger.

\end{newquote}
  
      \subsection{Provide flexibility for people trying to escape a hostile colleague or advisor}
\BPGtag{A} \BPGtag{P} 

Having a safe space to work is tremendously important to productivity and well-being. Maintain a policy of respecting requests to shuffle shared-office assignments due to conflicts between officemates, even in the middle of the academic year. Avoid absolute requirements for experimentalists to sit in the laboratory at all times; maintain some available office space so that graduate students and postdocs have another place to work on papers and analyses when the lab is a tense environment.

Sometimes the best solution to a hostile research environment is a change of group, which may impose significant professional and financial costs on a student. Be aware of possible university or department requirements that a student have an advisor in order to be considered in good standing. Maintain some temporary bridge funding for graduate students who have to switch groups due to serious conflicts, and who need funding to finish out a term or to try out a new group for one term. Encourage faculty members to take on ``orphaned'' graduate students, at least on a trial basis. Whether or not the switch is a good research fit, this is a service to the department, and should be recognized.

The solutions outlined here impose potentially significant costs on the harassed individual, who must change their working space or even their research area. While this may provide the quickest resolution to an immediate, intolerable environment, it is essential to impose significant consequences on the harasser, as well. Unprofessional behavior should not be tolerated in the department (Sec.~\ref{sec:serious_situations}).

      \subsection{Encourage good group practices}
\label{sec:good_groups}
\BPGtag{R} \BPGtag{P}
Encourage principal investigators (PIs) and group leaders to provide collaboration contracts or advising agreements for members of the group spelling out expectations (e.g. no harassment or hostile language; cleanliness in the lab; timely replies to emails; vacation time; etc). A good contract delineates rights and responsibilities on both sides -- PI and student or PI and postdoc -- and can prevent many misunderstandings, both trivial and major. Examples and guidelines may be found at various institutions, e.g.~\cite{sample_contract}.

Most research groups have regular or intermittent chores necessary to group success. Depending on 
the group, these include regular cleaning and maintenance of lab equipment and supplies,  
management of a group website or computational cluster, taking minutes for group meetings, or 
making laboratory purchases. Chore assignment has a large effect on group dynamics and is often 
vulnerable to undetected bias or expectations, even when done on a volunteer basis. Encourage PIs to 
institute chore schedules. Even in cases where chores cannot or should not be alternated, a formal 
assignment helps ensure that group members receive credit for their service.

\begin{newquote}{``Janet'', pseudonymous interviewee, a woman pursuing a PhD in Physics 
~\cite{barthelemy:2016}}{}
My place in the lab can at times feel uncomfortable ... I am basically the lab secretary.
\end{newquote}
  
\usechapterimagetrue
\chapterimage{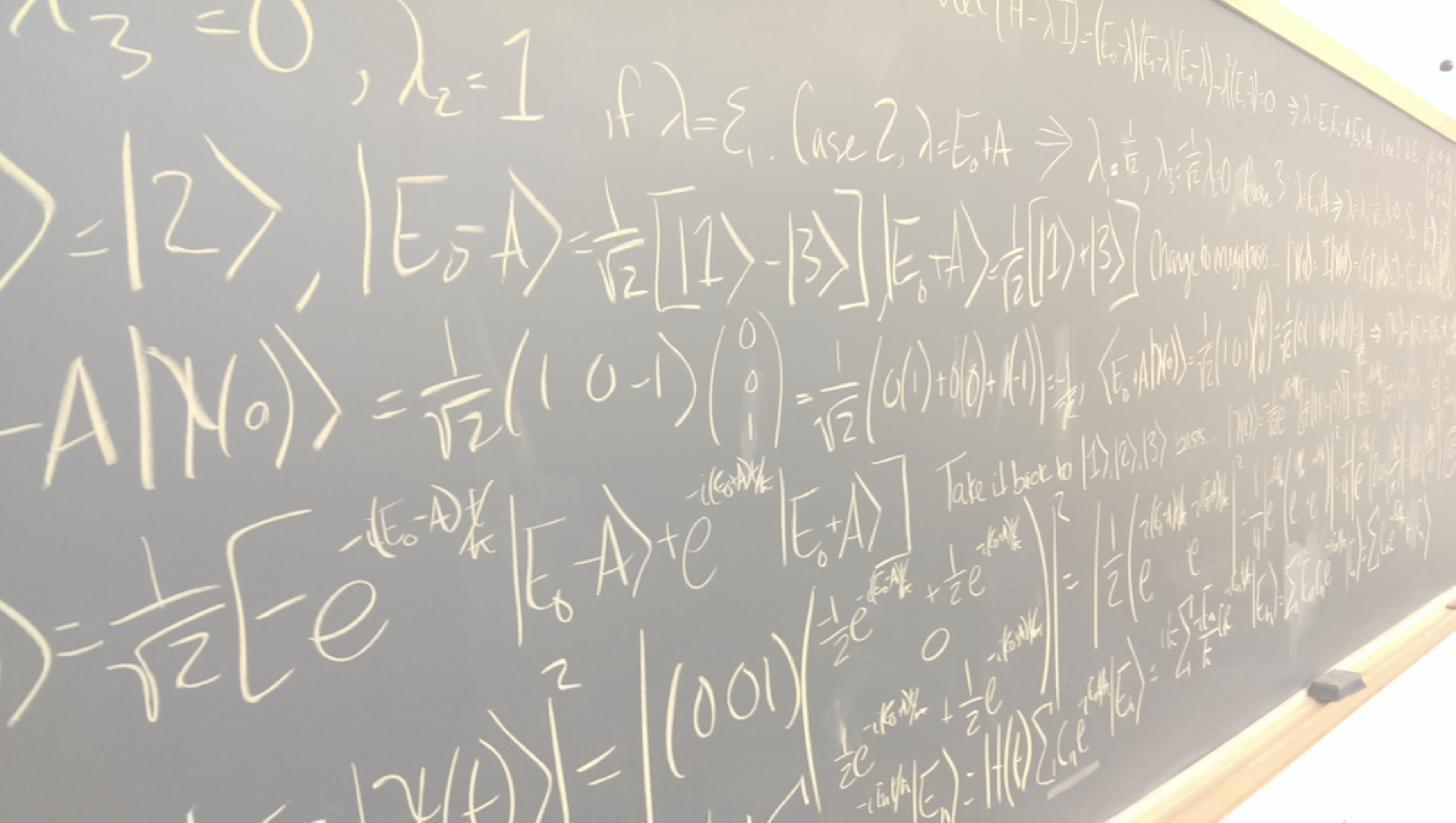}
\chapter{In the Classroom}
\label{ch:classroom}
\usechapterimagefalse

\begin{newquote}{B. E. Hughes in 2018 study of STEM retention of LGB students~\cite{hughes:2018}}{}
...{8\% fewer sexual minority students were retained in STEM fields than their heterosexual peers after 4 years in college, and this likelihood increases to nearly 10\% when controlling for other factors that support retention in STEM ... What was surprising was that sexual minority STEM aspirants participated in undergraduate research at higher rates, but were still less likely to be retained to the fourth year.} 
\end{newquote}

\medskip\noindent
Instructors and advisors have a responsibility to create a positive
environment in classes and research groups that facilitates learning
for \emph{all} students. Assumptions about gender
identity and sexuality in interpersonal interactions or course content
can easily contribute to a climate that is distracting or hostile
for students. Instructors play a key role in shaping student experiences
because they select material, model behavior, and manage TAs and other
supporting staff. This chapter gives suggested practices for instructors
to make their courses as inclusive as possible. 

\begin{newquote}{Anonymous interviewee, APS report on LGBT Climate in Physics~\cite{LGBTClimateInPhysics:2016}}{}
...{[}W{]}hen it comes to asking questions or completely having no idea about how to start a problem{[},{]} it becomes injurious to me in terms of being able to progress through a course because professors may have made the stereotype that I already came in with this low level of aptitude ... So in terms of bringing on the queer aspect to that ... it's kind of really difficult to deal with both at the same time ... I think in the long run it has definitely been very difficult for me to have confidence in my abilities. And I can for sure tell you that my grades have suffered because of that. 
\end{newquote}

\section{Importance of course climate in learning}
\label{sec:course_climate_importance}
\BPGtag{D} \BPGtag{P} 

Course climate refers to how welcoming a course as a whole is to students
of all backgrounds and identities. It is influenced by a number of
factors, including choice of subject matter, attitude and language
used by the instructor and TAs, and the nature of interactions among
students (see~\cite[Chapter 6]{abm+10} for a review). Each student
may perceive the climate differently, experiencing anything from overt
hostility or discrimination, to implicit marginalization, to an explicitly
welcoming environment \cite{dc94}. A student who identifies as belonging
to a marginalized group due to their race, gender, class, LGBT+ status, religion, disability status,
or nationality, is particularly likely to experience a negative climate
due to stereotypes and prior assumptions on the part of the instructor
about the students in the classroom \cite{sa95}. Additionally, the
student may overhear classmates using exclusionary language, be the direct
target of such remarks, or feel excluded by classmates during team
projects or group work. Over time, these experiences can have a corrosive
effect. 

Minority-identifying students may face additional hurdles to successful
learning in a negative climate because their natural reactions can
disrupt their cognitive processes. If the classroom climate is hostile,
they are less likely to ask questions, join study groups, or attend
faculty office hours, and are more likely to skip class sessions altogether;
these patterns can lead students to lag behind and underachieve in
the course. The effect can be exacerbated if the student has been
experiencing rejection outside the classroom as well (e.g., lack of
support from family or friends). Ultimately, affected students may
lose their motivation to continue with their chosen discipline and may
switch to one where they perceive the climate to be more congenial
\cite{mss+98}. Particular challenges exist when working to improve 
the climate for LGBT+ students, because they are less likely to be 
visible than other minorities, discrimination against LGBT+ individuals 
is still pervasive, and relatively few role models of LGBT+ scientists 
are presently available~\cite{LGBTClimateInPhysics:2016}.

Students who identify with multiple marginalized groups may have
different experiences from others identifying with one, and typically face additional
challenges. This is important, because diversity interventions intended to address
climatic or discrimination issues along one axis of identity may fail
to meet the needs of students with multiple marginalized identities. 

\section{Before the course}

\subsection{Include welcoming language and non-discrimination policies in your
syllabus}
\label{course-expectations} 
\BPGtag{D} \BPGtag{P} 

Explicitly establishing an inclusive classroom environment helps
both LGBT+ students and those with other marginalized identities to succeed,
and sets the standard for the students to treat each other with respect.
Instructors should set policies for course climate in their syllabi,
providing links to institutional and departmental policies where they
exist. Effective syllabus statements establish shared accountability
between the instructor, TAs and students, and clearly communicate expectations
rather than general principles. Typical statements address student
workload, academic integrity, classroom climate, academic accommodations,
Title IX, pronouns, and recording of classes. Many colleges provide recommended
language for syllabi; instructors should modify and supplement these
where necessary to match the particular context of their course. 

\subsection{Establish how your students wish to be identified}
\label{subsec:precoursecurvey}
\BPGtag{D} \BPGtag{P}

Ideally, an institution's student information
system should provide the course instructor the name that each student
and TA
should be called and the correct pronouns to use. Sometimes, this
information is either not provided to instructors or may contain misleading
information, e.g. some university record-keeping systems only provide
the name on the student's original application form. It is particularly
important to use the correct name for trans students, 
because calling them by an incorrect name 
in the classroom could out them (i.e., make their trans identity public
without their consent). Other groups of students also benefit from
this information being available to instructors: not all cultures
use gendered names and some students may use one name officially but
a different name informally. As discussed in Section~\ref{sec:pronouns},
departments should lobby to allow name changes in official records
and provide this information to instructors.

Where this information is not available to the instructor, an online
pre-course survey is a good practice. Such a survey might include
questions like \emph{``What name do you wish to be called in this
course?''} and \emph{``Which pronouns should the instructor use?''.
}For the latter, it is important to include a write-in box. Pre-course
surveys can also be used to gain information on students' preparation,
background and goals for the course; these can be used by the instructor
to better frame the material. 

\subsection{Educate faculty and TAs on climate issues}
\label{sec:train-faculty-climate}
\BPGtag{D} \BPGtag{S} \BPGtag{P}
Departments should provide opportunities for instructors and TAs to
educate themselves about the impact of course climate on marginalized
students and how to make their classrooms more welcoming for them.
This might include discussing the issue at a department meeting or
teaching seminar, incorporating a suitable book or article into journal
clubs, or inviting an education researcher to give a colloquium. Some
possible topics to discuss include the contents of the Classroom chapter of
this Guide, language used in the classroom, breadth
of role models available to students, welcoming language in course
materials, and prior assumptions implicit in questions. Departments
should also encourage instructors to share examples with their colleagues
and TAs of how they cultivate a welcoming climate, for example by
integrating positive diverse role models of physicists and astronomers
into their classes. This can help build a more inclusive teaching
culture in the department as a whole.

Prior to the start of their course, instructors should ensure TAs
are familiar with course non-discrimination policies and provide relevant
TA and course-specific training. This training could include grading
procedures and policies to reduce implicit bias, appropriate conduct
for interactions with students and pedagogical methods relevant to
the course structure. It is, as will be discussed in Sec.~\ref{class:monitorclimate}, 
important to continuously monitor climate and support TAs during
the course. 

\begin{newquote}{Anonymous interviewee, APS report on LGBT Climate in Physics~\cite{LGBTClimateInPhysics:2016}}{}
A gay student was openly mocked by a professor in front of the class, most of whom laughed in agreement. 
\end{newquote}

\subsection{Design for inclusion}
\BPGtag{D} \BPGtag{P}

Instructors can create an inclusive learning environment within individual
courses using a variety of research-based techniques. For example, interactive pedagogical methods such as Peer Instruction can both
increase the degree to which all students learn and alleviate gender
gaps in student performance in introductory physics classes~\cite{lor06}.
Other techniques are more specifically aimed at countering stereotype threats
and other barriers to the success of students from marginalized
groups. University Teaching Centers or LGBT+ Resource Centers may
be able to provide suitable training sessions (e.g., for Safe Zone
programs, see Sec.~\ref{safe-spaces}) or even fellowship programs to help instructors learn these
techniques. Departments should support instructors using these techniques
and facilitate sharing of best practices. 

\section{In the classroom}

\subsection{Communicate expectations on the first day of class}
\label{class:firstday}
\BPGtag{D} \BPGtag{P}

The first day of class communicates to students the objectives of
the course, its tone, and the instructor's teaching style, and establishes
classroom behavioral norms~\cite{provitera2001successful}. It is therefore important to use
the first class to set explicit expectations for a positive climate.
One way to do this might be to include a slide such as the following,

\begin{newexample}
In this course, I strive to provide an inclusive
climate in which each student 
feels welcome and free to question, contributes to the discussion, thrives, and learns, 
independent of gender identification, race, sexual orientation, ethnicity, disability,
economic background, national origin, or religious affiliation. 
I expect students to contribute actively to this learning environment
through open and respectful verbal and written communication. 
Discrimination or harassment of any form will not be tolerated. 
I also welcome any suggestions for improving the learning environment. 
\end{newexample}

\subsection{Do not tolerate offensive language}
\label{offensive-language} 
\BPGtag{D} \BPGtag{P}

Institutions and departments should adopt a policy that racist, sexist,
homophobic, transphobic, ableist, and ethnic slurs and jokes are unprofessional
and will not be tolerated by any course participant. Base the policy
on the institution's non-discrimination statement and on the need
for a departmental climate in which all are welcome and encouraged
to do great science. Remind department members about this policy regularly,
and look online for guides to identify words commonly used that \href{http://onlinelibrary.wiley.com/doi/10.1002/whe.10235/full}{may be exclusionary to some  people}.
This information should also be made available to TAs as part of anti-bias
training. 

In personal interactions, point out offensive language and ask that
it stop. Make clear that such language is unprofessional and unwelcome
in your department. Many people let homophobic language slide because
they do not know what to say. Saying something (even if it is challenging
to do so) is always better than saying nothing. In many cases, the
offensive connotations of the word or phrase may simply not be known
to the speaker.

In the classroom, if a course participant makes an inappropriate comment
or remark, make it clear that the language is unacceptable and refer
to the classroom expectations established in the syllabus and first
day of class. Remember that this should not be limited to offensive or hateful language, but also to comments that assume everyone is cisgendered and heterosexual.  These types of statements can lead to your LGBT+ students feeling excluded in your classroom and physics in general.  Provide opportunities for other students to speak, and
additionally a means (e.g. by a web form) for other students to express
their views anonymously. Handling such incidents is often challenging
for instructors, because it requires swift action, so departments
and institutions should provide training on how to do so. 

\begin{newquote}{Anonymous interviewee, APS report on LGBT Climate in Physics~\cite{LGBTClimateInPhysics:2016}}{}
I have witnessed hostile commentary made about LGBTQ people as being unnatural during an SPS [Society of Physics Students] campus chapter meeting. It made other students feel uncomfortable.  
\end{newquote}

\subsection{Intervene in problematic scenarios}
\label{class:intervene}
\BPGtag{D} \BPGtag{P}

We have already stressed the need to establish a classroom climate
that does not tolerate offensive language or behavior, but there
may be many more subtle scenarios that require intervention. This
is especially true in small-group discussions or team-based learning
approaches, in which students spend substantial time interacting with
each other. While small-group discussions have significant pedagogical
value, they may also be places where marginalized students repeatedly
experience microaggressions \cite{due_who_2014}. Further, while teamwork
skills are highly valued by employers and important in everyday life, they are rarely explicitly
taught in physics classes. Even where group work is a component of
courses, feedback on improving teamwork skills may not be explicitly
given by the instructor. 

Instructors using group work should therefore view teamwork skills
as an important auxiliary goal of the course and provide recommendations,
best practices, support and feedback to improve them. The importance
of climate should be an explicit part of these. A good practice is
to use student self-evaluations and include in these an assessment
of the team's dynamic and climate. These could be achieved by an online
survey post-activity incorporating questions such as ``how did each
member contribute?'', ``how well did the team work together?''
and ``how comfortable was the team's climate?''. It is important
that this is required for each team member individually since the
nature or tone of conversation may be different if it is filled out
by the group. Instructors should act on this information by providing
individualized and general feedback to students to improve their teamwork
skills and modifying grades where necessary. 

\subsection{Use correct names and pronouns}
\label{sec:classroom_pronouns}
\BPGtag{D} \BPGtag{S} \BPGtag{P}

As discussed in Sec.~\ref{sec:pronouns}, it is important to use correct pronouns to avoid
misgendering people. This is especially true in the classroom, where
an instructor may be addressing a student in front of tens or hundreds
of their peers. 

Instructors should guard against making assumptions about gender:
For example after a student asks a question, an instructor might later
refer to ``her question...'' or ``as he asked,''
using masculine or feminine pronouns based on perceptions of the student's
appearance. If the student does not use these pronouns, they may find
the situation uncomfortable or distressing, and other students may unknowingly
proliferate the error. How students wish to be identified, including
pronouns, is best determined prior to the course as described in Sec.~\ref{subsec:precoursecurvey}. 
This ensures that the instructor knows the pronouns of the students
in their class and avoids requiring any student to publicly ``out''
themselves.

\begin{newquote}{Anonymous student, as told to one of the authors}{}
When I was a first year undergraduate I was still very insecure in
my gender identity ... The first time I was asked for my pronouns was
in a public setting, where everyone in our building said their name,
pronouns, and where they were from. Part of me was glad I had a chance
to assert my identity, but coming out in front of a room full of strangers
was so overwhelming that I ran outside and ugly cried for about half
an hour afterwards.

A better experience was in my first undergraduate physics class, when
the professor asked for pronouns in an online survey. I was both relieved
I wouldn't have to out myself in front of strangers again and grateful
that she cared enough to ask. When she called on me in class, she
used my correct pronouns without skipping a beat, as if it were obvious
that was what I was supposed to be called. I never once felt uncomfortable.
Most of the other students just followed suit without question. \end{newquote}

\subsection{Use inclusive examples and analogies}
\label{class:inclusive_examples}
\BPGtag{D} \BPGtag{P}

While physics itself does not directly relate to gender, sexual orientation,
race or religion, examples and analogies found in textbooks or created
by instructors often have a cultural context that may invoke some of
these concepts. The impacts may range from confusing some students
to creating an unwelcome environment for others. Instructors should
critically evaluate the examples, analogies, and images that they
use in their classes and presentations. It is often helpful for colleagues
to review each other's notes to provide additional perspective. 

Below, we provide a rubric for reviewing such materials:

First, look for examples or analogies that should be removed. While race
or religion rarely appear in instructional examples in physics, references to gender
and sexual orientation occur with a surprising frequency. For instance,
Coloumbic attraction is sometimes explained through analogies to romantic
attraction. Not only does this imply an incorrect binary and fixed
view of gender, it also implies an exclusively heterosexual view of
romance. Any analogy, joke, or comment related to dating, sex, or
gender is unsuitable for an inclusive physics classroom and may create
a hostile climate for students.

Second, look for ways to use a greater variety of real-world contexts.
Instructors often use examples that they are familiar with, but those
will be most relevant to students who have similar identities.
Sports are one common case: baseball will be familiar to one subset
of students while hockey will be familiar to another. Often, however,
these sports have geographical and class connotations and international students
from cultures where a sport is not familiar may not understand the
terminology. While no set of examples will work for everyone, most
physics concepts can fortunately be presented in vastly different
contexts. Good instructors use a diverse range of analogies, explain
them clearly for students who lack the immediate context to understand
the analogy, and discuss the strengths and limitations of the analogy. 

\begin{newquote}{Lisa, bisexual undergraduate student in engineering~\cite{cech:2011}}{}
One of my friends who is a mechanical engineer was describing the body as a mechanical engine that only functions under various strains and stresses and relationships. And he didn't think that gayness was one of those relationships ... basically, `the man is the plug and the woman is the outlet and if there are two plugs, how is [anything] going to charge?' 
\end{newquote}

\subsection{Watch for implicit bias}
\label{class:implicitbias}
\BPGtag{D} \BPGtag{P}

Implicit bias \textemdash attitudes or stereotypes that affect our
understanding, actions, and decisions in an unconscious manner \textemdash has
been shown to influence auditions, hiring, and grading \cite{hofer_studying_2015}.
Studies on implicit bias have typically focused on race and gender,
but LGBT+ people have been shown to face bias in some situations \cite{badgett_bias_2009},
indicating that LGBT+ students may also face implicit bias in the
classroom. 

Instructors should take anti-bias training to assess their own biases
and increase the equity with which they treat students in interpersonal
interactions. One example scenario is that instructors should pay
particular attention to the range of students they call upon to ensure
that a few students do not dominate the discussion. This can be handled
sensitively by using language like \emph{``I want to make sure we
hear from everyone. I want to hear from someone who hasn't spoken
yet''} and by waiting for additional students to indicate their willingness
to answer. Instructors should provide positive reinforcement for all
answers, including those that aren't correct, and encourage students
to build upon and improve other students' answers where helpful. 

Due to the risk of implicit bias, grading should be performed anonymously
where possible. This could be facilitated by asking students to submit
work using student ID numbers rather than names \cite{john_m._malouff_risk_2013}.
Where possible, work should be graded by more than one person. 

\subsection{Encourage equitable participation in groups}
\label{sec:study_groups}
\BPGtag{D} \BPGtag{P}
Students in marginalized groups often report 
exclusion or isolation from their peers. Study groups, an
essential educational and support structure in many physics classes, are
a place where this isolation becomes both obvious and educationally
damaging~\cite{rosa:2016}. Students may directly exclude
marginalized peers from study sessions or intimidate peers from
participating fully~\cite{heller:1992}. Such exclusionary behavior
may be deliberate or inadvertent, but in either case it is harmful.

\begin{newquote}{Anonymous interviewee, APS report on LGBT Climate in Physics~\cite{LGBTClimateInPhysics:2016}}{}
...I don't know if it was based on my gender or my sexual orientation. But I was very out [as an undergraduate]. So it could have been either or both. I know that all of the other students, literally all of them, studied together and did their homework together and all of that. And I tried to participate in these things and was often, you know, given the runaround on the times, and I just stopped trying after a while and stopped interacting with them socially.
\end{newquote}

When assigning groups, teams, or partners, make sure to facilitate positive behaviors and interactions, especially for students who may be marginalized.  Many educators recommend forming groups in a way that does not isolate marginalized students, such as making sure there is either zero or two (or more) women or person of color in each group.  This advice could apply to LGBT+ students as well, however, this may be limited by the overall demographics in the classroom.  Some classroom situations may lend themselves to letting students form their own groups.  Keep in mind that students with shared identities or who appear to be friends may gravitate towards each other because they expect to be excluded from other groups.  This may be especially relevant in laboratory experiences where one student may "hog" the lab equipment and the other is denied a valuable learning opportunity.  Ideally, structure laboratory experiences and classes in a way that each student gets equal opportunity to have hands-on practice with the equipment.    

Promote equitable study-group participation among all students in
your classroom. Possible strategies for achieving this include assigning
study-groups or lab partners, perhaps changing assignments multiple
times during the term; providing ways for interested students to exchange
contact information (within the limits of FERPA regulations); using
online tools such as \emph{Piazza} where students can pose and answer
each other's questions; encouraging students to make use of departmental
or university study spaces at the same time as their classmates; and
providing study and practice problems, such as exams from past years,
to all students in the class. Study groups that are heterogeneous
in classroom performance and in demographic factors have been shown
to facilitate equal participation and to perform very well in problem
solving compared to homogeneous groups~\cite{heller:1992,lewis:2016}.

\begin{newquote}{Anonymous interviewee, an African-American woman with a PhD
in Applied Physics~\cite{rosa:2016}}{}
People would ... tell me, oh, they're not studying and {[}later I{]}
find out they're studying together, so I was studying on my own and
having a hard time. So yes, I was excluded; especially in graduate
school, I was excluded ... It was very hard for me because I was struggling
and I was feeling I was stupid. I couldn't get it, and they're getting
it and {[}I'm{]} not understanding how they're getting it. Or they're
getting it because they had access to previous tests, homework solutions,
you know, from previous years from previous students. \end{newquote}

\section{Monitor climate}
\label{class:monitorclimate}
\BPGtag{A} \BPGtag{D} \BPGtag{S} \BPGtag{P}

It is important to monitor course climate because, although the above
practices will help to promote a positive climate, they are not exhaustive
and there may be specific issues for a particular instructor, course
or institution. Instructors can obtain feedback on climate in several
ways:
\paragraph{Anonymous feedback mechanisms}
Instructors should provide
an anonymous feedback form for students to report concerns about the
course climate. These should make clear that they are not intended
to replace official mechanisms (e.g. the Title IX officer). They should
be promoted in the syllabus and described on the first day of class. 

\paragraph{Midsemester surveys}
Midsemester surveys are a helpful way
for an instructor to take the ``pulse'' of a class and make adjustments
based on the feedback. Climate questions should be included on such
surveys such as ``How comfortable do you feel with the climate in
this class?''.
It is important to include a
question like ``Please identify any observations or concerns you
have regarding the course climate'' and include a free-response box. Instructors
should read and act on the responses to midsemester surveys and, where
possible, explain to the class any action undertaken in response. 

\paragraph{Course evaluations}
While these are typically set by the
university, they should include climate questions as for midsemester
surveys. While they don't allow the instructor to make immediate adjustments,
the responses can be very useful for departmental monitoring and improvement
of subsequent courses. 

\paragraph{Peer observation}
Good instructors have their colleagues
observe their classes for a variety of purposes. Instructors should
ask their observer to pay attention to possible climatic issues, e.g.
range of students called on, level of engagement among different groups,
which can yield valuable data for improvement. Videorecording the class
and reviewing it with the observer can also be helpful. It is important
to remember that observers necessarily have only their frame of reference
and it is important to use as diverse a range of observers as possible.  
If possible, ask your students for permission before recording or bringing in an outside observer.  
At the minimum, notify them ahead of time and explain the motivation for doing so. 

\paragraph{Climate audits}
University Teaching Centers and professional
organizations such as the American Physical Society sometimes offer
the possibility of more structured feedback. This may take the form
of a skilled facilitator leading a discussion on climate in a class
with the instructor absent, or interviewing course participants. They
may be specific to a particular course or at a larger scale, and can be taken
in response to specific concerns or simply to assess the current climate.
Such audits should be safe spaces for students to disclose and describe 
concerns regarding instructors and advisors, without fear of retaliation.
Departments should strongly consider undertaking such an exercise
at regular intervals. 
Institutional Diversity Offices, Women's Resource
Centers, or LGBT+ Resource Centers can be helpful in conducting and
interpreting these audits. 
	
\usechapterimagetrue
\chapterimage{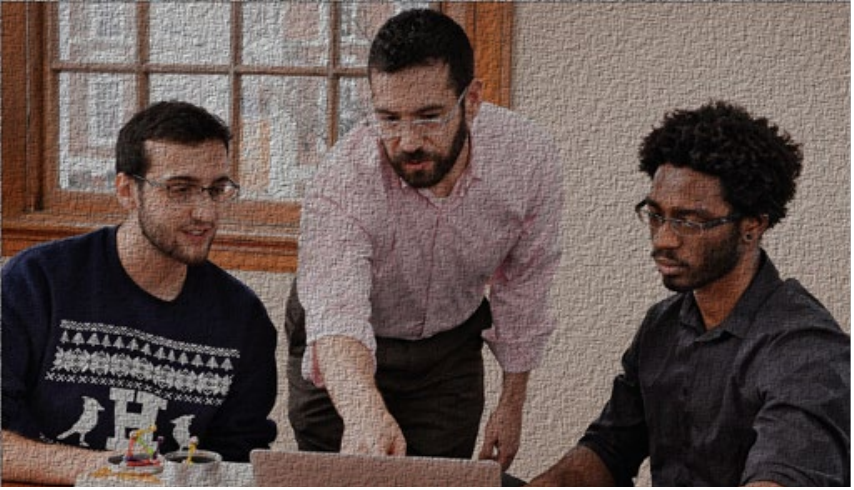}
\chapter{Mentoring and advising}
\label{ch:mentor-advise}
\usechapterimagefalse
\begin{newquote}{Eric Patridge in ``LGBTQ+ issues in STEM diversity'', Boston University Research, June 15, 2017~\cite{moran:2017}}{http://www.bu.edu/research/articles/lgbt-issues-stem-diversity/}
Our opportunity to pursue a career in science is tied not only to our success as scientists but to who we know and who knows us. It's usually a professor who buys into us as a person, who gives us the opportunity to publish. And so I would argue that our career advancement is intrinsically tied to social identities and personal relationships.
\end{newquote}

\medskip\noindent
Mentors and advisors -- whether focused on academics or on research -- are critical foci of 
activity, education, and physics identity, so it is essential that all advisees are able to learn 
and thrive as physicists to the best of their abilities. One size does not fit all in mentorship, 
and no advisee should be excluded or marginalized based on identity. The following sections give 
advice for mentoring and advising members of the LGBT+ community.\secretfootnote{
Original photograph by Greg Dohler for the APS report on LGBT Climate in Physics~\cite{LGBTClimateInPhysics:2016}. Modified and reproduced with permission of APS.}

    \section{Promoting quality mentorship}
      \subsection{Be aware of general issues in mentoring LGBT+ students}
\label{sec:general_mentoring}
\BPGtag{D} \BPGtag{T}

All students benefit from high-quality mentoring and advising, but LGBT+ students may have 
additional challenges that advisors should be prepared for.  Faculty advisors are not expected to 
have all of the answers, but must be able to interact with the student in a respectful way that 
benefits their learning and scientific career.  Make faculty familiar with organizations like 
\href{http://mentornet.org/}{mentornet.org}, where students can potentially find a mentor from the 
LGBT+ community.  Remember that students' LGBT+ identities cannot be separated from their 
scientific identities, and that you may not be aware that a student identifies as LGBT+.  They may 
be uncomfortable or unable to travel to certain locations for conferences, summer programs, or 
graduate school. Likewise, remember that platonic, affectionate touches, such as hugs or pats on the shoulder, 
may not be well-received by some students, regardless of their gender identities or LGBT+ status; always 
check that any touching is welcome.

Like all students, LGBT+ students may need to be directed to appropriate campus resources.  For 
instance, a study of LGB (but not T) students found that they are at a higher risk for psychological distress and mental health problems, 
but almost $\nicefrac{2}{3}$ of LGB students do not utilize any mental health services \citep{dsr+17}. 
The US Transgender Survey reported that 53\% of respondents (18 to 25 years old) reported experiencing current serious 
psychological distress~\cite{ustranssurvey:2016}.  Faculty should familiarize themselves with the mental health services on 
campus, especially regarding student access, confidentiality, and costs - concerns that prevent many 
students from utilizing on-campus services~\cite{dsr+17}.  Also take advantage of campus resources 
and training related to suicide prevention, such as ``Question, Persuade, Refer'' (QPR~\citep{qpr}).
All students will benefit from being quickly connected with appropriate resources. In conversations touching on these issues, it is respectful and helpful for faculty members to make a clear statement to advisees about the extent of their obligation to report the conversation (for example, if they are concerned about imminent harm).

Unfortunately, LGBT+ students may have had negative experiences with health services on campus or 
elsewhere.  One third of the respondents in the US Transgender Survey reported at least one negative 
interaction (related to being transgender) with a doctor or other health care provider~\cite{ustranssurvey:2016}.  Where 
possible, advocate for your students by voicing support for LGBT+ training of all staff, especially 
in student support services.  Find out if there are specific staff members, trained in LGBT+ issues, 
to whom you can direct students.

\begin{newquote}{``LGBT chemists seek a place at the bench'', Chemical and Engineering News, October 17, 2016~\cite{wang:2016}}{https://cen.acs.org/articles/94/i41/place-bench.html}
    At another U.S. university, a chemistry graduate student who is in the process of transitioning from male to female is considering whether to come out to her research group. ``After a year of being relatively depressed and not really knowing what I should do, I decided that if this is going to make me happy and be more productive and a better individual, I should just go forward with it,'' she says. However, a neutral reaction from her adviser has made this student question whether to come out to the rest of the lab.
\end{newquote}

      \subsection {Recruit LGBT+ students actively}
\label{recruit-students}
\BPGtag{D} \BPGtag{I} \BPGtag{P}

Departments should actively recruit LGBT+ students. Such efforts not only build a more diverse body of students within the department by including marginalized students, but also increase the number of undergraduate majors by inviting students who might not otherwise think of physics and astronomy as hospitable careers.  Send fliers to the LGBT+ center on campus, with a line specifically welcoming LGBT+ students, and make departmental representatives visible at LGBT+ student events.  Invite willing and out students, faculty and staff to take on mentoring roles in the department.  Get in touch with (or become) the advisor of the campus oSTEM chapter.  Send a representative to recruit students at Out to Innovate or the oSTEM national conference.  To broaden your department's reach, include information on inclusiveness and resources available to LGBT+ students in materials provided to prospective graduate students.

      \subsection {Increase protections for postdocs}
\label{post-docs}
 \BPGtag{D} \BPGtag{I} \BPGtag{R} \BPGtag{P}

By its very nature, a postdoctoral fellowship is a vulnerable position.  Postdocs typically work at the pleasure of their advisors and are paid out of the advisors' funding.  An advisors may feel tension between their obligation to advance the careers of their postdocs and their desire to complete a particular project before the money runs out.  To improve the oversight and advancement of its postdocs, the Space Telescope Science Institute recently began a postdoctoral mentorship program, assigning to each postdoc a mentor who is \emph{not} their advisor~\cite{peeples:2018}.  The program has sponsored workshops on topics ranging from ``How to Find a Job'' to ``How to Negotiate a Salary''.  Such mentoring programs can be replicated at your institution.

\clearpage
Postdocs may not receive the same job protections offered to other employees.  For 
example, while the Family and Medical Leave Act (FMLA) provides certain employees with up to 12 
weeks of unpaid, job-protected leave per year, postdocs are often not considered employees and thus 
are not necessarily guaranteed FMLA protections.  Paid medical leave would be extremely helpful for 
postdocs taking time off to recover from gender-affirming surgery, or to care for a new child.

\begin{newquote}{Savannah Garmon, as told to Physics Today~\cite{feder:2015}}{
http://physicstoday.scitation.org/do/10.1063/PT.5.9034/full/}
In the career of a physicist, there is the postdoc period, which I think in many ways is a 
particularly vulnerable period. You work at the whim of your mentor. So a good practice to implement 
at a department level -- and this is in the Best Practices Guide [by the group LGBT+ Physicists] -- is to 
assign postdocs an alternate mentor, someone to keep tabs on them, and give them advice on 
approaching the job market. Then, if someone feels some uncertainty, and feels they can't talk to 
their primary mentor, they automatically have someone else to talk to. This could be helpful for all 
postdocs, not just LGBT people.\\
\end{newquote}
      
      \subsection {Discuss climate with advisees}
\label{discuss_climate_advisees}
 \BPGtag{D} \BPGtag{P} 

Students and junior scientists who face a hostile climate -- in the classroom, in the laboratory, or in the department as a whole -- are often hesitant to raise the problem with an authority figure. They may worry that the climate is somehow their fault, that they will be blamed for a particular incident, or that reporting might make things worse; that even terrible harassment is insufficient cause to bother a busy, more powerful person; or that there is no possibility for positive change. They may feel pressure, external or internal, not to get a labmate or classmate into trouble, or they may believe that it is up to them to manage any harassment they face on their own. This type of problem almost never goes away by itself, and an instructor, supervisor, department chair, or other identified liaison is a natural person to intervene in an unhealthy climate.

Encourage instructors, advisors, supervisors, and mentors to ask about climate in their regular meetings with students, trainees, or other junior scientists. Helpful questions include: Do they feel welcomed in the department or lab group? Is the climate at homework help sessions conducive to their learning? Have they or others experienced harassment or belittlement by peers or by more junior or senior individuals? What kind of changes are needed? It is important to ask questions like these of everyone, since people may be targeted on many different grounds -- sexual orientation, gender identity or gender expression, race, ethnicity, nationality, sex, disability, religion, etc. By expressing a sincere interest in junior scientists' experiences, mentors can indicate that they will listen sympathetically and are willing to take active steps to solve any problems that arise. Complaints must be taken seriously and a complainant should be able to expect an improvement in their circumstances. A harasser may or may not intend to harm, but that does not make their actions less damaging. No one should feel that they must hide their identity in order to study or work in physics or astronomy.

Ensure that there are also avenues for complaints outside the normal advising hierarchy. Undergraduate- and graduate-student liaisons are natural candidates for this role, as are university trainers in diversity and ethics. Many departments have annual or semi-annual orientations, social events, or mailers, which are good opportunities to remind students, faculty and staff of the expectations of the department and the resources that are available to them.
      
    \section{Advancing the careers of LGBT+ students}
      \subsection{Advise on personal statements and applications}
\label{personal-statements}
 
Many students need significant direction when writing their first applications for summer programs and graduate school.  While faculty have significant expertise at writing about research experience, few have written about their identities.  Students may ask whether they should disclose their LGBT+ identity in personal statements and applications.  There is no single correct answer to this question, and faculty should not pressure students to include or remove statements about their identity -- although they should share any relevant information they may have about how inclusive the program actually is.  Faculty may find it helpful to connect students with mentors who can share how they have navigated similar experiences.

      \subsection{Advise on graduate school}

Undergraduate LGBT+ students will benefit from advice regarding grad school that all students receive, but they have some unique concerns too.  They may be particularly concerned about the location of graduate programs due to issues of safety or finding a supportive community.  Local and state laws, and institutional policies, may not prevent discrimination (in housing or employment, for instance)~\cite{tracker-employmentdiscrimination}, and many localities have discriminatory bathroom laws~\cite{tracker-bathroombill}. Transgender students may rule out programs due to insufficient health insurance coverage.  Remember that LGBT+ students may not be out to you, so be respectful of students ruling out programs that outwardly seem like a good match. Encourage your students to consider all aspects of a program, including intangibles, when making their decisions.

      \subsection{Advise on summer programs, study abroad, and other travel}

Educational experiences away from a student's home campus may be concerning for LGBT+ students.  Questions to which they would like to know the answers could include: What are the legal protections at the program site, especially if in another country?  What health services (mental and physical) will be available?  Students may feel uncomfortable asking these questions to the program directly, especially if it means they must out themselves.  Advocate for the student by seeing if there is an office on campus that can get the information for the student while preserving the student's privacy.

      \subsection{Write recommendation letters inclusively}
\label{rec_letters}
 \BPGtag{D} \BPGtag{P} 

\begin{newquote}{Anonymous transgender student, as told to the authors}{}
Once, when requesting a recommendation letter when applying to graduate school, I had to come out to a research advisor I had previously been closeted to. I had only worked with her one summer for an REU program, and she was nice enough, but I knew she was very Catholic so I was hesitant to reveal my trans status. Staying closeted wasn't an option, though, since I wanted all of my recommendation letters to have the same pronouns. 

There was no reply to my first email. I sent her another a couple weeks later to confirm she had received the first. No response to that either. I started to panic, wondering who I was going to get my letter from if she wasn't going to send hers. Finally, on the day of the deadline, she submitted her letter, apparently with the pronouns I requested. She never once acknowledged my coming-out email, but she has been otherwise courteous. The whole situation was just nerve-wracking and awkward. I shouldn't have to worry about whether or not someone will write me a letter because of my gender identity.
\end{newquote}
In writing letters of recommendation, it is important to use identifying language that matches the preference and usage of the person you are writing for, to avoid creating confusion or ambiguity as to whom you are writing about (and to avoid outing some aspect of someone's identity inadvertently).  Since it is common to ask the applicant to provide a CV for you to look at while you compose your letter, you might combine that request with one for details about how they wish to be referred to in your letter.  For instance, upon receiving a request to write a letter of recommendation, you might respond:

\begin{newexample}
I would be happy to write a letter of recommendation for you.  Please email me your CV and any 
drafts or research/personal statements you will be submitting for this award/position, so I can make 
sure my letter dovetails appropriately with your application.  Please also let me know how you will 
refer to yourself in your materials so my usage will match yours; e.g., the exact format of your 
name, preferred pronouns, and appropriate honorific (e.g.~Dr., Mx., Ms., Mr.)
\end{newexample}
Remember that it is not only unethical to disclose an applicant's transgender status, but, for students, it is also a 
violation of FERPA~\cite{dearcolleagueletter2017}. Especially if you are not using the name or pronouns you may be accustomed to using for a particular student, it is wise to review your letter carefully after composing it, using the search and replace function to protect against an error that might endanger the applicant. 

Just as you may refer to someone's leadership in athletics or various campus activities in recommendation letters, it may be appropriate to reference their leadership in LGBT+-related groups, such as Gay-Straight Alliance, Project Safezone, or Gender and Sexuality Alliance -- but only with the applicant's permission. If you are not particularly familiar with these activities, you might ask the applicant to request that the group's director or faculty adviser write \textit{you} a letter about their involvement. You could then draw quotes from this letter and make reference to it in your own letter of recommendation. This allows the applicant to buttress a professional, scientific letter from you with information about their advocacy work, showing how that work demonstrates that they are driven, mature and self-motivated.

\begin{newquote}{Anonymous postdoc, as told to the authors}{}
I recently ... sought letters of reference from people who knew me before and after a name change, and I am so grateful to the person who asked what name I wanted in their letter. It helped to alleviate the burden of figuring out who knew how much already.
\end{newquote}
         
\usechapterimagetrue
\chapterimage{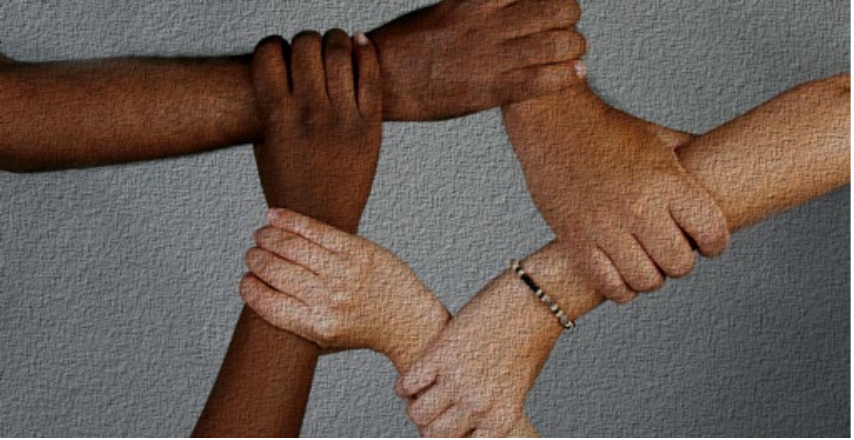}
\chapter{Hiring and promotions}
\label{ch:hire-promote}
\usechapterimagefalse
\begin{newquote}{Anonymous interviewees in Bilimoria and Stewart, 2009~\cite{bilimoria:2009}}{}
{[}O{]}ne individual was encouraged ``to temper how gay she looks'' in her job interview, and was explicitly advised to be more conservative in appearance, not to bring her partner to departmental events, and not to adopt children. Another was advised not to communicate with faculty and staff about partner benefits issues on campus, because some people objected. 
\end{newquote}

\medskip\noindent
Fairness and equity in promotions and in hiring -- whether of staff members, researchers, or faculty members -- will help your department not only recognize top talent, but attract it, too. The following sections give advice for avoiding bias and projecting openness in searches and promotion proceedings.\secretfootnote{\href{https://creativecommons.org/licenses/by/2.0/}{Creative Commons} photograph courtesy of \texttt{Wonder woman0731} on Flickr. Original image available at \url{https://www.flickr.com/photos/wildrose115/27623264486/}.}
%
Although not normally considered an aspect of the hiring and evaluation of employees, 
we note that both graduate admissions and the process of
assigning and managing teaching and research assistants share many of the same challenges and opportunities. As with the hiring and evaluation of any employee, those on graduate admissions committees, or in a position to evaluate teaching or research assistants, should ensure a welcoming working environment for, and fair evaluation of, LGBT+ students.
    \section{Toward an inclusive job search}
      \subsection [Include non-discrimination statements in job announcements]{Include non-discrimination statements in job announcements}
\label{nondisc-statement}
\BPGtag{D} \BPGtag{I} \BPGtag{R} \BPGtag{S} \BPGtag{P} 


Including a brief statement on the Equal Employment Opportunity (EEO) policy of the employer in the job announcement serves several goals. It clarifies the legal situation that a potential employee enters, but it also signals to potential employees that the employer is aware of the issues facing LGBT+ people. If competing institutions lack protections or partner benefits for LGBT+ people, qualified LGBT+ applicants may be attracted to a non-discriminating institution they might otherwise have overlooked.
Employers can include a brief EEO statement stating:
\begin{newexample}
This employer prohibits discrimination based on sexual orientation, gender identity, and gender expression. 
\end{newexample}
In individual job postings, employers can include language like the following:
\begin{newexample}
We encourage applications from eligible candidates regardless of gender, race, national origin, age, religion, marital status, political views, sexual orientation, gender identity, gender expression, or disability.
\end{newexample}

\begin{newquote}{Manil Suri in ``Why Is Science So Straight?'', op-ed in New York Times 4 September 2015~\cite{nyt:sciencesostraight}}{https://www.nytimes.com/2015/09/05/opinion/manil-suri-why-is-science-so-straight.html}
After years of stubborn resistance, Exxon finally added sexual orientation and gender identity to its anti-discrimination policy in January [2015], reportedly in part to better attract top STEM talent.
\end{newquote}

      \subsection{Provide bias training for search committees}
\label{sec:biastraining}
\BPGtag{D} \BPGtag{I} \BPGtag{S} \BPGtag{P} 

Many institutions have implemented bias training for tenure-track faculty searches, typically focused on dismantling biases that prevent the successful hiring of women and/or people of color.  These trainings can help individuals identify and deconstruct their stereotypes and biases, including those towards LGBT+ individuals, and set up equitable hiring structures.  Even if the bias trainings do not explicitly address LGBT+ people, LGBT+ people may also have other marginalized identities.

If your institution provides bias training, utilize it before starting a new faculty search.  If 
not, utilize the materials (such as videos) that have been published online by other institutions.  
Consider implementing bias training in other application evaluation situations, such as hiring staff 
and non-tenure track instructors, awarding prizes, considering promotions, and even admitting 
graduate students. 

\begin{newquote}{Anonymous anecdote about being ``out'' at a job interview, recounted on LGBT+ 
Physicists blog}{http://blog.lgbtphysicists.org/2014/05/should-i-be-out-during-job-interview.html}
... I had a rather humorous exchange over dinner in a fancy French restaurant during my 
first faculty interview. One senior male faculty member casually turned to me to ask ``So, are you 
married?'' Without missing a beat a young female faculty member jumped up, nearly knocking over the table, and said ``Out of bounds! You can't ask that!'' Everybody laughed about it, and we moved on with dinner. I was offered the position, and on my second interview I explained to the chair that I 
wanted moving expenses reimbursed for my same-sex partner although we lived in different cities at 
the time. This was dealt with professionally and satisfactorily since at that point the chair's job 
was to enact the will of the faculty and get me to come take the job.
\end{newquote}
      \subsection {Avoid assumptions}
\label{assumptions}
\BPGtag{D} \BPGtag{I} \BPGtag{S} 
It is always embarrassing to use the wrong title, name, nickname or pronoun to address or refer to someone. When the individual in question is LGBT+, this type of mistake can be particularly hurtful.
To avoid such errors, beware of assigning pronouns to people you have not met (e.g. use singular they/them until you learn otherwise). In conversations and deliberations, consider each applicant by name until as late as possible in the process. This precaution also helps reduce potential gender bias in the hiring process. When you make direct contact with an applicant for the first time (e.g., in a telephone or in-person interview), ask ``How would you prefer to be addressed?" and then communicate this information to other department members involved in the hiring process. This simple question accommodates a wide range of situations, from gender expression to nicknames.

      \subsection {Be open to name changes for job and tenure applicants}
\label{job-name-change}
\BPGtag{D} \BPGtag{I} \BPGtag{S} \BPGtag{P} 
Anyone who has changed names may encounter challenges when applying for employment, awards, or promotions. Employers are increasingly likely to require a background check as part of the application process; this entails providing all of the legal names one has held to the agency doing the background check. Evaluation committees typically require that one submit a list of publications as part of a job, award, or promotion application. It is conventional to provide the names of the authors so that the evaluators can note their relative seniority and/or ordering as part of assessing the candidate's relative contributions to the work. People change their names for many personal reasons, including witness protection, entering a life partnership (usually a change of last name) or undergoing a gender transition (usually a change of first name). Since these reasons are not generally relevant to job qualifications and tend to reveal information that the employee may prefer to keep private, employers should minimize the instances in which employees must reveal the history of their names.

Departments can take steps to balance employees' reasonable privacy concerns against the requirements of employment and evaluation processes. In the case of background checks, the department should already be ensuring that only those directly involved in performing the check see any information that the individual submits. In the case of evaluations for awards or tenure, the department could explicitly establish a convention of including only the last (family) names of all authors on publication lists. This would still enable experts in the field to evaluate the author ordering and seniority of collaborators, while protecting transgender individuals from being forced to out themselves. Since this would still show where an individual has changed their family name, raising issues of gender bias, the department could issue a statement such as:
\begin{newexample}
	In publication lists for award or tenure applications, please list all authors' last names only and put your own last name(s) in bold type to make it easy for the readers to find. {Name changes are not relevant to our decision and will not be considered in the evaluation}. Please also include a brief statement at the start of the publication list that notes the author ordering convention in your sub-field (e.g., alphabetical, students first, primary author first, etc.).
\end{newexample}
This establishes that the department will only consider professionally relevant information and offers a practical way for individuals to handle several name-related issues that frequently arise.

Finally, it should be noted that addressing naming issues in one's CV or publication list does not cover all eventualities. Evaluators who look up a journal article may still discover that someone has changed their name. Individuals may wish to contact their publishers to investigate the possibility of updating their name on past publications.

    \section{Retention and promotion}
      \subsection {Provide help for all dual-career couples}
\label{dual-career}
\BPGtag{D} \BPGtag{I} \BPGtag{S} \BPGtag{P} 

Similar to different-gender couples, 43\% of academics in same-gender relationships have partners who are also academics \citep{schiebinger2008dual}.  For any dual-career partners, decisions about employment opportunities can be affected by uncertainty about the career prospects for the partner who is not the one primarily being recruited. 
Even if the institution has a robust non-discrimination policy, partners' employment may be limited by a lack of local or state non-discrimination policies.

Discussing dual-career issues before an offer of employment has been made is challenging.  The 
potential employer is legally barred from inquiring about the personal life of the job candidate. 
Additionally, the candidate may not wish to raise these issues, lest they influence the likelihood 
of receiving an offer.  Therefore, the department chair should make it standard practice to inform 
all job candidates or finalists about general university resources related to work-life balance; the 
chair should state clearly that this one-way flow of information (from chair to candidate) is 
standard practice.  For instance, the chair might provide a copy of a university work-life resource 
guide, links to the local Higher Education Recruitment Consortium (HERC \citep{herc}) website,
or the contact information of the university's point person for dual-career issues. 

Note that even when a potential employee is comfortable discussing dual-career issues with potential employers (e.g., after a formal offer is in hand), a satisfactory solution may be impeded or precluded by legal barriers.
      \subsection {Consider LGBT+ persons when developing family-friendly policies}
\label{families}
\BPGtag{D} \BPGtag{I} \BPGtag{S} \BPGtag{P} 

More and more adults are juggling child and/or elder care with their professional responsibilities.  In 
response, many employers are developing family-friendly policies to help employees balance their 
work and personal lives.  When developing such policies, be mindful of non-traditional families.  
For example, explicitly include adoption in parental-leave policies, domestic partners in 
family-leave policies, and LGBT+ couples in dual-career accommodation practices (Sec.~\ref{dual-career}). Parental leave 
policies for many institutions are listed on the AstroBetter web site \citep{astrobetter:leavepolicies}.

      \subsection{Assess teaching performance fairly}
\label{teaching-evaluations}
\BPGtag{C} \BPGtag{D} \BPGtag{I} \BPGtag{P}

When evaluating faculty or teaching assistants, it is important for administrators to recognize that teaching evaluations are often negatively affected by the instructor's identity. For instance, it has been well documented that female instructors will receive lower evaluations of their teaching compared to their male colleagues~\cite{flaherty:2016, macnell:2015, boring:2016}. Similarly, bias against instructors has also been documented based on race~\cite{reid:2010, merritt:2012}. Investigations of whether LGBT+ instructors also receive poor evaluations have not been as extensive. However, as a group that has historically suffered discrimination, it is not surprising that the relevant studies have generally found bias~\cite{cesario:2003, russ:2002}.

When evaluating the teaching effectiveness of instructors, avoid relying solely on student evaluations of teaching. Instead, use direct observations of the instructor by the administrator or another faculty member who is an expert teacher.

\begin{newquote}{Ron Buckmire, as told to Physics 
Today~\cite{feder:2015}}{http://physicstoday.scitation.org/do/10.1063/PT.5.9034/full/}
I had issues in some of my [teaching] evaluations. Students put bizarre comments that were 
related to my sexual orientation, like, `Why does he have to be so gay?' That complicated my tenure 
process.
\end{newquote}

\begin{newquote}{Anonymous professor of physics, as told to the authors}{}
After I transitioned from male to female, my chair, dean and I thought that the impact on our students would be minor as the course I was to teach was a repeat of an introductory physics course that students only took once. However, the first day of class I noticed a few of the female students repeatedly asking me questions by my new name. ... When I got back to my office, I found that I had a call from my dean telling me that an issue had arisen with regard to some of my students and my gender transition. When I called, he told me that a few of my students had complained about the change in the gender presentation of their instructor.  We suspected that these were students who were repeating the course from me and obviously noticed the change. ... When I called the Dean [of Students] he told me that they had handled it by telling the students that the change in their instructor's gender presentation was a non-issue as far as the university was concerned and that if they had any more problems with it that they should go to the counseling center. I was very relieved at the outcome but the behavior of the students was troubling and disappointing.
\end{newquote}

\usechapterimagetrue
\chapterimage{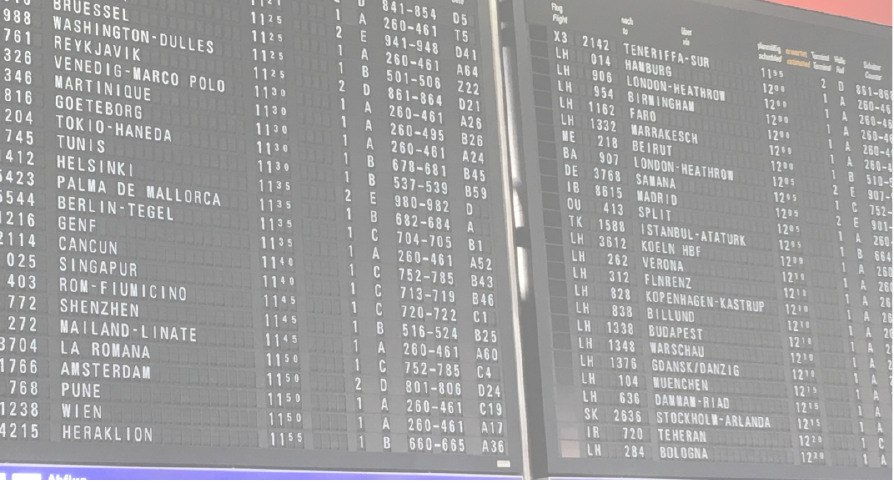}
\chapter{Travel and hosting}
\label{ch:travel-host}
\usechapterimagefalse
\begin{newquote}{Lisa, bisexual undergraduate student in engineering~\cite{cech:2011}}{}
If I got fired, it wouldn't be under the pretense that I'm gay; I would most likely be fired because I screwed up because I'm gay. Like, if I was sent to another country where they weren't accepting of homosexuality, and I started to be free to act as I choose, then they would be like, you screwed up the sale and our relationship with the customer ... I'm sure that it would come around to me in a way that I could not take legal action against the company.
\end{newquote}

\medskip \noindent Traveling to another location -- for a research conference, a sabbatical, a seminar visit, or an experiment -- can produce a lot of headaches for LGBT+ people, whose legal protections and local encounters may vary widely by location. The advice in the following section will help you facilitate productive travel by LGBT+ department or group members, and will help you welcome LGBT+ guests at your department and elsewhere. Please note that travel should not be mandatory: while there are often good reasons to travel as an academic, there are also often good reasons not to. People should always be free to decline travel without needing to defend the decision.
      \section{Support attendance at LGBT+ related conferences}
\label{lgbt-conferences}
\BPGtag{A} \BPGtag{D} \BPGtag{I} \BPGtag{R} \BPGtag{P} 

\subsection{Institute departmental and institutional policies}
Professional conferences specifically for LGBT+ people exist, and may be of use to LGBT+ students, postdocs, and faculty. Some, like the annual oSTEM conference, are specifically relevant to scientists. Departments, administrators, and mentors can help encourage people to attend these conferences, and provide travel funding.

\subsection{Remember intersectionality}
A department or funding pool which provides support for LGBT+-related conferences can and should also provide support for conferences related to race, ethnicity, disability, etc. As always, supporting LGBT+ people must be one component of a larger commitment to supporting marginalized scientists. (Not least because many LGBT+ students are also marginalized in other ways).

\subsection{Advertise support}
If funding is available, it must be advertised, not simply available. Requesting funding may present a barrier to LGBT+ students who are not sure of support from administrators or advisors. In addition, advertising funding sends a message that the department or college strongly supports LGBT+ people. Be sure to also advertise opportunities for students that would not require direct departmental or mentor intervention for their participation. Such opportunities are especially important for those students or post-docs who are only questioning or prefer to not out themselves yet.

An example of advertising conference support could be a note on a department web page devoted to diversity, equity, and inclusion, e.g. ``Every year, we provide support for marginalized students to attend conferences and professional development events related to social identity in STEM. Some examples of conferences students have attended in the past include [list here]. To apply for funding, [instructions here].'' Flyers for diversity-focused conferences should be posted in the same place as other conference flyers.

\subsection{Monitor discretion and safety}
Administrators or mentors who would like to nominate a student for support to attend a conference should first check in with that student to confirm their interest in the conference, and again before advertising their attendance by name in newsletters, websites or funding reports. Failure to do so could inadvertently out someone.

\subsection{Establish connections with existing campus organizations}
Many universities have LGBT+ student organizations, which may send, or be interested in sending, their members to conferences. Departments and colleges can reach out to student organizations and offer financial support for such existing efforts.

      \section{Ensure inclusive lodging and travel arrangements}
\label{sec:lodging}
\BPGtag{D} \BPGtag{I} \BPGtag{S} \BPGtag{P} 

LGBT+ scientists may face serious dilemmas when required to stay in shared lodging, e.g. for conferences, workshops, summer student internships, or research programs. Living with strangers can be risky, due to the prevalence of homophobia and transphobia. Even living with known individuals may cause an LGBT+ person to be outed against their will. (How will a person in a same-gender relationship call their significant other without being overheard? How will a transgender person on hormone replacement therapy keep their medications private?) Gender-segregated housing can be particularly problematic for transgender people, who are sometimes placed into gender-inappropriate lodging due to incorrect documentation or other misunderstandings. They may also be placed into gender-appropriate lodging which is nonetheless potentially unsafe.

\begin{newquote}{Anonymous interviewee, APS report on LGBT Climate in Physics~\cite{LGBTClimateInPhysics:2016}}{}
One of my undergraduate mentees was beaten by his roommate because of his sexual orientation. He has endured lasting physical and emotional injury that will likely persist for years to come.  
\end{newquote}

\subsection{Allow travelers to choose their modes of transport}
\label{alt-transport}
Scientists traveling for work should be free to use alternate modes of transportation without needing to defend the decision, especially if the cost is negligible. For instance, many trans individuals encounter trouble while flying, because of the way TSA handles security screenings. Train travel or driving may be less problematic. On the other hand, some people find road trips problematic, because public restrooms along the way may be unsafe. Individual preferences on this point vary greatly.

\subsection{Be flexible about how travel arrangements are made}
\label{flex:travelarrange}

While it is often convenient for travel arrangements to be made centrally through a department administrator, this system can present some travelers with difficult choices. When purchasing tickets for air travel, in particular, the purchaser must know the traveler's full name and gender marker as they appear on the identification that will be used to fly -- and this name and gender marker may not match the way in which the traveler presents in a professional setting. Do not require visitors or department members to obtain tickets through the department's administrative staff -- although the staff should treat this information confidentially, it is never appropriate to require a traveler to out themselves. Provide additional options, such as going through a contracted travel agent who is not employed by the university, or purchasing a ticket directly and being reimbursed later. All travelers should be made aware of their options, as in the example below.

\begin{newexample}
If you would like me to arrange your travel, please provide me with the identifying information necessary for your ticket (name, gender, and date of birth as listed on the identification you will use to travel). If you would prefer to arrange your own travel, please send me your arrival and departure information once you have done so. I would also be happy to connect you with a travel agent contracted with the university, if you prefer.
\end{newexample}

\subsection{Do not require people to have roommates}
There are many situations where it may be tempting to assign roommates, or require individuals to seek roommates, in order to save on space or cost of housing. But it should always be possible to opt out of shared housing, without fear of judgment and without the need to justify one's decision. The option to decline a roommate or housemate should be made clear to all parties.

This is not purely an LGBT+ issue. It may also affect someone if they regularly face prejudice due to their race and/or religion, have a confidential medical condition, need privacy/security/quiet for mental health, or for other reasons.

\subsection{Provide private facilities in shared housing}
In general, providing more privacy is better. In approximate order from minimal requirements to ideal cases, privacy measures may include:

\begin{itemize}
\item Single-occupancy locking shower (this is critical)
\item Single-occupancy locking bathroom, including a sink, mirror, and counter
\item Nearby areas for phone conversations out of earshot
\item Personal storage space with lock
\item Private bedroom with locking door
\end{itemize}

\subsection{Identify a contact person for confidential lodging concerns}
Make sure there is a contact person with whom problems or concerns relating to lodging may be raised. In particular, this person:

\begin{itemize}
\item should not be an immediate coworker or supervisor of any resident.
\item should be prepared to handle sensitive issues, including bias and harassment incidents.
\end{itemize}

Regarding the latter point, if the lodging contact person is a legal mandatory reporter, this should be made clear to all residents. As usual, it is best to provide multiple options for incident reporting, so that a complainant can choose the level of confidentiality they wish to maintain.

      \section{Host inclusive conferences}
\label{hosting-conferences}
\BPGtag{D} \BPGtag{I} \BPGtag{P} 

\subsection{Pay attention to names, biographies, and pronouns}
 
For individuals or institutions that will be hosting or organizing gatherings such as conferences, workshops or internships, accommodations regarding names used by the participants should be implemented for those who have not yet been able to change their names legally. Registration for such events should allow for registrants to indicate a preferred name and pronouns to be used for the public aspects of a conference or internship. For example, the preferred name and pronouns could be used on lists of speakers, programs, proceedings, identification badges, and for all other social or networking activities. The legal name would be associated with the preferred name and will be used for any reasons for which the use of a legal name is necessary (for example, visa support documents).
 
Organizers of conferences or those who will publish professional biographies of authors or speakers should allow these biographies to be written using the preferred pronouns of the author, speaker or participant. 
Social hours or ``ice-breakers'' for departments or at conferences can assist those who are transgender or gender non-conforming by encouraging the use of name tags by all participants, not just those who are transgender or gender non-conforming, that indicate the preferred pronouns of the person wearing the name tag.
 
\subsection{Provide equal access to restrooms}

For more information about restroom access in general, please see Sec.~\ref{restroom_access_dept}.

\paragraph{Event venue selection}
When organizing a conference or event, especially one which will last for more than a few hours, make sure that non-gendered accessible restrooms are available.

\paragraph{Converting a restroom (temporary)}
If a non-gendered restroom is not available at the venue for your event, plan to convert a set of gendered restrooms for the duration. Temporary signs reading "all gender restroom" or similar may be placed over the existing ones. You will probably need to contact the venue to discuss this, especially if the space is shared with people who are not part of your event. It will likely be easiest to do this with single-occupancy restrooms.

\paragraph{Signage}
Distribute a map of the venue to attendees and be sure that non-gendered restrooms (temporary or permanent) are clearly marked. 

\begin{newquote}{``LGBT chemists seek a place at the bench'', Chemical and Engineering News, October 17, 2016~\cite{wang:2016}}{https://cen.acs.org/articles/94/i41/place-bench.html}
Many LGBT chemists say they won't attend scientific meetings in a state that doesn't offer them protections. ``I would never go to a conference now in Arkansas, Tennessee, or North Carolina because I would not be able to go to the bathroom without fear that I would be fined or go to jail,'' says {[}a trans postdoc{]}.
\end{newquote}

      \section{Be inclusive in gender-based outreach and events}
\label{travel_gender_outreach}
\BPGtag{A} \BPGtag{D} \BPGtag{I} \BPGtag{S} \BPGtag{P} 

Many departments, student groups, and individuals engage in important outreach aimed at girls and young women. These efforts should also be aware of LGBT+ people among their target demographic.

It is important to be explicitly inclusive of transgender women and girls, who may otherwise be afraid to engage in these opportunities. Organizers should also follow inclusive practices regarding lodging and restroom access, where applicable. It is also important to think about nonbinary people and their place in gender-based outreach efforts. Some organizations and initiatives with a history of serving women in STEM have chosen to expand their focus to women and nonbinary people (i.e., to anyone who is not male in a male-dominated field).

Organizers of gender-based outreach events should also be conscious of heterosexism, an issue which affects LGB+ women and girls. Speakers should avoid the assumption that all participants have or want a boyfriend, for example. Most gender-based outreach efforts prioritize breaking down gender stereotypes and the dominance of traditional gender roles; that should include assumptions about relationships and attraction. Likewise, speakers and organizers should not assume that all women and girls have the same anatomy.

\begin{newquote}{Anonymous undergraduate student, as told to the authors}{}
I was at a women's conference. There were students and professionals at each table. One of the professionals at my table asked the students if we had boyfriends or girlfriends. It made me so happy. She didn't look around and then add ``girlfriends'' as an afterthought. ``Girlfriends'' rolled off her tongue as easily as ``boyfriends'' did. Although I didn't answer her question, the fact that she didn't assume everyone would have a boyfriend made me feel comfortable at the conference.
\end{newquote}
      
      \section {Invite LGBT+ speakers to campus}
\label{lgbt-speakers}
\BPGtag{D} \BPGtag{I} \BPGtag{P}
One way to help those belonging to marginalized populations become more integrated into the academic community is to recognize them publicly for their professional accomplishments. This recognition provides other members of the community with role models with whom they can identify. The APS has, for years, publicized speakers' lists of women and minority physicists in order to encourage departments to diversify their colloquium and seminar series.  Similarly, when a department invites a speaker from the LGBT+ community to make a research presentation, it simultaneously showcases that individual's work, provides them with networking opportunities in the department, and offers local students (and even faculty) a role model. It also enables the department to demonstrate publicly its commitment to inclusive excellence.

When {inviting a speaker to campus}, it is always good practice to arrange for them to meet with 
individuals or groups with whom they have common interests. For example, an invited speaker who is 
an expert in instructional methods may wish to meet with fellow educators. If you are hosting 
someone whom you know to be a public advocate for LGBT+ concerns, ask whether they are willing to 
meet with interested student or faculty groups, either formally or informally.  Sharing pizza and 
conversation with the local chapter of oSTEM or the campus WISE group will be a valuable experience 
for all concerned.  With the speaker's permission, {provide a mini-bio that references their 
interest in LGBT+ issues} as well as their scientific accomplishments; this will encourage a wider 
range of individuals to come to their talk or seek to meet with them. For all speakers and visitors, 
be sure to find out what pronouns they use, and use these pronouns in seminar advertisements, 
introductions, etc.

When arranging travel for any speaker, be conscious of the privacy complications that can arise, and be sure to provide them with options for their travel arrangements (see Sec.~\ref{alt-transport} and~\ref{flex:travelarrange}).
\usechapterimagetrue
\chapterimage{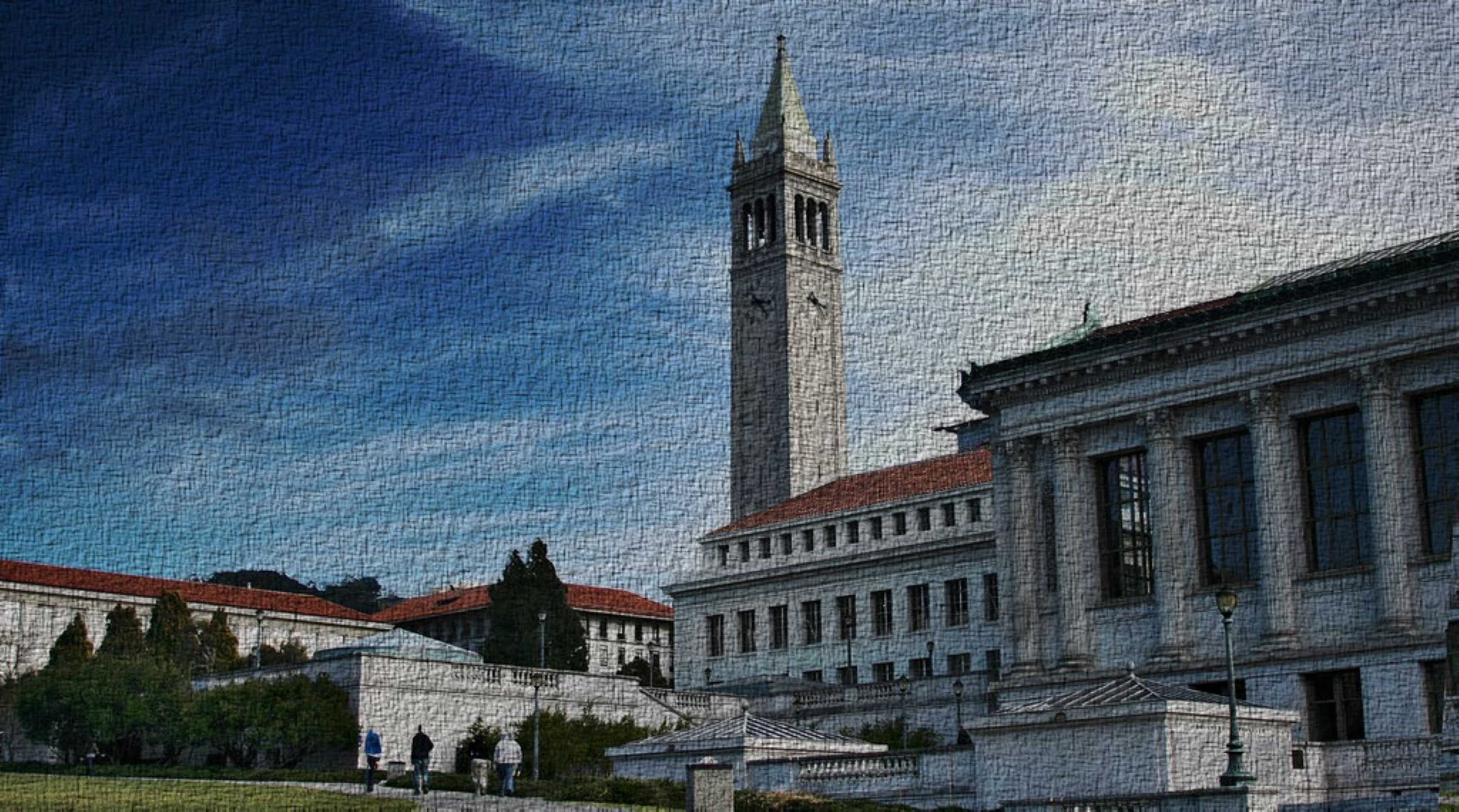}
\chapter{Toward a welcoming institution}
\label{ch:university}
\usechapterimagefalse

\begin{newquote}{Anonymous interviewee, APS report on LGBT Climate in Physics~\cite{LGBTClimateInPhysics:2016}}{}
Physicists need to kind of take it down a notch and pull more of the human element because no matter what answers you search for by using a laser, doing all different types of experiments with detectors, you still have people doing these things and you have to remember if you want them to work at their best you have to be able to treat people right, you have to make a climate that's hospitable.
\end{newquote}

\medskip\noindent
As the superstructure within which large-scale decisions are made, the institution is the unit with the greatest power to shape the future. Owing to its broader scope and larger resources, a long-term vision toward improving diversity and inclusion is best encoded by Institutional policies. A welcoming institution is aware of the needs of its marginalized members, provides resources to support them, establishes inclusive policies with sufficient flexibility to accommodate its diversity, and is sensitive to its place as an institution of higher education within the local cultural context. In this section, we provide suggestions by which institutions may guide their diversity work, derive visions for the future of their departments, remove barriers against access by members of diverse communities, and promote a healthy environment for thriving, productive communities. A well-situated institution will champion Awareness and advocacy in a Caring environment that engenders Equity, or ACE. Here, we will refer to the broadest institutional context as the ``institution'', referring simultaneously to liberal arts colleges, community colleges, national research centers, universities, and any other organizations where research, teaching, or learning occur.

    \section{Awareness and Advocacy}
      \subsection {Identify and consult your LGBT+ students}
\label{univ-identify}
\BPGtag{S} \BPGtag{P} 
In 2013, the {\em New York Times} reported that a small but growing number of institutions is including questions about sexual orientation and gender identity in their undergraduate applications~\cite{nyt_gayquestion:2013}.  The principal goal is to make prospective LGBT+ students feel welcome, but the information allows institutions to consider these students for scholarships aimed at increasing diversity on campus, to provide them with information about support services for LGBT+ students, and to track their success relative to their heterosexual and cisgender peers.  Protecting students' privacy is key, so institutions should not transfer this information to the student's permanent file.  Still, it is difficult to evaluate how well the institution is serving its LGBT+ students if it does not know who they are.

Despite the institutional value of identifying individual students as LGBT+, many students will judge that they should not answer such questions frankly. Students may not be out to family members who are reviewing their undergraduate applications; they may worry that this information may follow them on their permanent files and even on transcripts; they may need more time to become confident in their sense of their identity. For this reason, when surveying LGBT+ students to evaluate whether the institution is meeting their needs, it is essential not to rely on a pre-existing list. Instead, distribute the survey to all students, asking them to take it if they identify as LGBT+, assuring them of the confidentiality of their responses, and explaining why the information is being collected and how it will be used. It is essential to the integrity of such surveys that anonymous responses be allowed: otherwise, those students who experience the most problems on campus will feel the least comfortable responding.

      \subsection{Be aware of Title IX and safe incident reporting}
\label{univ-titleix}
\BPGtag{D} \BPGtag{R} \BPGtag{S} \BPGtag{P} 

Institutions receiving Federal financial assistance from the U.S. Department of Education (ED) are 
required to conform to Title IX statutes, which state ``No person in the United States shall, on 
the basis of sex, be excluded from participation in, be denied the benefits of, or be subjected to 
discrimination under any education program or activity receiving Federal financial assistance''. 
Schools and their STEM departments should strive to:

\begin{enumerate}
 \item Provide Title IX information to all incoming students, staff, and scholars, including, but 
not limited to:
\begin{enumerate}
 \item the meaning, scope, and purpose of Title IX,
 \item the name and contact information of the school's Title IX coordinator,
 \item information on the school's sexual harassment and discrimination policy,
 \item a description of the roles of responsible employees~\cite{titleix-memo},
 \item and clarification on confidentiality and mandatory reporting as applicable at the institution 
under Federal and local law.
\end{enumerate}
\item Establish a means for incident reporting without fear of retribution.
\item Establish a means for collecting anonymous reports.
\item Perform regular surveys and release anonymized statistics on reported incidents.
\item Designate, train, and advertise resources provided by a point person or team to act as a 
liaison or liaisons between the Title IX office and the staff, students, and scholars.
\item Actively train all current staff, students, and scholars regarding their roles and duties as 
potential responsible employees in the context of Title IX.
\item Provide a mechanism for applicants and interviewees to communicate perceived climate issues 
to the Department or University in search of clarification or redressal.
\end{enumerate}

Further information about Title IX may be obtained from the US Department of Education~\cite{titleix-depted} or the 
non-profit National Women's Law Center~\cite{titleix-natwomenlawctr}. The University of California system provides an example form that allows anonymous reporting~\cite{titleix:samplecomplaint}.
The 2016 Astronomy Climate Survey at the University of California, Berkeley~\cite{berkeley-climatesurvey-2016} provides a good example of a departmental climate survey.
For an example of departmental liaisons to the Title IX office, see the Climate Advisor model established at UC Berkeley Astronomy~\cite{titleix-berkeley-reporters}.

      \subsection {Participate in surveys exploring LGBT+ experiences}
\label{univ-surveys}
\BPGtag{D} \BPGtag{R} \BPGtag{S} \BPGtag{P} 
Encourage participation in {national or regional surveys} that address LGBT+ issues. For example, Campus Pride produces a national index of LGBT+-friendly colleges and universities~\cite{campusprideindex}, a valuable resource for students and administrators. An official authorized to represent the college or university on LGBT+ issues may contact Campus Pride to take part in the assessments for the Index.

      \subsection {Become an advocate}
\label{become-advocate}
\BPGtag{A} \BPGtag{D} \BPGtag{R} \BPGtag{S} \BPGtag{P} 
If your institution's policies are not inclusive, lobby to change them.  Advocate for inclusion of the words ``sexual orientation'' and ``gender identity or gender expression'' in your institution's non-discrimination policy.  (See Appendix~\ref{resources} for resources).  If you lose existing or prospective faculty, staff or students because of institutional policies that are unwelcoming to LGBT+ persons, complain to your top administrators.

      \subsection{Support students in diversity initiatives}
\label{student-diversity-initiatives}
\BPGtag{A} \BPGtag{D} \BPGtag{H} \BPGtag{R} \BPGtag{S} \BPGtag{P} 

Students are increasingly aware and engaged in on-campus work related to Diversity, Equity, \& Inclusion (DEI) around the country~\cite{studentdiversity:michigan,studentdiversityactivism:parnell}. The student body thus forms an untapped and dynamic resource to support top-level efforts to improve DEI on college campuses. In particular, systematically including students in this work increases the staff capacity to implement DEI strategic plans; provides recognition for students already engaged in diversity labor; harnesses student expertise; establishes transparent communication channels across the academic substructure of the institution; and reduces the burden on marginalized faculty, students, and staff (Section \ref{cultural-taxation}). DEI connections forged by students can also lay important groundwork for sections related to broader impact in grant proposals.

Encourage your institution to establish incentives to promote and reward diversity work by the students. For example, your institution might:

\begin{itemize}
	\item Establish awards for outstanding contributions to community and diversity-related activities.
	\item Establish paid internships or paid DEI undergraduate or graduate student staff positions. Existing position titles include Program Assistants, Alumni Engagement \& Fundraising Assistants, Summer internships, Peer Mentors, and Peer Counselors.
	\item Maintain an online list of institutional and community DEI opportunities to help students looking for positions.
	\item Encourage faculty to highlight students' diversity work in letters of recommendation.
	\item	Value and weigh diversity work while evaluating candidates for admissions and hiring, as well as for promotions.
\end{itemize}

Examples of paid internships at a range of colleges and universities may be found in references~\cite{studentdiversityinternships:berkeley, studentdiversityinternships:ohsu, studentdiversityinternships:bradley, studentdiversityinternships:aquinas, studentdiversityinternships:wesleyan, studentdiversityinternships:delaware, studentdiversityinternships:georgetown, studentdiversityinternships:lynchburg}.

    \section{Caring}
      \subsection {Help trans students deal with Selective Service}
\label{univ-trans}
\BPGtag{A} \BPGtag{D} \BPGtag{S} \BPGtag{P} 
US citizens and immigrants to the US who were assigned male at birth are required to register with the Selective Service System within thirty days of their eighteenth birthday -- regardless of whether they have transitioned before or since then. People who were assigned female at birth are not required to register regardless of their current gender or transition status.  These requirements are important for all college students, because many government benefits, including federal financial aid and federal employment, are contingent upon Selective Service registration, but they can be particularly daunting for trans students.  The National Center for Transgender Equality has helpful information for trans people dealing with the Selective Service System~\cite{selectiveservice:transequality}.  Universities should be aware that transgender students may have difficulty providing proof of registration.

      \subsection {Appoint a contact person for dual-career couples}
\label{dual-career-contact}
\BPGtag{D} \BPGtag{H} \BPGtag{S} \BPGtag{P} 

For any potential employee, it may be important to acquire information about dual-career resources in confidence and without affecting the job search.  As noted in Section~\ref{dual-career}, the issues can be even more complex for same-gender dual-career couples. The university should provide means for job candidates to acquire this information as early as possible in the process, to ensure that there is clarity about the prospects when an offer of employment is made.  It is helpful to have a visible, comprehensive web page where resources related to dual-career issues and work-life balance are collected.  Best of all is for the university to appoint a {single contact person for all dual-career issues, regardless of gender of the couple}. This person (e.g., a vice-provost for academic human resources), who is far removed from the search committee, can assist in assessing the situation while the hiring process is on-going.  Having an external, confidential, and well-informed person from whom potential employees may obtain advice about the details of a dual-career situation can help make all job candidates feel welcome.
      
      \subsection{Advocate for intersectional inclusion at institutional resource centers}
\label{sec:inclusion-universitycenters}
\BPGtag{D} \BPGtag{P}  
Colleges and universities typically strive to support various student, staff and faculty populations through official resource centers and officially recognized student groups. Often, each resource is targeted to a specific demographic marker, e.g.\, LGBT+ resource centers; women-in-physics groups; specific ethnic, cultural, or religious associations; or offices or centers for international students. This specialization has the frequent side effect of excluding people with multiple marginalized identities, who often find that they must suppress part of their identity in order to fully engage with a group that is meant to bolster another aspect of their identity~\cite{marsh:2016}. For example, a student may find that a major event at the LGBT+ resource center is scheduled at the same time as an important religious or cultural holiday. The international center may be ill-equipped to assist with the complexities of gender transition for international scholars. A lesbian physics student may be frustrated by women-in-science panels that consistently focus on dual-career problems and child-care inequities as experienced by heterosexual couples.

\begin{newquote}{Anonymous interviewee, APS report on LGBT Climate in Physics~\cite{LGBTClimateInPhysics:2016}}{}
I think I grappled more with the race element than I do with the sexuality because the deal is ... that's what they see first. I can't actually closet my race because I'm -- evidently I'm brown -- my hair looks different, so it's just there. That said, I think there's already a prejudgment there on the basis of how high my aptitude is, just in general. It doesn't necessarily have to be specific to physics but anything that requires some level of critical thinking is always kind of under examination or assumed to be mediocre or subpar.
\end{newquote}
Diversity efforts should be planned to support the whole person, honoring the ways in which different aspects of identity intersect. Information on campus resource centers and groups should be distributed to all students, staff and faculty, regardless of whether they appear to belong to the relevant demographic. At the department level, diversity panels should strive to welcome a range of voices and raise a range of questions, representing the demographics of the audience (and remembering that some marginalized statuses may be invisible to the department). At the university level, advocate for partnership between the various centers and groups. These provide a crucial support network for students, staff, and faculty, and should be able to support everyone.

      \subsection{Ensure sensitivity training for campus first-responders}
\label{firstresponders}
\BPGtag{P}  
\begin{newquote}{Beyond Coming Out report from CHEE at OSU \cite{strayhorn:2015}}{}
100\% of Black gay [undergraduate] males in our studies have contemplated or attempted suicide at least once.
\end{newquote}
Suicides, attempted suicides, and interpersonal violence are tragic realities on campus, and LGBT+ people are at elevated risk for suicide and hate crimes. National studies of college students have found that gay, lesbian, bisexual and queer college students are $2-3$ times more likely than straight college students to attempt suicide, and that 20\% experience fear for their physical safety because of their gender identities or their sexual orientations~\cite{strayhorn:2015}. The 2015 US Transgender Survey found that 53\% of $18-25$-year-old transgender respondents were in current serious psychological distress at the time they were surveyed, compared to 10\% of the general U.S. population in that age group~\cite{ustranssurvey:2016}\footnote{Serious psychological distress was assessed by self-reporting via the widely used Kessler Psychological Distress Scale (K6)~\cite{kessler:2002}}. This vulnerability is amplified for students who also belong to additional minority groups. Insensitive treatment of a vulnerable individual, such as negative or dismissive comments about a victim's gender expression or sexual orientation, will compound that person's emotional pain surrounding the crisis. Members of ethnic minority groups, especially African-Americans, also suffer a well-documented, additional risk of police violence, even during episodes of mental crisis. Tragedies are compounded when a fear of violence prevents vulnerable people from seeking help.

Advocate with the university to make sure that campus police, medics, and other first-responders 
take advantage of racial, cultural-awareness and LGBT+-specific sensitivity training, to help them 
safely and appropriately assist students and other community members in distress. The university 
should maintain a mechanism for receiving complaints about the conduct of first-responders on 
campus, and should take such complaints seriously.

\begin{newquote}{2015 US Transgender Survey~\cite{ustranssurvey:2016}}{}
Forty percent (40\%) of [transgender] respondents have attempted suicide in their lifetime, nearly nine times the rate reported in the U.S. population (4.6\%) ... Lifetime suicide attempts were also higher among people of color, with American Indian (57\%) respondents reporting the highest rates, followed by multiracial (50\%), Black (47\%), Latino/a (45\%), and Middle Eastern (44\%) respondents, in contrast to white (37\%) respondents ... People with disabilities (54\%) in the sample were more likely to report attempting suicide.
\end{newquote}

    \section{Equity}
      \subsection {Use gender-neutral and inclusive language}
\label{univ-language}
\BPGtag{D} \BPGtag{S} \BPGtag{P}
Encourage your institution to include ``sexual orientation'' and ``gender identity and gender expression'' in its public statements about diversity and multiculturalism.  Include LGBT+ issues and concerns in grievance procedures, housing guidelines, application materials, health-care forms, and alumni materials and publications.  Include representations of LGBT+ people in these publications.  Until a university-wide policy is in place, your department can take an active role in improving the language that is used at your institution by pointing out instances of discrimination through language and work with other departments (including human resources, benefits, and public relations) to adopt gender-neutral and inclusive wording.

      \subsection{Facilitate name and gender changes on institutional records while preserving privacy}
\label{institutional-paper-trail}
\BPGtag{D} \BPGtag{S} \BPGtag{P}

Paper trail issues regarding name and gender are of critical importance to many transgender people. The ability to present an accurate name and gender to the world is a major part of being able to live authentically. In some cases, paperwork can even out an individual against their will, exposing them to risks that could otherwise have been avoided.

Unfortunately, legal documentation changes can be expensive, time-consuming, and invasive. Therefore, many transgender people are stuck for long periods of time (or indefinitely) with an inaccurate legal name and/or gender marker. Even after a legal change, outdated information may persist in institutional records. A person's ability to live authentically and safely can be severely compromised if their previous (or current) legal name or gender marker is made public, even by accident. On the other hand, some transgender people may be out within the environment of the institution, but not in all other contexts. In that case, sadly, an individual may need to avoid mention of their preferred name or pronouns on public-facing records (such as web pages). This may be a situation with very concrete and serious risks; for instance, in the case of a student who is financially dependent on unsupportive family members. Such situations will require coordination between instructors, advisors, or department administration and the person who has requested that they be called by a new name. 

Fortunately, institutions and their sub-units can do much to assist the preservation of a transgender person's privacy and make the transition process easier for all.

\subsubsection{Handle name changes flexibly}

The use of a person's preferred name (rather than legal name) should be permitted in a wide range of situations, including the partial list below. Ideally, people should have the ability to use their preferred name exclusively in daily life, and to keep their legal name private except where absolutely required. However, the choice must always be made by the individual in question, so that they can control how their identity is shared -- especially when the record is visible to the public.

There should also be a clearly documented procedure for changing one's name after a legal update, and this change should propagate across all records systems where legal names are stored, including department records (Sec.~\ref{name-changes}) and computer usernames (Sec.~\ref{univ-username}). Human Resources should serve as an interface between the individual and the various sub-units involved.

For privacy reasons, university policy should spell out which names are to be used when records are requested. Any record of a person's preferred name should be kept separate from those files and records that might fall under mandatory-disclosure rules, unless the person gives explicit, specific permission to include it.

Here are some places where names are used around a typical university, which should be included in a policy on preferred names and legal name changes:

\begin{itemize}
	\item Diploma and transcript
	\item Class roster
	\item Course instructor listing
	\item Email address, if it is based on a person's name
	\item Unit website or department directory entry
	\item Institution identification card
	\item Library identification (if different from ID card)
	\item Dean's Lists
	\item Office door signs
	\item Commencement programs
\end{itemize}

Flexible name change policies also benefit those who are not transgender as there are other situations (marriage, estrangement from family, etc.) in which a name change is desired and helpful. Preferred name policies also benefit people who simply go by a different name than their legal one, for whatever personal or professional reason.

\subsubsection{Handle gender markers with caution and flexibility}

Institutions also record legal gender markers for a variety of purposes. This information should be handled carefully.

As in the case of names, there should be a clearly documented procedure for changing one's gender marker after a legal update. This change should propagate across university systems. Historical records, showing outdated names or gender markers, should be expunged or kept in the strictest confidence.

Gender markers should not be shown where they are not strictly necessary. If gender markers are displayed to members of the university community, or shown on documents that may be seen by those outside the university (e.g. transcripts), this practice should be carefully reviewed.

For privacy reasons, institutional policy should spell out where, when, and to whom, legal gender markers are visible.

Finally, be aware that some US states have recently adopted a third category of legal gender marker. University records systems will need to adapt to this fact. Some countries outside the US also allow for a third category of gender marker. \footnote{Many systems in the US, legal and otherwise, are unfortunately not able to accommodate a third option; this may raise issues especially with regard to travel, where identification regulations are often strict, and available gender options are typically limited to male or female. See \ref{flex:travelarrange} and \ref{alt-transport} for relevant recommendations on giving people travel flexibility.} \footnote{At the time of writing, three states (Oregon, California, and Washington) and Washington, DC, allow for third gender options (typically an ``X'' marker) on various forms of legal documentation. Legislation and/or court cases on the issue are pending in several other jurisdictions.}

      \subsection {Allow rare changes to computer usernames}
\label{univ-username}
\BPGtag{S} \BPGtag{P}

Institutional computer services -- including email, web publishing, printing, course registration, teaching, and library services -- are usually accessed with a short username, which is often based on the affiliate's name. Depending on university policy, a student named John Q.~Doe might be assigned a username based on first name (\texttt{johnd}), on surname (\texttt{jdoe}), or on initials (\texttt{jqd}). If John or Doe are common names at that institution, the username might include numbers as well. At some universities, John would be able to choose from a variety of usernames, but most require a relationship to John's name in order to promote professionalism. A serious problem arises if John's name changes at some point during their affiliation with the university. For example, John might change their first name as part of a gender transition, or they might change their surname due to a marriage, a divorce, or a severed relationship with their parents. (The latter issue is more common for LGBT+ students than for straight/cis students). In these circumstances, a legacy username can be a painful or annoying stressor every time someone logs onto email or authenticates with the library to access a scientific paper online. When a username based on an outdated first name is visible to the public (as in an email address or webpage) it also runs the risk of outing the user as trans.

Find out whether your institutional information technology (IT) services permit users to change their usernames. If they do not, advocate with the university to allow username changes in rare circumstances on diversity grounds. This issue also disproportionately affects women entering or leaving different-gender marriages, where there is a cultural norm of a surname change. Harvard University, for example, allows username changes in connection with legal name changes~\cite{harvard:username}. Note, however, that many people face real financial, administrative, and even medical obstacles to legal name changes, depending on where they live, where they were born, and their reasons for the change; a name change due to a gender transition is usually more difficult than one due to a change in marital status. The University of California-Berkeley allows one lifetime username change for each user~\cite{username:calnet}. We recommend a policy that allows users to select a username from several automatically generated possibilities, allowing the choice to emphasize or de-emphasize a first name or surname, and in which $1-2$ lifetime username changes are permitted for each user. The new username should not be publicly associated with the old one, which should be purged from public-facing directories.

If IT services do not permit username changes, and advocacy with the university is not effective, consider ways in which the department might provide support for affected students, faculty and staff. For example, the department might be able to provide email addresses (e.g., \texttt{janedoe@phys.example.edu}) or web hosting that can bypass the official, generic university services. 

      \subsection {Provide inclusive health insurance}
\label{health-insurance}
\BPGtag{H} \BPGtag{P} 

LGBT+ students, faculty and staff are often unable to take full advantage of university or college health benefits. The impact on individuals who are unable to access quality healthcare is difficult to overstate. Healthcare is also a priority for many job-seekers, so offering inclusive health insurance is a competitive advantage when hiring candidates who are LGBT+, or who have LGBT+ beneficiaries. 

The issue of inclusive health care is often particularly relevant to transgender people. Many health plans exclude ``procedures related to being transgender''. As documented by the Transgender Law Center, this exclusion not only applies to transition-related medical services -- such as hormone treatments, gender-affirming surgeries, or therapy for those who require it -- but also has been used to deny treatment for pathologies associated with the sex assigned at birth (e.g., uterine cancer in a transgender man) and for such non-gendered problems as the flu or a broken arm. The problem extends to medical practices, as well, independent of insurance~\cite{ohara:2015}. The impact of such policies on an individual's physical and mental health cannot be overstated. Advocate for removing the transgender exclusion from your college or university's health plan (this has been done successfully by the University of California system).  As a smaller step, advocate for including specific coverage for certain procedures (such as therapy, hormone treatments, and gender-affirming surgeries) for transgender students, faculty, staff and family members.

Whenever discussing healthcare for transgender people, it is also important to be aware of the diversity of needs within the transgender community. There are a wide variety of procedures and services that fall under the general umbrella of ``medical transition'' or ``gender-affirming health care'', and transgender people make an equally wide variety of choices about which ones (if any) to pursue. Healthcare which assumes an all-or-nothing model of medical transition poorly serves many members of the transgender community. 

Fertility is another area of healthcare that may be particularly important to LGBT+ people. It is important to make sure that any coverage of fertility care is not exclusionary toward LGBT+ people, directly or indirectly. Policies on this subject are often oriented strictly toward heterosexual, cisgender couples, and may include specific requirements for coverage that LGBT+ people are unable to meet. For instance, fertility coverage may be contingent upon diagnosis of a medical condition leading to infertility. This can cause LGBT+ people to be denied reproductive options available to others, either because they are not technically infertile (as in the case of a couple who are both individually able to reproduce, but not with each other) or because they are not yet infertile (as in the case of someone seeking fertility preservation in advance of gender-affirming care which will impact their fertility). Inclusive language matters, too: since many non-binary people and trans men can become pregnant, it is not only women who may need prenatal care.

As with other advocacy projects, pushing for improved benefits should ideally be undertaken in cooperation with the affected population (e.g. in consultation with local LGBT+ community organizations). The landscape of care access and quality varies so greatly between institutions and locations that local expertise is indispensable. Additionally, advocacy undertaken in partnership with labor unions can be particularly fruitful. Unions are tasked with advocating for their members' healthcare benefits, and are typically experienced in doing so. In fact, it may not even be possible to address healthcare benefits for university employees without the involvement of the relevant labor union, where one exists, for legal reasons surrounding collective bargaining and representation.

\begin{newquote}{Anonymous professor of physics, as told to the authors}{}
My university had exclusionary language regarding health insurance for transgender-related care. To file the appeals and finally get the approval to override the exclusions, the effort was the equivalent of writing two full NSF proposals in fields outside of my expertise. I argued to my President that he'd probably rather have me spend my time on research rather than trying to get my health care covered by insurance.
\end{newquote}


      \subsection {Provide benefits for same-gender partners}
\label{partner_benefits}
\BPGtag{P} 

Same-gender couples now have the legal right to marry in the US, but this does not mean that the problem of same-gender partner benefits is entirely solved. 
First, not all physicists are US citizens, nor are they all citizens of other nations that recognize same-gender unions. Second, even in the US, same-gender couples marry at a lower rate than different-gender couples~\cite{gallup:lgbtmarriage:2017}. Reliance on marriage as a requirement for access to partner benefits therefore disadvantages LGBT+ people.

The reasons for the discrepancy in marriage rates are not fully understood. But in broad terms, it is clear that legalization of same-gender marriage did not establish comprehensive equality for same-gender couples. Social stigma persists, which may lead people to eschew the visibility that typically comes with marriage, even when it would appear to benefit them financially and/or logistically. For instance, non-discrimination laws are not uniform across the US; notably, in some jurisdictions in the US, same-gender couples have the right to marry, but lack full legal protection from discrimination for being in a same-gender relationship in the first place. Furthermore, legal protection does not necessarily mitigate familial and social pressures which can lead people to hide same-gender relationships, regardless of how long-term and committed they may be.

Fortunately, there are alternatives to providing partner benefits based solely on marital status. A number of academic institutions cover unmarried partners using alternative frameworks, e.g. providing insurance coverage for an ``other qualified adult'' or a ``legally domiciled adult.'' These policies, which mostly originated in the era preceding Obergefell vs Hodges, remain useful today as a method of improving access to same-gender partner benefits. They are, of course, also beneficial to people with unmarried partners of a different gender.

\begin{newquote}{Anonymous interviewee, a STEM faculty member, in Bilimoria and Stewart, 2009~\cite{bilimoria:2009}}{}
Partner benefits were widely advertised on the Web site, and it was obvious that the university was really positive about it. That made a huge difference.
\end{newquote}

      \subsection {Provide equal access to restrooms}
\label{restroom_access_uni}
\BPGtag{S} \BPGtag{P}
Lack of access to restrooms can seriously disrupt a person's ability to learn, teach, research, or 
participate in events. Consider the impact on a student who must go to another building in order to 
access a safe restroom, or a colleague who is constantly dehydrated at conferences as part of a 
strategy to avoid public restrooms altogether. These concerns can be alleviated by providing proper facilities, ideally with the help and 
support of the university administration. For further discussion of this issue, see section 
\ref{restroom_access_dept}.

\begin{newquote}{Anonymous transgender student, as told to the authors}{}
Bathrooms have been an ongoing issue both as an undergraduate and a graduate student. At the university where I currently work, the only gender neutral restroom in the (rather gigantic) facility is on the complete opposite side of the building, one floor down from my office. This has led to me drinking less fluids during the day to avoid the trek. 
Even worse, half the time the restroom is occupied. I've found myself standing outside waiting for 10+ minutes while people go in and out of the adjacent gendered restrooms.
\end{newquote}
\usechapterimagetrue
\chapterimage{Introduction.pdf}
\chapter*{Conclusions}
\label{ch:conclusions}
\usechapterimagefalse

The issue of LGBT+ Inclusion in Physics and Astronomy (and by extension, to other areas within the broad Science, Technology, Engineering, and Mathematics, or STEM fields) is fundamentally a question of cultural transformation. We seek to create an atmosphere where all persons are encouraged to grow and thrive regardless of their genders, orientations, or the associated complex intersections of these with the rest of their whole selves; where each individual is respected and recognized for their unique perspective, and where equity is the indelible norm. Ultimately, such an atmosphere is not only beneficial to the execution of science - it is an imperative of a society that holds as its fundamental tenet the inherent value of all human beings.

The establishment of such an environment requires a paradigm shift in the thinking and operation of our society and of the scientific venture within it, and like all transformations, this change cannot happen overnight. Beginning with visibility and supported by active agency, the endeavor to make science a more equitable and just enterprise proceeds through you. No matter your role -- faculty, staff, administrator, or even student -- equipped with this Guide, YOU can be an instrument of change. Bring these issues to discussion at your institutions. Listen actively to your LGBT+ colleagues’ and students’ concerns. At educational institutions, build your response on the principles of student support and faculty development. Educate yourself about the matters facing LGBT+ scientists today, elucidate the goals and philosophies that govern your responses and those of your institution, and encourage your peers and colleagues to take them seriously. 

This Guide is simply a starting point toward a broader discourse. We recognize that the issues raised herein are part of an ongoing conversation, and that norms, practices, laws, and society are always evolving. We have attempted to include a broad cross-section of concerns, but will inevitably have left out items of importance. If you have any feedback, please correspond with the editors so we may include it in future editions of the Guide.

The process of cultural transformation that will converge toward the ideal of a more inclusive world remains long and arduous, and we are glad to have your support. Our bold vision is that someday the completion of this transformation will make guides like this one obsolete.

\medskip
\noindent The Authors\\ 
April 2018

\chaptermark{Conclusions}
\addcontentsline{toc}{chapter}{Conclusions}

\appendix
\pagestyle{simple}
\chapter{Resources}
%

\label{resources}		
\normalsize			

\section{Resources for LGBT+ Scientists and Science Students}
\label{resources:scientists}

\begin{tabular*}{\textwidth}{@{\extracolsep{\fill}}lr}
	\textbf{LGBT+ Physicists} & \href{http://lgbtphysicists.org}{lgbtphysicists.org}	
\end{tabular*}
This website was created to collect resources for and address the issues of LGBT+ people in physics. It contains information on joining LGBT+ Physicists, an Out List, current and past events, and links to other resources.

\vspace*{\baselineskip}
\noindent\begin{tabular*}{\textwidth}{@{\extracolsep{\fill}}lr}
	\textbf{SGMA} & \href{https://sgma.aas.org}{sgma.aas.org/}
\end{tabular*}
SGMA (the Committee for Sexual-Orientation \& Gender Minorities in Astronomy), an official committee of the American Astronomical Society, works to support LGBT+ people in astronomy and astrophysics through development of resources, training and mentoring. The successor of WGLE (Working Group on LGBTIQ Equality), SGMA often works jointly with other committees focusing on promoting women and racial/ethnic minorities in the astronomy community.

\vspace*{\baselineskip}
\noindent\begin{tabular*}{\textwidth}{@{\extracolsep{\fill}}lr}
	\textbf{Astronomy and Astrophysics Outlist} & \href{https://astro-outlist.github.io/}{https://astro-outlist.github.io/}	
\end{tabular*}
The Astronomy and Astrophysics Outlist gives names and contact information for both LGBT+ astronomers and allies.

\vspace*{\baselineskip}
\noindent\begin{tabular*}{\textwidth}{@{\extracolsep{\fill}}lr}
    \textbf{Inclusive Astronomy:} & \href{https://tiki.aas.org/tiki-index.php?page=Inclusive_Astronomy_The_Nashville_Recommendations}{tiki.aas.org/tiki-index.php?page=Inclusive\_Astronomy}	\\
    \textbf{The Nashville Recommendations} & \\
\end{tabular*}
Beginning in June 2015, astronomers, sociologists, policy makers and community leaders have convened meetings to discuss issues faced by members of one or more marginalized communities in astronomy. This set of recommendations is a living document, compiled to address these challenges.

\vspace*{\baselineskip}
\noindent\begin{tabular*}{\textwidth}{@{\extracolsep{\fill}}lr}
	\textbf{LGBT+ Physical Sciences Network} & \href{http://www.iop.org/policy/diversity/lgbt-network/page_68474.html}{iop.org/policy/diversity/lgbt-network/page\_68474.html}	
\end{tabular*}
The LGBT+ Physical Sciences Network is an international network in the UK and Ireland, connecting physicists, astronomers, and chemists. It is a joint project by the Institute of Physics, the Royal Astronomical Society, and the Royal Society of Chemistry.

\vspace*{\baselineskip}
\noindent\begin{tabular*}{\textwidth}{@{\extracolsep{\fill}}lr}
	\textbf{NOGLSTP} & \href{http://www.noglstp.org}{noglstp.org}	
\end{tabular*}
The National Organization of Gay and Lesbian Scientists and Technical Professionals, Inc., is a national professional society in the US. It educates STEM communities about the needs of their LGBT+ members and supports lesbian, gay, bisexual, transgender, and queer people in STEM fields, especially via mentoring, networking, and advocacy.

\vspace*{\baselineskip}
\noindent\begin{tabular*}{\textwidth}{@{\extracolsep{\fill}}lr}
	\textbf{oSTEM} & \href{http://www.ostem.org}{ostem.org}	
\end{tabular*}
Out in Science, Technology, Engineering, and Mathematics is a national society dedicated to the organization and professional development of LGBT+ students in STEM. The group consists of affiliate chapters throughout the US.

\vspace*{\baselineskip}
\noindent\begin{tabular*}{\textwidth}{@{\extracolsep{\fill}}lr}
	\textbf{Out to Innovate} & \href{http://www.noglstp.net/outtoinnovate}{noglstp.net/outtoinnovate}	
\end{tabular*}
Sponsored by NOGLSTP, Out to Innovate is a two-day career summit for LGBT+ students, faculty and professionals in science, technology, engineering, and mathematics, held every other year.

\vspace*{\baselineskip}
\noindent\begin{tabular*}{\textwidth}{@{\extracolsep{\fill}}lr}
	\textbf{WISE} & \\
\end{tabular*}
Women in Science and Engineering is active on many college campuses. Check for a chapter near you.


\section{Resources for LGBT+ People at Large National \\and International Laboratories}
\label{resources:labs}

\vspace*{\baselineskip}
\noindent\begin{tabular*}{\textwidth}{@{\extracolsep{\fill}}lr}
	\textbf{LGBTQ CERN} & \href{http://lgbtqcern.com/}{lgbtqcern.com}
\end{tabular*}
CERN-recognized Informal Network supporting and welcoming members of the LGBT+ community and allies.

\vspace*{\baselineskip}
\noindent\begin{tabular*}{\textwidth}{@{\extracolsep{\fill}}lr}
	\textbf{Spectrum} & \href{http://blogs.anl.gov/spectrum/}{blogs.anl.gov/spectrum/}\\
	\textbf{Argonne National Laboratory} & \\
\end{tabular*}
Spectrum is an employee resource group for LGBTQ+ people and allies at Argonne National Laboratory, also serving the U.S. Department of Energy, the University of Chicago, and the Argonne Credit Union.

\vspace*{\baselineskip}
\noindent\begin{tabular*}{\textwidth}{@{\extracolsep{\fill}}lr}
	\textbf{Prism} & \href{https://www.lanl.gov/careers/diversity-inclusion/erg/lgbtq/index.php}{lanl.gov/careers/diversity-inclusion/erg/lgbtq/index.php} \\
	\textbf{Los Alamos National Laboratory} & \\
\end{tabular*}
Prism is the LGBTQ+ Staff and Allies Employee Resource Group at Los Alamos National Laboratory, promoting diversity and providing a safe space for LGBT+ employees and allies.


\section{Resources for LGBT+ People on Campus}
\label{resources:campus}

\vspace*{\baselineskip}
\noindent\begin{tabular*}{\textwidth}{@{\extracolsep{\fill}}lr}
	\textbf{Campus Climate Index} & \href{http://www.campusprideindex.org}{campusprideindex.org}	
\end{tabular*}
This index was put together by CampusPride and ranks colleges and universities by how friendly they are to LGBT+ students. It contains information on how ranking is done, as well as how you can add your institution to the list.

\clearpage
\noindent\begin{tabular*}{\textwidth}{@{\extracolsep{\fill}}lr}
	\textbf{CampusPride} & \href{http://www.campuspride.org}{campuspride.org}	
\end{tabular*}
CampusPride is a leading organization in research on LGBT+ people in colleges and universities. They put together the 2010 State of Higher Education for LGBT People, organize LGBT+ job fairs, and compile lists of those colleges and universities that excel in LGBT+ issues.

\vspace*{\baselineskip}
\noindent\begin{tabular*}{\textwidth}{@{\extracolsep{\fill}}lr}
	\textbf{Out for Work} & \href{http://www.outforwork.org}{outforwork.org}	
\end{tabular*}
Out for Work is a  national educational non-profit that hosts career fairs for LGBT+ undergraduate students.

\section{Resources for LGBT+ People in Broader Communities}
\label{resources:communities}

\vspace*{\baselineskip}
\noindent\begin{tabular*}{\textwidth}{@{\extracolsep{\fill}}lr}
	\textbf{CenterLink} & \href{http://www.lgbtcenters.org}{lgbtcenters.org}	
\end{tabular*}
CenterLink, the Community of LGBT Centers, is a member-based coalition of local LGBT community centers.  It represents over 200  centers in the U.S. and abroad. 


\vspace*{\baselineskip}
\noindent\begin{tabular*}{\textwidth}{@{\extracolsep{\fill}}lr}
	\textbf{Gay, Lesbian, and Straight Education Network} & \href{http://www.glsen.org}{glsen.org}	
\end{tabular*}
The leading national education organization focused on ensuring safe schools for all students.  Although focused on K-12 education, GLSEN's research provides many insights into LGBT+ students, including those soon to become college freshmen. 

\vspace*{\baselineskip}
\noindent\begin{tabular*}{\textwidth}{@{\extracolsep{\fill}}lr}
	\textbf{GLAAD Media Reference Guide} & \href{http://www.glaad.org/reference}{glaad.org/reference}	
\end{tabular*}
Although created for journalists, this guide provides information on terminology used with LGBT+ communities as well as a list of current national issues in the United States faced by the community.

\vspace*{\baselineskip}
\noindent\begin{tabular*}{\textwidth}{@{\extracolsep{\fill}}lr}
	\textbf{Human Rights Campaign} & \href{http://hrc.org/resources}{hrc.org/resources}	
\end{tabular*}
The HRC website provides extensive resources on a variety of topics.  Of particular relevance are articles on advocating for LGBT equality in the workplace and domestic-partner benefits.

\vspace*{\baselineskip}
\noindent\begin{tabular*}{\textwidth}{@{\extracolsep{\fill}}lr}
	\textbf{Immigration Equality} & \href{http://immigrationequality.org}{immigrationequality.org}	
\end{tabular*}
This is a nonprofit advocacy group focused on U.S. immigration issues as applied to members of the LGBT+ community. Their website provides up-to-date resources for issues frequently faced by transgender immigrants, binational same-gender couples, other LGB+ individuals, and people who are HIV-positive. They may also provide legal help or recommend a private attorney.

\vspace*{\baselineskip}
\noindent\begin{tabular*}{\textwidth}{@{\extracolsep{\fill}}lr}
	\textbf{The National Center for Transgender Equality} & \href{http://transequality.org}{transequality.org}	
\end{tabular*}
A social-justice organization dedicated to advancing the equality of transgender people through advocacy, collaboration and empowerment.  Their web site provides information on a variety of issues that affect trans people.

\vspace*{\baselineskip}
\noindent\begin{tabular*}{\textwidth}{@{\extracolsep{\fill}}lr}
	\textbf{Marriage Equality FAQ}  	
	 & \href{https://marriageequalityfacts.org/}{marriageequalityfacts.org/}	
\end{tabular*}
The American Civil Liberties Union, Freedom to Marry, Gay \& Lesbian Advocates \& Defenders, Human Rights Campaign, Lambda Legal and National Center for Lesbian Rights have jointly compiled a useful set of fact sheets about various aspects of marriage equality and marriage-related legal issues -- from wedding planning to government spousal benefits -- after the \textit{Obergefell v.~Hodges} decision.

\vspace*{\baselineskip}
\noindent\begin{tabular*}{\textwidth}{@{\extracolsep{\fill}}lr}
	\textbf{The TONI Project} & \href{http://transstudents.org}{transstudents.org}	
\end{tabular*}
Organized by the National Center for Transgender Equality (NCTE), the TONI Project is a student-oriented space for sharing college and university practices and policies of particular interest to trans students. Campus-by-campus information may be useful to prospective students choosing a school, or to people hoping to improve policies at their own institutions.

\vspace*{\baselineskip}
\noindent\begin{tabular*}{\textwidth}{@{\extracolsep{\fill}}lr}
	\textbf{The Transgender Law and Policy Institute} & \href{http://www.transgenderlaw.org}{transgenderlaw.org}	
\end{tabular*}
This institute is a non-profit organization dedicated to engaging in effective advocacy for transgender people in society. The TLPI brings experts and advocates together to work on law and policy initiatives designed to advance transgender equality. Of particular interest to colleges and universities, this institute has put together a list of policies that affect transgender students and the institutions which implemented them 

\vspace*{\baselineskip}
\noindent\begin{tabular*}{\textwidth}{@{\extracolsep{\fill}}lr}
	\textbf{The Trevor Project} & \href{http://www.thetrevorproject.org}{thetrevorproject.org}
\end{tabular*}
The Trevor Project is the leading national organization providing crisis intervention and suicide prevention services to LGBT+ young people ages 13-24.  Their 24/7 confidential lifeline number is 1-866-488-7386.  They also provide training for adults who work with LGBT+ youth.

\chapter{Sample survey questions}
%

\label{survey}		
\normalsize			

\paragraph{Demographic information}
Information about the demographics of your department can be useful for tailoring
programs to be LGBT+-friendly. We suggest that questions requesting demographic information
should be as inclusive as possible, with multiple selections allowed, and including options for the respondent
to decline to answer or to state an identity not already in the list. Since gender identity and sexual orientation are related but distinct concepts,
it is important to address issues related to both separately.

Below, we reproduce sample demographic questions dealing with aspects of LGBT+ identity, prepared by the Human Rights Campaign~\cite{hrc-surveyqns}. These questions are meant to supplement, not replace, demographic questions about racial and ethnic background, national origin, age, disability, student or veteran status, and other categories that may be important to understanding a department or institution's demographics. As with all demographic questions, these should be optional, and it is worth reassuring respondents of their privacy, and reminding them of the relevant non-discrimination policies, in order to encourage frank responses. Please note that the language around LGBT+ identities evolves over time.

\begin{newexample} 
\noindent \textit{Source: \href{https://www.hrc.org/resources/collecting-transgender-inclusive-gender-data-in-workplace-and-other-surveys}{Human Rights Campaign}~\cite{hrc-surveyqns}}

\noindent \textbf{Do you consider yourself a member of the Lesbian, Gay, Bisexual and/or Transgender (LGBT) community?}
\begin{itemize} 
\item Yes
\item No
\item No, but I identify as an Ally
\item Prefer not to say 
\end{itemize}
\end{newexample}

\newpage
\begin{newexample}
\noindent \textbf{What is your gender?}
\begin{itemize} 
\item Female 
\item Male
\item Nonbinary/ third gender
\item Prefer to self-describe \_\_\_\_\_ 
\item Prefer not to say 
\end{itemize}

\noindent \textbf{Do you identify as transgender?}
\begin{itemize} 
\item Yes
\item No
\item Prefer not to say 
\end{itemize}

\noindent \textbf{What is your Sexual Orientation?}
\begin{itemize} 
\item Straight/Heterosexual
\item Gay or Lesbian
\item Bisexual
\item Prefer to self-describe \_\_\_\_\_ 
\item Prefer not to say
\end{itemize}

\end{newexample}

\paragraph{Harassment and Discrimination}
When asking questions about discrimination, it may be helpful to ask respondents to specify what aspects of identity were targeted in the incident. For example, a trans woman of color might experience harassment or discrimination based on her status as a woman, her status as a trans person, her racial or ethnic background, her sexual orientation, or some combination.
In addition, it may be helpful to separate questions relating to overt discrimination (harassment, denial of service, insults) from those relating to less overt discrimination (such as micro-aggressions), and to ask the respondent to rate the severity of the incident. Even when collecting general information on discriminatory incidents or harassment (as opposed to specific accusations), departments may wish to inquire whether an aggressor was associated with the department; with the institution; or with another institution.
\chapter{Definitions of sex and gender}
      \label{defs_sex_gender}		
\normalsize			

Gender and sex are terms with somewhat different meaning, but the distinction is a nuanced and actively debated one. Due to the centrality of this topic, we wished to address it directly; but due to its complexity, we found it best suited to an appendix rather than a glossary entry.

In broad terms, sex is a category into which organisms are divided on the basis of their reproductive roles. On the level of individual humans, this can be a complicated designation, as it includes a myriad of physical traits (chromosomes, hormones, internal organs, primary and secondary sex characteristics) which may or may not fit easily into a sex assignment of ``male'' or ``female''~\cite{Ainsworth:2015}.

Gender, while sometimes colloquially used interchangeably with sex, broadly refers to a category of social identity, culturally linked to sex but not synonymous with it~\cite{traxler:2016}. In this context, people often refer to ``gender roles'' and ``gender stereotypes,'' but the cultural significance of gender extends beyond role prescription. For instance, gender is marked in speech, used as a category of social and political analysis, and may deeply influence interpersonal relationships. It should not be understood as solely a matter of traditional gender roles.

In addition to these different uses, sex and gender can also refer to a legal category, which is typically assigned as based on external genitalia at birth, but may be legally changed in most jurisdictions (requirements vary). While legal sex/gender categories have often been limited to ``male'' and ``female,'' some countries and states now legally recognize third categories of sex/gender.

It is important to note that there is an ambiguity in the distinction between ``biological'' sex and ``cultural'' gender. For instance, sex assignment is an act of categorization carried out by humans for cultural purposes, almost always on the basis of incomplete information about an individual's physical traits. Gender, on the other hand, can also refer to a deeply-held sense of self which is arguably innate and/or biologically linked, regardless of alignment with assigned sex.

As a result of this complexity, views on the distinction between sex and gender, their biological and cultural origins, and appropriate legal and social role, vary greatly. In this guide, we will typically refer to gender rather than sex, and will take the view that any given individual is the final authority on their own gender. We do not pretend that there is a consensus on the nature of either gender or sex, nor do we provide here any kind of literature review on the subject.

\chapter{Glossary}
%

\label{glossary}	

\textit{Some of the following terms may be unfamiliar, and some familiar terms may have unfamiliar 
definitions, but their use has evolved out of the literature and debates on gender and 
sexual-diversity issues over the last few decades. We also note that these terms are not universal: 
various communities and language groups have evolved distinct nomenclature. It is a best 
practice to respect the identifiers that persons use for themselves.}\\

\noindent\textbf{Ally} (noun)\\
Someone who may not be part of the LGBT+ community, but works to ensure equal rights and opportunities 
for LGBT+ people.\\

\noindent\textbf{AFAB/AMAB} (adjective)\\
Terms referring to 
the sex a person is assigned at birth, without making assumptions about gender.
Abbreviated as Assigned Female At Birth (AFAB), Assigned Male At Birth (AMAB), 
or alternatively, Designated Female/Male At Birth (DFAB/DMAB). 
Note that birth certificates can often be changed to 
have a gender marker different than that assigned at birth. 
The laws and regulations are inconsistent across states; some states do not allow any changes, while
others may only issue a partial change that still makes note of the sex assigned at birth.\\

\noindent\textbf{Asexuality} (noun)\\
The lack of sexual attraction or low or absent interest in sexual activity. 
Asexual people may or may not still experience romantic attraction; for 
example, an heteroromantic asexual man would be romantically attracted to women. 
Asexuality should not be confused with celibacy or sexual abstinence, which are choices, not sexual 
orientations. \\

\noindent\textbf{Bisexuality} (noun)\\
Sexual orientation characterized by attraction to two or more genders. (Note that bisexuality does 
not necessarily exclude attraction to people outside the gender binary).\\

\noindent\textbf{Cisgender or Cis} (adjective)\\
Term referring to a person whose gender corresponds with the sex they were assigned at 
birth.  For example, a woman who was assigned female at birth and identifies as a woman is cisgender.\\

\noindent\textbf{Gay} (adjective)\\
Term referring to a person who is attracted to people of the same gender.  Formally referring to a 
man who is attracted to other men, this term also functions as an inclusive identity used, among others,
by lesbian women, queer, bisexual, and nonbinary persons.\\

\noindent\textbf{Gender}
See Appendix~\ref{defs_sex_gender}.\\

\noindent\textbf{Gender binary} (noun)\\
A model of gender which includes only two categories (man and woman) in a strict binary. This model 
is commonly assumed, but is harmful to people who are neither men nor women (often identifying as 
nonbinary).\\

\noindent\textbf{Gender expression} (noun)\\
How a person represents or expresses themselves in relation to gender -- through the clothes they 
wear, their hairstyle, language, communication style and/or their mannerisms. A person's gender 
expression may not always match the social expectations for their gender, and 
can change from situation to situation or from day to day.\\ 

\noindent\textbf{Gender identity} (noun)\\
An individual's private sense and personal experience of their gender. 
A person's gender identity may not necessarily align with the social expectations for the sex 
they were assigned at birth. \\

\noindent\textbf{Gender nonconforming} (adjective)\\
Referring variously to either (i) persons whose gender expression does not conform to gendered expectations; or (ii) an umbrella term synonymous with non-binary and/or genderqueer. \\


\noindent\textbf{Heteronormative} (adjective)\\
Term describing the often unstated assumption or expectation that people are heterosexual 
and cisgender, and conform to traditional gender roles.
Heteronormativity can manifest as overt hostility, or as a more subtle form of 
exclusion. For example, a joke that a young man must be looking for a girlfriend implicitly 
excludes and alienates persons who aren't interested in having a girlfriend,
implying that they are not normal and not fully welcome. \\

\noindent\textbf{Heterosexuality} (noun)\\
Sexual orientation characterized by attraction to individuals of another gender; for example, a man who is attracted to women.\\

\noindent\textbf{Homosexuality} (noun)\\
Sexual orientation characterized by attraction to individuals of the same gender; for example, a woman who is attracted to women. Note, however, that many gay people resent being referred to as ``homosexuals'', and prefer the term ``gay.''\\

\noindent\textbf{Intersex} (adjective)\\
A general term for a variety of conditions in which a person is born with a reproductive, sexual,
or genetic makeup that doesn't seem to fit the typical definitions of female or male 
(see~\cite{intersex-faq}).\\

\noindent\textbf{Lesbian} (adjective or noun)\\
Term referring to a woman who is attracted to women.\\

\noindent\textbf{LGBT+} (adjective)\\
Lesbian, Gay, Bisexual, Transgender. The plus is meant to include 
people who are not cisgender and heterosexual, but who do not 
fit neatly into LGBT. For many LGBT+ individuals, the label also connotes 
membership in a community of gender and sexual minorities.
Often ``LGBTQ+'', to include Queer/Questioning.
In this document, we use this term to represent the entire community of lesbian, gay, bisexual, 
transgender, queer, questioning, intersex, asexual, and other persons with marginalized gender 
identities and sexual orientations.
\\

\noindent\textbf{Nonbinary} (adjective)\\
An umbrella term for people whose gender does not fit neatly into the categories of man or woman. 
``Genderqueer'' is also a commonly used term; different people may prefer different words. Some 
nonbinary/genderqueer people identify as transgender, while others do not. Some use pronouns other 
than ``she'' or ``he,'' such as ``they'' or ``ze.''\\

\noindent\textbf{Out (of the Closet)} (adjective phrase) / \textbf{To Out Someone} (verb phrase)\\
Having openly identified oneself or having been identified by someone else as LGBT+. Being ``out'' 
as LGBT+ can depend on context. For example, a person may be out to some people but not to others 
(e.g., out at school but not to family members, or vice versa). The decision to come out is highly 
personal and others must respect how, when and to whom they are out. To out someone without their 
explicit consent is unethical and to do so can be harmful and even dangerous to the 
individual's personal safety.\\

\noindent\textbf{Pansexuality} (noun)\\
Sexual orientation characterized by attraction to people regardless of their gender. Some people identify as both bisexual and pansexual, while others consider pansexuality distinct from bisexuality.\\

\noindent\textbf{Pronoun} (noun)\\
A pronoun is a part of speech, often gendered, that stands in for a noun, like ``he'', ``she'', 
``they'', or ``ze''. You can determine what pronouns to use when referring to a specific person by 
asking them, e.g. ``What are your pronouns?'' or ``What pronouns should I use to refer to you?'' \\

\noindent\textbf{Queer} (adjective)\\
Despite its use as a slur, this term has been reclaimed by some members of the LGBT+ community as an 
identity that may be used in place of or in conjunction with other identities from the LGBT+ 
spectrum. Like all reclaimed slurs, it should be treated with caution, particularly by persons outside of the community.\\

\noindent\textbf{Questioning} (adjective)\\
An individual who is not yet certain of their sexual orientation or gender. Questioning is 
considered to be a legitimate identity in itself, and those who are questioning should not be 
coerced to ``make up their minds.'' \\

\noindent\textbf{Romantic orientation} (noun)\\
Term referring to how one defines the set of people one is romantically attracted to. 
Heteroromantic, homoromantic, biromantic, and aromantic are examples of romantic orientation. 
Note that an aromantic individual may still identify as heterosexual, homosexual, bisexual, etc., and may use terms like ``gay'' or ``straight'' that do not distinguish between sexual and romantic attraction.
\\

\noindent\textbf{Sex}
See Appendix~\ref{defs_sex_gender}.\\


\noindent\textbf{Sexual orientation} (noun)\\
Term referring to how one defines the set of people one is sexually attracted to. Heterosexual, 
homosexual, bisexual, and asexual are examples of sexual orientation. Note that an asexual person may still be heteroromantic, homoromantic, biromantic, etc., and may use terms like ``gay'' or ``straight'' that do not distinguish between sexual and romantic attraction. \\

\noindent\textbf{Transgender (sometimes Trans or Trans*)} (adjective)\\
Term referring to a person whose gender differs from the sex they were assigned at birth.  
For example, a trans woman is someone who was assigned male at birth but whose gender is 
female. When using ``trans" before a term of gender, such as ``trans woman" or ``trans man," the 
modifier ``trans" should describe the person's gender, not their birth assignment. 
Regarding usage, the word trans or transgender should only be used as a modifier and not a noun; 
for example, a person could be referred to as a transgender woman but she is not a transgender. 
Avoid the adjective ``transgendered.''

The related term ``transsexual'' should be used with caution, and should not be used synonymously with transgender. While some people do refer to themselves as transsexual, others object to it.

Additional resources for terms and their definitions can be found at:
\begin{itemize}
  \item 
\href{www.uwsuper.edu/genderequity/resources/lgbtqipa.cfm}{
www.uwsuper.edu/genderequity/ resources/lgbtqipa.cfm}
  \item
\href{
https://campusclimate.berkeley.edu/students/ejce/geneq/resources/lgbtq-resources/definition-terms}{
campusclimate.berkeley.edu/students/ejce/geneq/resources/ lgbtq-resources/definition-terms}
\end{itemize}

\backmatter
      
\chaptermark{References}

%

\addcontentsline{toc}{chapter}{References}

\begin{thebibliography}{107}
\providecommand{\natexlab}[1]{#1}
\providecommand{\url}[1]{\texttt{#1}}
\expandafter\ifx\csname urlstyle\endcsname\relax
  \providecommand{\doi}[1]{doi: #1}\else
  \providecommand{\doi}{doi: \begingroup \urlstyle{rm}\Url}\fi

\bibitem[Patridge et~al.(2014)Patridge, Barthelemy, and Rankin]{patridge:2014}
E.~V. Patridge, R.~S. Barthelemy, and S.~R. Rankin.
\newblock Factors impacting the academic climate for {LGBQ STEM} faculty.
\newblock \emph{J. Women Minorities in Sci. Eng.}, 20:\penalty0 75, 2014.

\bibitem[Yoder and Mattheis(2016)]{queerinstem:2016}
J.~B. Yoder and A.~Mattheis.
\newblock Queer in {STEM}: Workplace experiences reported in a national survey
  of {LGBTQA} individuals in science, technology, engineering, and mathematics
  careers.
\newblock \emph{J. Homosexuality}, 63:\penalty0 1, 2016.
\newblock \doi{10.1080/00918369.2015.1078632}.

\bibitem[Atherton et~al.(2016)Atherton, Barthelemy, Deconinck, Falk, Garmon,
  Long, Plisch, Simmons, and Reeves]{LGBTClimateInPhysics:2016}
T.~J. Atherton, R.~S. Barthelemy, W.~Deconinck, M.~L. Falk, S.~Garmon, E.~Long,
  M.~Plisch, E.~H. Simmons, and K.~Reeves.
\newblock Lgbt climate in physics: Building an inclusive community.
\newblock Technical report, American Physical Society, College Park, MD, 2016.
\newblock URL \url{http://www.aps.org/programs/lgbt/}.

\bibitem[Rankin(2003)]{rankin:2003}
S.~R. Rankin.
\newblock Campus climate for gay, lesbian, bisexual, and transgender people: A
  national perspective.
\newblock Technical report, National Gay and Lesbian Task Force Policy
  Institute, 2003.
\newblock URL
  \url{http://www.thetaskforce.org/downloads/reports/reports/CampusClimate.pdf}.

\bibitem[Rankin et~al.(2010)Rankin, Weber, Blumenfeld, and Frazer]{rankin:2010}
S.~R. Rankin, G.~Weber, W.~Blumenfeld, and S.~Frazer.
\newblock State of higher education for lesbian, gay, bisexual and transgender
  people.
\newblock Technical report, Campus Pride, 2010.
\newblock URL
  \url{https://www.campuspride.org/wp-content/uploads/campuspride2010lgbtreportssummary.pdf}.

\bibitem[pca(2012)]{pcast-engagetoexcel}
Engage to excel: Producing one million additional college graduates with
  degrees in science, technology, engineering, and mathematics.
\newblock Technical report, President's Council of Advisor on Science and
  Technology, 2012.
\newblock URL
  \url{https://obamawhitehouse.archives.gov/sites/default/files/microsites/ostp/pcast-engage-to-excel-final_2-25-12.pdf}.

\bibitem[Reason et~al.(2010)Reason, Cox, Quaye, and Terenzini]{reason:2010}
R.~D. Reason, B.~E. Cox, B.~R.~L. Quaye, and P.~T. Terenzini.
\newblock Faculty and institutional factors that promote student encounters
  with difference in first-year courses.
\newblock \emph{Rev. Higher Ed.}, 33:\penalty0 391, 2010.
\newblock \doi{10.1353/rhe.0.0137}.

\bibitem[Smith et~al.(2010)Smith, Parr, Woods, Bauer, and Abraham]{smith:2010}
H.~Smith, R.~Parr, R.~Woods, B.~Bauer, and T.~Abraham.
\newblock Five years after graduation: Undergraduate cross-group friendships
  and multicultural curriculum predict current attitudes and activities.
\newblock \emph{J. College Student Dev.}, 51:\penalty0 385, 2010.
\newblock \doi{10.1353/csd.0.0141}.

\bibitem[Love(1997)]{love:1997}
P.~Love.
\newblock Contradiction and paradox: Attempting to change the culture of sexual
  orientation at a small catholic college.
\newblock \emph{Rev. Higher Ed.}, 20:\penalty0 381, 1997.
\newblock \doi{10.1353/rhe.1997.0009}.

\bibitem[mas(1993)]{massachusetts-cgly:1993}
Making colleges and universities safe for gay and lesbian students: Report and
  recommendations.
\newblock Technical report, Massachusetts Governor's Commission on Gay and
  Lesbian Youth, 1993.
\newblock URL \url{http://www.mass.gov/cgly/Education_Report.pdf}.

\bibitem[cam()]{campusprideindex}
{Campus Pride Index: National Listing of LGBTQ-Friendly Colleges and
  Universities}.
\newblock URL \url{http://www.campusprideindex.org}.

\bibitem[Hanna(2016)]{hanna:2016}
A.~Hanna.
\newblock Being transgender on the job market.
\newblock \emph{Inside Higher Ed}, July 2016.
\newblock URL
  \url{https://www.insidehighered.com/advice/2016/07/15/challenge-being-transgender-academic-job-market-essay}.

\bibitem[Neunzert(2017)]{pronoun_spec}
A.~Neunzert.
\newblock Specifying your pronouns: an introductory resource, 2017.
\newblock URL
  \url{http://blog.lgbtphysicists.org/2017/05/specifying-your-pronouns-introductory.html}.

\bibitem[Moran(2017)]{moran:2017}
B.~Moran.
\newblock Is science too straight? {LGBTQ+ issues in STEM diversity}.
\newblock \emph{Boston University Research}, June 2017.
\newblock URL
  \url{http://www.bu.edu/research/articles/lgbt-issues-stem-diversity/}.

\bibitem[Flanagan(2017)]{flanagan:2017}
P.~Flanagan.
\newblock Becoming a woman.
\newblock \emph{Chronicle of Higher Education}, November 2017.
\newblock URL
  \url{http://www.chronicle.com/article/Becoming-a-Woman/241766?cid=wcontentgrid_hp_2}.

\bibitem[Bilimoria and Stewart(2009{\natexlab{a}})]{bilimoria:2009}
D.~Bilimoria and A.~J. Stewart.
\newblock ''don't ask, don't tell'': The academic climate for lesbian, gay,
  bisexual, and transgender faculty in science and engineering.
\newblock \emph{NWSA Journal}, 21\penalty0 (2):\penalty0 85,
  2009{\natexlab{a}}.
\newblock URL \url{http://www.jstor.org/stable/20628175}.

\bibitem[Cech and Waidzunas(2011)]{cech:2011}
E.~A. Cech and T.~J. Waidzunas.
\newblock Navigating the heteronormativity of engineering: the experiences of
  lesbian, gay, and bisexual students.
\newblock \emph{Eng. Studies}, 3:\penalty0 1, 2011.
\newblock \doi{10.1080/19378629.2010.545065}.

\bibitem[Stout and Dasgupta(2011)]{stout:2011}
J.~G. Stout and N.~Dasgupta.
\newblock When \textit{He} doesn't mean \textit{You}: Gender-exclusive language
  as ostracism.
\newblock \emph{Personality and Social Psychology Bulletin}, 36:\penalty0 757,
  2011.
\newblock \doi{10.1177/0146167211406434}.

\bibitem[nam({\natexlab{a}})]{namechanges:michigan}
{Name Changes at the University of Michigan}, {\natexlab{a}}.
\newblock URL \url{http://www.itcs.umich.edu/itcsdocs/r1461/}.

\bibitem[nam({\natexlab{b}})]{namechanges:vermont}
{Name changes at the University of Vermont}, {\natexlab{b}}.
\newblock URL
  \url{http://www.uvm.edu/~rgweb/?Page=policiesandprocedures/p_preferredname.html}.

\bibitem[Woodford et~al.(2012)Woodford, Howell, Silverschanz, and
  Yu]{woodford:2012}
M.~R. Woodford, M.~L. Howell, P.~Silverschanz, and L.~Yu.
\newblock ``that's so gay!'': {Examining} the covariates of hearing this
  expression among gay, lesbian, and bisexual college students.
\newblock \emph{J. Am. College Health}, 60:\penalty0 429, 2012.
\newblock \doi{10.1080/07448481.2012.673519}.

\bibitem[Feder(2015)]{feder:2015}
T.~Feder.
\newblock {LGBT physicists: The interviews}.
\newblock \emph{Physics Today}, 2015.
\newblock \doi{10.1063/PT.5.9034}.

\bibitem[Marsh et~al.(2016)Marsh, S\'err\'ano, MacDonald-Dennis, McCoy, Lewis,
  and Pierce]{marsh:2016}
J.~Marsh, B.~C. S\'err\'ano, C.~MacDonald-Dennis, S.~McCoy, T.~Lewis, and
  I.~Pierce.
\newblock Recommendations for supporting trans and queer students of color.
\newblock Technical report, {Consortium of Higher Education LGBT Resource
  Professionals}, 2016.
\newblock
  \url{https://lgbtcampus.memberclicks.net/assets/tqsoc%20support%202016.pdf}.

\bibitem[lgb()]{lgbtphys_outlist}
{LGBT$+$ Physicists Outlist}.
\newblock URL \url{http://lgbtphysicists.org/outlist.html}.

\bibitem[ast({\natexlab{a}})]{astro-outlist}
{The Outlist of Lesbian, Gay, Bisexual, Transgender, Intersex, Queer, and Ally
  Astronomers}, {\natexlab{a}}.
\newblock URL \url{https://astro-outlist.github.io/}.

\bibitem[Suri(4 September 2015)]{nyt:sciencesostraight}
M.~Suri.
\newblock Why is science so straight?
\newblock \emph{New York Times}, 4 September 2015.
\newblock URL
  \url{https://www.nytimes.com/2015/09/05/opinion/manil-suri-why-is-science-so-straight.html}.

\bibitem[maa()]{maa-awards-bestpractices}
Avoiding implicit bias: Guidelines for {MAA} selection committees.
\newblock URL
  \url{https://www.maa.org/sites/default/files/pdf/ABOUTMAA/avoiding_implicit_bias.pdf}.

\bibitem[Padilla(1994)]{pad94}
Amado~M. Padilla.
\newblock Ethnic minority scholars, research, and mentoring: Current and future
  issues.
\newblock \emph{Educational Researcher}, 23\penalty0 (4):\penalty0 24--27,
  1994.
\newblock URL
  \url{http://journals.sagepub.com/doi/pdf/10.3102/0013189X023004024}.

\bibitem[Bilimoria and Stewart(2009{\natexlab{b}})]{bs09}
Diana Bilimoria and Abigail~J. Stewart.
\newblock "don't ask, don't tell": The academic climate for lesbian,
  gay,bisexual, and transgender faculty in science and engineering.
\newblock \emph{Feminist Formations}, 21\penalty0 (2):\penalty0 85--103,
  2009{\natexlab{b}}.

\bibitem[June(2015)]{jun15}
Audrey~Williams June.
\newblock The invisible labor of minority professors.
\newblock \emph{The Chronicle of Higher Education}, 62\penalty0 (11), 2015.
\newblock URL
  \url{http://www.chronicle.com/article/The-Invisible-Labor-of/234098}.

\bibitem[{Social Sciences Feminist Network Research Interest
  Group}(2017)]{ssfn17}
{Social Sciences Feminist Network Research Interest Group}.
\newblock The burden of invisible work in academia: Social inequalities and
  time use in five university departments.
\newblock \emph{Humboldt Journal of Social Relations}, 39\penalty0
  (39):\penalty0 228--245, 2017.
\newblock URL \url{http://www.jstor.org/stable/90007882}.

\bibitem[Blackwell(1988)]{bla88}
James~E. Blackwell.
\newblock Faculty issues: The impact of minorities.
\newblock \emph{The Review of Higher Education}, 11\penalty0 (4):\penalty0
  417--434, 1988.

\bibitem[Bollinger(2007)]{bol07}
L.~C. Bollinger.
\newblock Why diversity matters.
\newblock \emph{Chronicle of Higher Education}, 35\penalty0 (39):\penalty0 B20,
  2007.

\bibitem[dal()]{dallasnews:facultydiversification}
{UT Dallas can't diversity faculty fast enough to keep up with student body}.
\newblock URL
  \url{https://www.dallasnews.com/news/education/2015/07/31/ut-dallas-can-t-diversify-faculty-fast-enough-to-keep-up-with-student-body}.

\bibitem[Taylor et~al.(2010)Taylor, Apprey, Hill, McGRann, and Wang]{tah+10}
Orlando Taylor, Cheryl~Burgan Apprey, George Hill, Loretta McGRann, and
  Jaianping Wang.
\newblock Diversifying the faculty.
\newblock \emph{Peer Review}, 12\penalty0 (3), 2010.
\newblock URL
  \url{https://www.aacu.org/publications-research/periodicals/diversifying-faculty}.

\bibitem[Grollman(2015)]{grollman:invisiblelabor}
E.~A. Grollman.
\newblock {Invisible Labor: Exploitation of Scholars of Color in Academia},
  2015.
\newblock URL
  \url{https://conditionallyaccepted.com/2015/12/15/invisible-labor}.

\bibitem[Baez(1999)]{bae99}
Benjamin Baez.
\newblock Faculty of color and traditional notions of service.
\newblock \emph{Thought and Action: The NEA Higher Education Journal},
  15\penalty0 (2):\penalty0 131--138, 1999.
\newblock URL
  \url{http://www.nea.org/assets/img/PubThoughtAndAction/TAA_99Fal_16.pdf}.

\bibitem[Banks(2000)]{ban00}
William~M. Banks.
\newblock Race-related service and faculty of color: Conceptualizing critical
  agency in academe.
\newblock \emph{Higher Education}, 39:\penalty0 363--391, 2000.

\bibitem[Bird et~al.(2004)Bird, Litt, and Wang]{blw04}
Sharon Bird, Jacquelyn Litt, and Yong Wang.
\newblock Creating status of women rreport: Institutional housekeeping as
  `women's work'.
\newblock \emph{NWSA Journal}, 16\penalty0 (1):\penalty0 194--206, 2004.

\bibitem[Monroe et~al.(2010)Monroe, Ozyurt, Wrigley, and Alexander]{mowa10}
Kristen~Renwick Monroe, Saba Ozyurt, Ted Wrigley, and Amy Alexander.
\newblock Gender equality in academia: Bad news from the trenches, and some
  possible solutions.
\newblock \emph{The Sociological Quaterly}, 51\penalty0 (2):\penalty0 179--204,
  2010.

\bibitem[Joseph and Hirshfield(2011)]{jh11}
Tiffany~D. Joseph and Laura~E. Hirshfield.
\newblock Why don't you get somebody new to do it?' race and cultural taxation
  in the academy.
\newblock \emph{Ethnic and Racial Studies}, 34\penalty0 (1):\penalty0 121--141,
  2011.
\newblock \doi{10.1080/01419870.2010.496489}.

\bibitem[Matthews(2016)]{mat16}
Patricia Matthews, editor.
\newblock \emph{Written/Unwritten: Diversity and the Hidden Truths of Tenure}.
\newblock The University of North Carolina Press, 2016.
\newblock ISBN 146962771X.

\bibitem[y~Muhs et~al.(2012)y~Muhs, Niemann, Gonzalez, and Harris]{ymngh12}
Gabriella~Guierrez y~Muhs, Yolanda~Flores Niemann, Carmen~G. Gonzalez, and
  Angela~P. Harris, editors.
\newblock \emph{Presumed Incompetent: The Intersections of Race and Class for
  Women in Academia}.
\newblock Utah State University Press, 2012.

\bibitem[Sandstr{\"o}m and H{\"a}llsten(2008)]{sh08}
U.~Sandstr{\"o}m and M.~H{\"a}llsten.
\newblock Persistent nepotism in peer-review.
\newblock \emph{Scientometrics}, 74\penalty0 (2):\penalty0 175--189, 2008.

\bibitem[Evans(2002)]{evans:2002}
N.~J. Evans.
\newblock The impact of an {LGBT} safe zone project on campus climate.
\newblock \emph{J. College Student Dev.}, 43\penalty0 (4):\penalty0 522, 2002.

\bibitem[James et~al.(2016)James, Herman, Rankin, Keisling, Mottet, and
  Anafi]{ustranssurvey:2016}
S.~E. James, J.~L. Herman, S.~Rankin, M.~Keisling, L.~Mottet, and M.~Anafi.
\newblock \emph{{The Report of the 2015 U.S. Transgender Survey}}.
\newblock National Center for Transgender Equality, Washington, DC, 2016.
\newblock URL \url{http://www.ustranssurvey.org/report}.

\bibitem[{National Conference of State
  Legislatures}({\natexlab{a}})]{tracker-bathroombill}
{National Conference of State Legislatures}.
\newblock ``bathroom bill'' legislative tracking, {\natexlab{a}}.
\newblock URL
  \url{http://www.ncsl.org/research/education/-bathroom-bill-legislative-tracking635951130.aspx}.

\bibitem[{National Conference of State
  Legislatures}({\natexlab{b}})]{tracker-employmentdiscrimination}
{National Conference of State Legislatures}.
\newblock State laws on employment-related discrimination, {\natexlab{b}}.
\newblock URL
  \url{http://www.ncsl.org/research/labor-and-employment/discrimination-employment.aspx}.

\bibitem[Rosa and Mensah(2016)]{rosa:2016}
K.~Rosa and F.~M. Mensah.
\newblock Educational pathways of black women physicists: Stories of
  experiencing and overcoming obstacles in life.
\newblock \emph{Phys. Rev. Phys. Educ. Res.}, 12:\penalty0 020113, 2016.
\newblock \doi{10.1103/PhysRevPhysEducRes.12.020113}.

\bibitem[Barthelemy et~al.(2016)Barthelemy, McCormick, and
  Henderson]{barthelemy:2016}
R.~S. Barthelemy, M.~McCormick, and C.~Henderson.
\newblock Gender discrimination in physics and astronomy: Graduate student
  experiences of sexism and gender microaggressions.
\newblock \emph{Phys. Rev. Phys. Educ. Res.}, 12:\penalty0 020119, 2016.
\newblock \doi{10.1103/PhysRevPhysEducRes.12.020119}.

\bibitem[of~the Vice Provost~for Graduate~Education()]{sample_contract}
Stanford University~Office of~the Vice Provost~for Graduate~Education.
\newblock Advising \& mentoring.
\newblock URL
  \url{https://vpge.stanford.edu/academic-guidance/advising-mentoring}.

\bibitem[Hughes(2018)]{hughes:2018}
B.~E. Hughes.
\newblock {Coming out in STEM: Factors affecting retention of sexual minority
  STEM students}.
\newblock \emph{Sci. Adv.}, 4:\penalty0 eaao6373, 2018.
\newblock \doi{10.1126/sciadv.aao6373}.

\bibitem[Ambrose et~al.(2010)Ambrose, Bridge, M., C., Norman, and
  Mayer]{abm+10}
S.~A. Ambrose, M.~W. Bridge, DiPietro M., Lovett~M. C., M.~K. Norman, and R.~E.
  Mayer.
\newblock \emph{How Learning Works: 7 Research-based Principles for Smat
  Teaching}.
\newblock Jossey-Bass, 1st ed. edition, 2010.

\bibitem[DeSurra and Church(1994)]{dc94}
C.~DeSurra and K.~A. Church.
\newblock Unlocking the classroom closet: Privileging the marginalized voices
  of gay/lesbian college students.
\newblock Annual Meeting of the Speech Communication Association, 1994.

\bibitem[Steele and Aronson(1995)]{sa95}
C.~M. Steele and J.~R. Aronson.
\newblock Stereotype threat and the intellectual performance of african
  americans.
\newblock \emph{Journal of Personality and Social Psychology}, 69\penalty0
  (5):\penalty0 797--811, 1995.

\bibitem[Major et~al.(1998)Major, Spencer, Schmader, Wolfe, and
  Crocker]{mss+98}
B.~Major, S.~Spencer, T.~Schmader, C.~Wolfe, and J.~Crocker.
\newblock Coping with negative stereotype about intellectual performance: The
  role of psychological disengagement.
\newblock \emph{Personality and Social Psychology Bulletin}, 24\penalty0
  (1):\penalty0 34--50, 1998.

\bibitem[Lorenzo(2006)]{lor06}
M.~Lorenzo.
\newblock Reducing the gender gap in the physics classroom.
\newblock \emph{American Journal of Physics}, 74:\penalty0 118, 2006.
\newblock \doi{https://doi.org/10.1119/1.2162549}.

\bibitem[Provitera-McGlynn(2001)]{provitera2001successful}
A.~Provitera-McGlynn.
\newblock \emph{Successful Beginnings for College Teaching: Engaging Your
  Students from the First Day}.
\newblock Publicaffairs Reports. Atwood Pub., 2001.
\newblock ISBN 9781891859380.
\newblock URL \url{https://books.google.com/books?id=ASUlAQAAIAAJ}.

\bibitem[Due(2014)]{due_who_2014}
Karin Due.
\newblock Who is the competent physics student? {A} study of
  students{\textquoteright} positions and social interaction in small-group
  discussions.
\newblock \emph{Cultural Studies of Science Education}, 9\penalty0
  (2):\penalty0 441--459, June 2014.
\newblock ISSN 1871-1502, 1871-1510.
\newblock \doi{10.1007/s11422-012-9441-z}.
\newblock URL
  \url{https://link.springer.com/article/10.1007/s11422-012-9441-z}.

\bibitem[Hofer(2015)]{hofer_studying_2015}
Sarah~I. Hofer.
\newblock Studying {Gender} {Bias} in {Physics} {Grading}: {The} role of
  teaching experience and country.
\newblock \emph{International Journal of Science Education}, 37\penalty0
  (17):\penalty0 2879--2905, November 2015.
\newblock ISSN 0950-0693.
\newblock \doi{10.1080/09500693.2015.1114190}.
\newblock URL \url{https://doi.org/10.1080/09500693.2015.1114190}.

\bibitem[Badgett et~al.(2009)Badgett, Sears, Lau, and Ho]{badgett_bias_2009}
M.V. Badgett, Brad Sears, Holning Lau, and Deborah Ho.
\newblock Bias in the {Workplace}: {Consistent} {Evidence} of {Sexual}
  {Orientation} and {Gender} {Identity} {Discrimination} 1998-2008.
\newblock \emph{Chicago-Kent Law Review}, 84\penalty0 (2):\penalty0 559, April
  2009.
\newblock ISSN 0009-3599.
\newblock URL
  \url{https://scholarship.kentlaw.iit.edu/cklawreview/vol84/iss2/7}.

\bibitem[{John M. Malouff} et~al.(2013){John M. Malouff}, {Ashley J. Emmerton},
  and {Nicola S. Schutte}]{john_m._malouff_risk_2013}
{John M. Malouff}, {Ashley J. Emmerton}, and {Nicola S. Schutte}.
\newblock The {Risk} of a {Halo} {Bias} as a {Reason} to {Keep} {Students}
  {Anonymous} {During} {Grading}.
\newblock \emph{Teaching of Psychology}, 40\penalty0 (3):\penalty0 233--237,
  July 2013.
\newblock ISSN 0098-6283.
\newblock \doi{10.1177/0098628313487425}.
\newblock URL \url{https://doi.org/10.1177/0098628313487425}.

\bibitem[Heller and Hollabaugh(1992)]{heller:1992}
P.~Heller and M.~Hollabaugh.
\newblock Teaching problem solving through cooperative grouping. {Part 2:
  Designing} problems and structuring groups.
\newblock \emph{Am. J. Phys.}, 60:\penalty0 637, 1992.
\newblock \doi{10.1119/1.17118}.

\bibitem[Lewis et~al.(2016)Lewis, Stout, Pollock, Finkelstein, and
  Ito]{lewis:2016}
K.~L. Lewis, J.~G. Stout, S.~J. Pollock, N.~D. Finkelstein, and T.~A. Ito.
\newblock Fitting in or opting out: A review of key social-psychological
  factors influencing a sense of belonging for women in physics.
\newblock \emph{Phys. Rev. Phys. Educ. Res.}, 12:\penalty0 020110, 2016.
\newblock \doi{10.1103/PhysRevPhysEducRes.12.020110}.

\bibitem[Dunbar et~al.(2017)Dunbar, Sontag-Padilla, Ramchand, Seelam, and
  Stein]{dsr+17}
Michael~S. Dunbar, Lisa Sontag-Padilla, Rajeev Ramchand, Rachana Seelam, and
  Bradley~D. Stein.
\newblock Mental health service utilization among lesbian, gay, bisexual, and
  questioning or queer college students.
\newblock \emph{Journal of Adolescent Health}, 61\penalty0 (3):\penalty0
  294--301, 2017.
\newblock \doi{https://doi.org/10.1016/j.jadohealth.2017.03.008}.

\bibitem[qpr()]{qpr}
Qpr institute.
\newblock URL \url{http://qprinstitute.com}.

\bibitem[Wang(2016)]{wang:2016}
L.~Wang.
\newblock {LGBT chemists seek a place at the bench}.
\newblock \emph{Chem. \& Eng. News}, October 2016.
\newblock URL \url{https://cen.acs.org/articles/94/i41/place-bench.html}.

\bibitem[{Peeples}(2018)]{peeples:2018}
M.~{Peeples}.
\newblock {Postdoctoral Mentoring at the Space Telescope Science Institute}.
\newblock In \emph{American Astronomical Society Meeting Abstracts 231}, volume
  231 of \emph{American Astronomical Society Meeting Abstracts}, page 156.04,
  jan 2018.

\bibitem[dea()]{dearcolleagueletter2017}
Dear colleague letter.
\newblock URL
  \url{https://www2.ed.gov/about/offices/list/ocr/letters/colleague-title-ix-201709.pdf}.

\bibitem[Schiebinger et~al.(2008)Schiebinger, Henderson, and
  Gilmartin]{schiebinger2008dual}
Londa~L Schiebinger, Andrea~Davies Henderson, and Shannon~K Gilmartin.
\newblock \emph{Dual-career academic couples: What universities need to know}.
\newblock Michelle R. Clayman Institute for Gender Research, Stanford
  University, 2008.

\bibitem[her()]{herc}
Higher education recruitment consortium.
\newblock URL \url{http://hercjobs.org}.

\bibitem[ast({\natexlab{b}})]{astrobetter:leavepolicies}
Astrobetter leave policies, {\natexlab{b}}.
\newblock URL
  \url{http://www.astrobetter.com/wiki/tiki-index.php?page=Leave+Policies}.

\bibitem[Flaherty(2016)]{flaherty:2016}
C.~Flaherty.
\newblock Bias against female instructors, January 2016.
\newblock
  https://www.insidehighered.com/news/2016/01/11/new-analysis-offers-more-evidence-against-student-evaluations-teaching.

\bibitem[{MacNell} et~al.(2015){MacNell}, Driscoll, and Hunt]{macnell:2015}
L.~{MacNell}, A.~Driscoll, and A.~N. Hunt.
\newblock What's in a name: {Exposing} gender bias in student ratings of
  teaching.
\newblock \emph{Innov. High. Educ.}, 40:\penalty0 291, 2015.
\newblock \doi{https://doi.org/10.1007/s10755-014-9313-4}.

\bibitem[Boring et~al.(2016)Boring, Ottoboni, and Stark]{boring:2016}
A.~Boring, K.~Ottoboni, and P.~B. Stark.
\newblock Student evaluations of teaching (mostly) do not measure teaching
  effectiveness.
\newblock \emph{{ScienceOpen Research SOR-EDU}}, 2016.
\newblock \doi{http://dx.doi.org/10.14293/S2199-1006.1.SOR-EDU.AETBZC.v1}.

\bibitem[Reid(2010)]{reid:2010}
L.~D. Reid.
\newblock The role of perceived race and gender in the evaluation of college
  teaching on {RateMyProfessors.com}.
\newblock \emph{J. Diversity Higher Educ.}, 3:\penalty0 137, 2010.

\bibitem[Merritt(2012)]{merritt:2012}
D.~J. Merritt.
\newblock Bias, the brain, and student evaluations of teaching.
\newblock \emph{St. John's Law Review}, 82:\penalty0 235, 2012.
\newblock URL \url{http://scholarship.law.stjohns.edu/lawreview/vol82/iss1/6}.

\bibitem[Cesario and Crawford(2003)]{cesario:2003}
J.~Cesario and I.~Crawford.
\newblock The effect of homosexuality on perceptions of persuasiveness and
  trustworthiness.
\newblock \emph{J. Homosexuality}, 43:\penalty0 93, 2003.
\newblock \doi{10.1300/J082v43n02_06}.

\bibitem[Russ et~al.(2002)Russ, Simonds, and Hunt]{russ:2002}
T.~Russ, C.~Simonds, and S.~Hunt.
\newblock Coming out in the classroom... an occupational hazard?: The influence
  of sexual orientation on teacher credibility and perceived student learning.
\newblock \emph{Commun. Educ.}, 51:\penalty0 311, 2002.
\newblock \doi{10.1080/03634520216516}.

\bibitem[Stainburn(30 July 2013)]{nyt_gayquestion:2013}
Samantha Stainburn.
\newblock The gay question: Check one.
\newblock \emph{New York Times}, 30 July 2013.
\newblock URL
  \url{http://www.nytimes.com/2013/08/04/education/edlife/more-college-applications-ask-about-sexual-identity.html?_r=0}.

\bibitem[Office~for Civil~Rights({\natexlab{a}})]{titleix-memo}
U.S. Department of~Education Office~for Civil~Rights.
\newblock Questions and answers on {Title IX} and sexual violence,
  {\natexlab{a}}.
\newblock URL
  \url{https://www2.ed.gov/about/offices/list/ocr/docs/qa-201404-title-ix.pdf}.

\bibitem[Office~for Civil~Rights({\natexlab{b}})]{titleix-depted}
U.S. Department of~Education Office~for Civil~Rights.
\newblock Title {IX} and sex discrimination, {\natexlab{b}}.
\newblock URL
  \url{https://www2.ed.gov/about/offices/list/ocr/docs/tix_dis.html}.

\bibitem[Center()]{titleix-natwomenlawctr}
National Women's~Law Center.
\newblock Title {IX} information.
\newblock URL \url{http://www.titleix.info/}.

\bibitem[{University of California}()]{titleix:samplecomplaint}
{University of California}.
\newblock Discrimination, harassment, retaliation complaint form.
\newblock URL
  \url{http://ucop.edu/local-human-resources/_files/op-life/discrimination-complaint-form.pdf}.

\bibitem[Chiang et~al.(2016)Chiang, Gutierrez, Kriek, Laskar, Quataert,
  Trotter, and Veale]{berkeley-climatesurvey-2016}
E.~Chiang, Sara Gutierrez, Mariska Kriek, Tanmoy Laskar, Eliot Quataert,
  Lochland Trotter, and Melanie Veale.
\newblock 2016 astronomy climate survey.
\newblock Technical report, University of California, Berkeley, 2016.
\newblock URL \url{http://astro.berkeley.edu/ACS16_Report.pdf}.

\bibitem[Department~of Astronomy()]{titleix-berkeley-reporters}
Berkeley Department~of Astronomy, University of~California.
\newblock Resources to prevent and respond to sexual harassment and
  discrimination.
\newblock URL
  \url{http://astro.berkeley.edu/department-resources/reporting-harassment}.

\bibitem[Council(2017)]{studentdiversity:michigan}
The Multicultural~Leadership Council.
\newblock Pay students for diversity labor.
\newblock \emph{The Michigan Daily}, 2017.
\newblock URL
  \url{https://www.michigandaily.com/section/mic/pay-students-diversity-labor}.

\bibitem[Parnell(2016)]{studentdiversityactivism:parnell}
Amelia Parnell.
\newblock Affirming racial diversity: Student affairs as a change agent.
\newblock \emph{Higher Education Today}, 2016.
\newblock URL
  \url{https://www.higheredtoday.org/2016/06/29/affirming-racial-diversity-student-affairs-as-a-change-agent/}.

\bibitem[stu({\natexlab{a}})]{studentdiversityinternships:berkeley}
{Gender Equity Resource Center at UC Berkeley}, {\natexlab{a}}.
\newblock URL \url{https://campusclimate.berkeley.edu/students/ejce/geneq}.

\bibitem[stu({\natexlab{b}})]{studentdiversityinternships:ohsu}
{Internships at Center for Diversity and Inclusion at Oregon Health and Science
  University}, {\natexlab{b}}.
\newblock URL
  \url{https://www.ohsu.edu/xd/about/vision/center-for-diversity-inclusion/academic-resources/internships/internships-at-cdi.cfm}.

\bibitem[stu({\natexlab{c}})]{studentdiversityinternships:bradley}
{Diversity and Inclusion Internship Program, Bradley University},
  {\natexlab{c}}.
\newblock URL \url{https://www.bradley.edu/campuslife/diversity/internship/}.

\bibitem[stu({\natexlab{d}})]{studentdiversityinternships:aquinas}
{Diversity Assistants at Center for Diversity, Inclusion and Equity at Aquinas
  College}, {\natexlab{d}}.
\newblock URL
  \url{https://www.aquinas.edu/life-aq/center-diversity-inclusion/diversity-assistants}.

\bibitem[stu({\natexlab{e}})]{studentdiversityinternships:wesleyan}
{Dwight Greene '70 Internship for Equity and Inclusion, Wesleyan University},
  {\natexlab{e}}.
\newblock URL
  \url{http://www.wesleyan.edu/inclusion/internships/dg_internship_Equityand%20Inclusion.html}.

\bibitem[stu({\natexlab{f}})]{studentdiversityinternships:delaware}
{Office of Equity and Inclusion Internship Opportunities, University of
  Delaware}, {\natexlab{f}}.
\newblock URL \url{https://sites.udel.edu/oei/get-involved/}.

\bibitem[stu({\natexlab{g}})]{studentdiversityinternships:georgetown}
{Center for Multicultural Equity and Access, Georgetown University},
  {\natexlab{g}}.
\newblock URL \url{https://cmea.georgetown.edu/interns}.

\bibitem[stu({\natexlab{h}})]{studentdiversityinternships:lynchburg}
{Office of Equity and Inclusion -- Get Involved at Lynchburg College},
  {\natexlab{h}}.
\newblock URL
  \url{https://www.lynchburg.edu/aboutlc/office-of-equity-and-inclusion/get-involved/}.

\bibitem[{National Center for Transgender
  Equality}()]{selectiveservice:transequality}
{National Center for Transgender Equality}.
\newblock Selective service and transgender people.
\newblock URL
  \url{http://www.transequality.org/issues/resources/selective-service-and-transgender-people}.

\bibitem[Strayhorn et~al.(2015)Strayhorn, Johnson, Henderson, and
  Tillman-Kelly]{strayhorn:2015}
T.~L. Strayhorn, R.~M. Johnson, T.~S. Henderson, and D.~L. Tillman-Kelly.
\newblock Beyond coming out: New insights about {GLBQ} college students of
  color.
\newblock Technical report, Center for Higher Education Enterprise, The Ohio
  State University, 2015.
\newblock URL
  \url{https://chee.osu.edu/wordpress/wp-content/uploads/2016/04/beyond-coming-out.pdf}.

\bibitem[Kessler et~al.(2002)Kessler, Andrews, Colpe, Hiripi, Mroczek, Normand,
  Walters, and Zaslavsky]{kessler:2002}
R.~C. Kessler, G.~Andrews, L.~J. Colpe, E.~Hiripi, D.~K. Mroczek, S.~L.
  Normand, E.~E. Walters, and A.~M. Zaslavsky.
\newblock Short screening scales to monitor population prevalences and trends
  in non-specific psychological distress.
\newblock \emph{Psychol. Med.}, 32:\penalty0 959, 2002.
\newblock \doi{10.1017/S0033291702006074}.

\bibitem[{Harvard University Information Technology}()]{harvard:username}
{Harvard University Information Technology}.
\newblock Can {FAS} account names be changed?
\newblock URL
  \url{https://huit.harvard.edu/faq/can-fas-account-names-be-changed}.

\bibitem[University~of California()]{username:calnet}
Berkeley~Calnet University~of California.
\newblock Change calnet id.
\newblock URL
  \url{https://calnetweb.berkeley.edu/calnet-me/manage-my-account/change-calnet-id}.

\bibitem[O'Hara(2015)]{ohara:2015}
M.~E. O'Hara.
\newblock 'trans broken arm syndrome' and the way our healthcare system fails
  trans people.
\newblock \emph{The Daily Dot}, 2015.
\newblock URL
  \url{https://www.dailydot.com/irl/trans-broken-arm-syndrome-healthcare/}.

\bibitem[Jones(June 22, 2017)]{gallup:lgbtmarriage:2017}
J.~M. Jones.
\newblock {In U.S., 10.2\% of LGBT Adults Now Married to Same-Sex Spouse}, June
  22, 2017.
\newblock URL
  \url{http://news.gallup.com/poll/212702/lgbt-adults-married-sex-spouse.aspx}.

\bibitem[hrc()]{hrc-surveyqns}
Collecting transgender-inclusive gender data in workplace and other surveys.
\newblock URL
  \url{https://www.hrc.org/resources/collecting-transgender-inclusive-gender-data-in-workplace-and-other-surveys}.

\bibitem[Ainsworth(2015)]{Ainsworth:2015}
Claire Ainsworth.
\newblock Sex redefined.
\newblock \emph{Nature}, 518\penalty0 (7539):\penalty0 288--291, Feb 2015.
\newblock \doi{10.1038/518288a}.
\newblock URL \url{http://dx.doi.org/10.1038/518288a}.

\bibitem[Traxler et~al.(2016)Traxler, Cid, Blue, and Barthelemy]{traxler:2016}
Adrienne~L. Traxler, Ximena~C. Cid, Jennifer Blue, and Ram\'on Barthelemy.
\newblock Enriching gender in physics education research: A binary past and a
  complex future.
\newblock \emph{Phys. Rev. Phys. Educ. Res.}, 12:\penalty0 020114, Aug 2016.
\newblock \doi{10.1103/PhysRevPhysEducRes.12.020114}.
\newblock URL
  \url{https://link.aps.org/doi/10.1103/PhysRevPhysEducRes.12.020114}.

\bibitem[int()]{intersex-faq}
What is intersex?
\newblock URL \url{http://www.isna.org/faq/what_is_intersex}.

\end{thebibliography}

\end{document}